\documentclass[preprint,authoryear,12ptcitep]{article}

\usepackage{amssymb}
\usepackage{natbib}

\usepackage{subfig}
\usepackage{graphicx}
\usepackage{amsmath}
\usepackage{array}

\usepackage{authblk}
\title{This is some thing}
\author[1,2]{S. Carvalho}
\author[3]{S. Gomes}
\author[1,3]{T. Barata}
\author[1,3]{A. Lourenço}
\author[1,3]{N. Peixinho}
\affil[1]{CITEUC - Centre for Earth and Space Research, University of Coimbra, 3040–004 Coimbra, Portugal}
\affil[2]{CMUC - Centre for Mathematics, University of Coimbra, 3001–501, Coimbra, Portugal}
\affil[3]{OGAUC - Geophysical and Astronomical Observatory, University of Coimbra,  3040–004, Coimbra, Portugal}

\title{Comparison of automatic methods to detect sunspots in the Coimbra Observatory spectroheliograms}

\begin{document}
\maketitle



%
%
%
%

\begin{abstract}
The Astronomical Observatory of the University of Coimbra has a huge collection of solar images, acquired daily since 1926. From the beginning, only spectroheliograms in the $Ca\,\textsc{ii}\,K$ line has been recorded, and since 1989 in the $H_\alpha$ line also. Such dataset requires efficient tools to detect and analyze solar activity features. The objective of this work is to create a tool that allows to automatic detect sunspots, umbra, and penumbra, that can be applied to the entire dataset. To achieve this, two different approaches have been developed, one based on mathematical morphology and another based on the intensities of the digital levels of the pixels. Both approaches were applied to a subset of images with features identified visually by an experimented solar observer. The performance of both methods was compared through the metrics Precision, Recall and F-score. Another evaluation was made based on the catalogs from Heliophysics Features Catalog and the SILSO catalogue. 
\end{abstract}





\section{Introduction}
\label{int}
The records of sunspot observations constitute, probably, the longest series of scientific data \citep{Wittmann}.  The first known record of sunspots dates to 364 B.C. made by Chinese observers \citep{Lin}. In the 17th century, Keppler, with the use of camera obscura, starts a new way of observing the Sun, but the revolutionary perspective of solar observations occurs latter, with the use of telescopes. From a very extensive list of observers \citep{doi:10.1002/asna.201512292}, Harriot (first datable sunspots observations, 1610), David and Johann  Fabricius in 1611 (first publication concerning sunspots, 1611), Scheiner (first continuous observations during several weeks, 1611), Marius (observations between 1617-1619), and Malapert (observations in 1612) are noteworthy. Also, in 1612, during the summer months, Galileo made systematic observations which were published in Istoria e Dimostrazioni Intorno Alle Macchie Solari e Loro Accidenti Rome.  In 1633, Malapert registered the observations made in October and December 1620, by a Guilemus Wely (born ca. 1600) a sunspot observer for Coimbra \citep{doi:10.1002/asna.201512292}. 

Traditionally, the record of sunspots was handmade drawn, and some details were also recorded, such as the number, position, and area of each sunspot. This systematic way of registering sunspots and their characteristics lead to the creation solar catalogs and databases. The first catalog was published by the Royal Greenwich Observatory, UK, between 1874 and 1976 \citep{baranyi1}. Other examples of solar observation catalogs are: Mount Wilson’s, USA \citep{Lefevre}, Debrecen's, Hungary \citep{baranyi2}, the Spanish observatories catalogs (\citet{aparicio}, \citet{Curto2016}), and the Coimbra Observatory catalog, Portugal \citep{Carrasco}.  

During the last century, and until today, with the success of several solar missions, the solar observations increased enormously. At the same time, image processing techniques have also increased, techniques that when applied to solar images allow getting information on solar activity in a prompt and efficient way (\citet{Gill}, \citet{doi:10.1029/2009SW000537}). It is, therefore, natural that digital catalogs would also emerge, being the EGSO (European Grid of Solar Observations) a good example of that (\citet{Fuller}, \citet{Zharkova2005a}). Another pioneer example is the Solar Monitor, which labels active regions on the Sun using NOAA's (National Oceanic and Atmospheric Association) numbers and heliographic positions \citep{Higgins}. 

Despite having continuously more data from new instruments, including space missions, it is nonetheless very important to maintain many older instruments working and keep using their data for several important reasons (\citet{hill}, \citet{ayres}). One of them is the long–term observations they have been performing- of  at least several decades-, which are crucial to determine the number and distribution of sunspots over the time, to understand the solar cycle, to predict the solar cycle and its implications on climate changes, and to monitor and forecast solar activity- allowing to obtain results much needed for Space Weather research (\citet{2000HvaOB..24..195V},\citet{2016ASPC..504..247V}). Moreover, ground–based observations allow us to preserve and extend consistent data sequences, and consistency is a key element when dealing with long datasets. 

There are several advantages in applying image processing techniques to solar observations, namely: precision, objectivity, and statistical significance, as pointed in the review work \citet{aschwanden}. On his work, Aschwanden refers the most common approaches to process solar images in order to detect solar features, including pre-processing techniques to correct effects due image acquisition, atmospheric effects common on ground-based images and during image registration. Concerning sunspots, the processing approaches more oftenly used are threshold techniques (\citet{Zharkov2005a}, \citet{Jewalikar}, \citet{6133028}), edge detection (\citet{Zharkov2005a}, \citet{Mohammed}), region growing \citep{Zharkova2005b}, mathematical morphology transforms (\citet{Zharkov2005b}, \citet{Curto2008}, \citet{carvalho2015}, \citet{Zhao}, \citet{Deepa}), neural networks \citep{Colak}, fuzzy sets (\citet{Fonte2009}, \citet{Gafeira2014}), and classification schemes (\citet{Nguyen}, \citet{Qahwaji}, \citet{Colak}). Hybrid methods, that include different approaches, have also been developed and can be found in \citet{dorotovic2014}, \citet{manish} and \citet{Qahwaji}. Another example of the integration of different methods is the work of \citet{6896184}, which combines morphological operators and region growing techniques to automatic detect sunspots and to differentiate the umbra and the penumbra. The performance of such techniques has also been the subject of analysis. \citet{Zharkova2005b} compares automatic approaches with manual analysis and proves the efficiency of the automatic techniques. 

The application of these techniques and their robustness contributes to the development of completly automatic tools, able to detect the sunspots and to extract their geometric characteristics (for instance, area and McIntosh classification). The Automatic Solar Activity Prediction (ASAP) software, the first automatic tool developed, by \citet{Colak2009}, is a good example \citep{7766421}.  The ASAP tool is also used to perform the automatic segmentation of umbra and penumbra \citep{2015JSWSC...5A..15A}. The Sunspot Tracking and Recognition Algorithm (STRARA), developed in 2008 by \citet{Watson2009}, also detects sunspots on long term observations. Another example is the Automatic Solar Synoptic Analyzer (ASSA), a software developed in 2013 (first version) by the Korean Space Weather Center of the Radio Research Agency and SELab (Republic of Korea). This tool identifies automatically the sunspot groups and classifies them according to the McIntosh classification. The comparison of different automatic methods was analyzed by \citet{Verbeeck2013} and \citet{carvalho2015}.  

This paper intends to contribute for an automatic detection of umbra and penumbra of sunspots acquired at the Geophysical and Astronomical Observatory of the University of Coimbra during the cycle 24. Two different approaches are presented here, one based on morphological transforms (MM) and another based on pixel intensity (PI). The evaluation of the performance of both methods is made using statistical metrics. The main objective is to define and choose one method to apply to the entire historical data set. The following section introduces the data used in this work. Section \ref{morph} describes the automatic method based on mathematical morphology, and the approach based on pixel intensities is described in section \ref{pint}. Data analysis and discussion of the results are made in section \ref{analysisdiscussion}. Finally, the conclusions are presented in the section \ref{conclusions}. 

\section{The spectroheliograms of Coimbra}
\label{dataCoimbra}
\begin{figure}[!ht]
	\centering
	\includegraphics[width=0.8\textwidth]{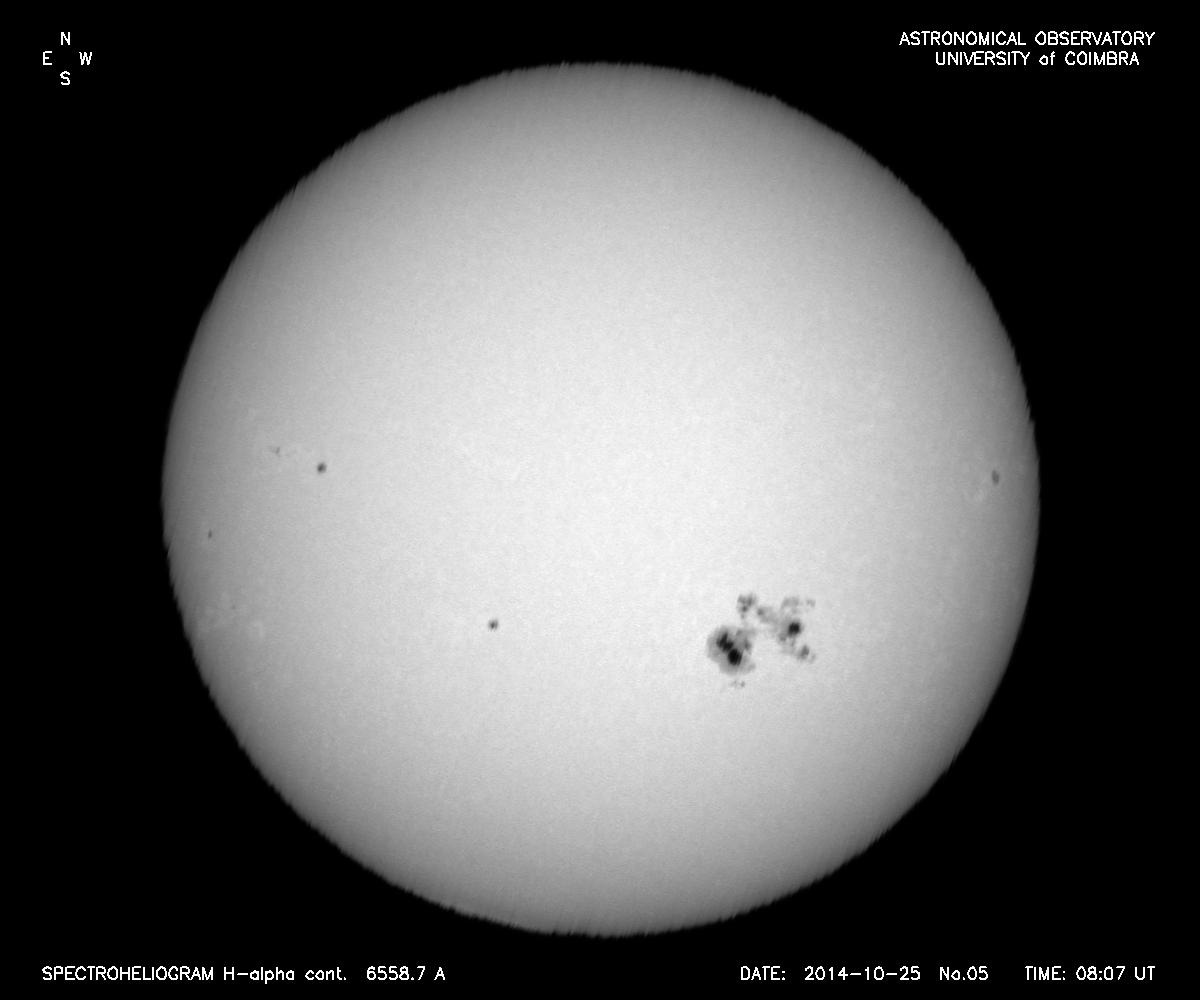}
	\caption{Example of a $H_\alpha$ spectroheliogram with sunspots (some of them with umbra and penumbra), acquired on the $25^{th}$ of October 2014.} \label{fig1}
\end{figure}

The Astronomical Observatory of the University of Coimbra- renamed Geophysical and Astronomical Observatory in 2013- has a collection of solar observations on a daily basis that spans near nine decades until today. Regular observations of the full solar disk in the spectral line of $Ca\,\textsc{ii}\,K$ started in 1926 and those in the $H_\alpha$ started in 1989 \citep{2010CEAB...34...47G}. This extense collection acquired with the same instrumental apparatus is presently entirely available in digital format. The image acquisition instrument, a spectroheliograph based on Deslandres principles, consists of coelostat, with a primary mirror of 0.4 m of diameter and a secondary one of 0.3 m, which sends the sunlight into an optical system with a slit, filters, collimators and a diffraction grid. The entire solar disk is swept, mechanically, across a slit (which takes 80 seconds to do the scan), being, therefore, not instantaneously recorded as a whole but recorded in “slices” onto a CCD (onto a photographic plate before 2007). 
An example of a $H_\alpha$ spectroheliogram acquired at the Observatory of Coimbra is shown in Fig. (\ref{fig1}). The image consists of the solar disk and some overwritten information related with the acquisition: orientation, place, spectral line, and date. 

Sunspots are temporary manifestations of magnetic field effects on the Sun. They are seen as dark areas in the photosphere due a strong magnetic field concentration, which diminishes convection hence decreasing the surface temperature. A nice example of sunspots can be seen on Fig. \ref{fig1}. The sunspots are constituted by two main regions: umbra and penumbra. The darker region on the central part of a sunspot is the umbra, where the magnetic field is stronger, being surrounded by a less dark region called penumbra. 
This work is based on a $H_\alpha$ data subset and aims to automatically detect sunspots and to automatically differentiate the umbra and the penumbra. For both methods, the data set used consists of 144 spectroheliograms, which are 8 bits digital images with 1200 x 1000 pixels, taken at the $H_\alpha$ continuum line. The set comprises images of the solar cycle 24 and chosen to represent the whole cycle: taken in different years and in different seasons. Additionally, for each image, an observer, with about 40 years of experience, delineated, manually, the umbra and penumbra regions in order to build the ground-truth data set used to validate the results obtained.
Prior to any automatic detection of sunspots on the recorded images, it is not only necessary to correctly identify the solar disk, evidently, but also to remove all the overwritten text since it hampers any automatic processing algorithm. Another aspect to consider is that one spectroheliogram results from the juxtaposition of multiple “slices” of the solar disk, which leads to a heterogeneous background. Therefore, the background does not have the same digital value over all pixels, although visually it seems that the solar disk is sitting on a homogeneous black background. It is also possible to have pixels inside and outside the solar disk with the same value. To overcome these problems, a pre-processing was applied to our set of images. It is also true that the solar disk is not a perfect circle, being slightly flatted over its rotation axis, but, relatively to the hundreds of pixels of the solar disk diameter, and not thousands, the flattening effect be neglected. So, the solar disk can be taken as perfectly circular, which facilitates the construction of the algorithms of sunspots detection. Both methods take these issues in consideration in the pre-processing step.

\section{Automatic detection of sunspots based on morphological transforms}
\label{morph}

In the mid 1960s, George Matheron and Jean Serra, from the École des Mines de Paris, France, wanted to describe geometric features in porous media \citep{Matheron1967}. The resolution of this problem gave rise to a new image analysis theory: the Mathematical Morphology. Since then, news developments allow to construct a solid framework (\citet{Matheron1975}, \citet{Serra}) with successful applications in different scientific fields (see \citet{Soille}, for a review), including solar physics \citep{aschwanden}. 

The essence of mathematical morphology consists in comparing features to be analyzed with some known object/shape, called the structuring element. The power of mathematical morphology resides on its versatility, like its applicability to both binary or greyscale images, the fact that operators can be applied  in one go or applied sequentially (to obtain more elaborated morphologic transformations, for specific ends), and its capability to deal with the geometry of complex shapes extracting, nonetheless, quantitative measurements like area, length, and sinuosity. All of the above make the application of mathematical morphology attractive for the detection of sunspots.

\subsection{Pre-processing data}
\label{morph:preproc}
The pre-processing of our 1200 x 1000 pixels’ images starts by applying the basic morphological operation \textit{closing} (or \textit{close}) over the original image using as structuring element a disk of 10 pixels in diameter- see Fig. \ref{remove_text_0} which essentially removes small holes, hence uniting some objects/shapes. The choice of the structuring element was done in order to preserve the circular nature of the solar disk. The resulting image, Fig. \ref{remove_text_1}, became more homogeneous, but the text was not removed, so the basic morphological operation \textit{opening} (or \textit{open}) using as structuring element a disk of 20 pixels in diameter was performed. The result of this open operation-which essentially removes small objects/shapes-is shown in Fig. \ref{remove_text_2} and, analyzing it, one can see that the text disappeared, as desired, but the digital levels inside the solar disk were not preserved. Therefore, another set of operations must be done until reaching the desired final image. First, we use the original image as a \textit{mask}, and the subtraction of the image in Fig. \ref{remove_text_3} by that mask (the original image) as a \textit{marker}. Then, a morphological reconstruction on that marker is made (Fig. \ref{remove_text_4}) after which an adaptative threshold filter is applied with a lower cutting-value of 30 and an upper cutting-value of 124. This last operation allows to recover the solar disk as a binary image. The result is shown in Fig. \ref{remove_text_5}. Analyzing the image, we detected the existence of a \textit{hole} (a black spot) inside the solar disk. The presence of holes at this stage of the pre-processing happens for some images of our data set. In order to suppressed it, a \textit{fill hole} operation is performed (Fig. \ref{remove_text_6}). To make the solar disk a perfect circle a few more pre-processing steps are needed: the center and the radius of the solar disk of the Fig. \ref{remove_text_6} is calculated with the same algorithm used in the method based on intensities levels of the pixels, which is explain in section \ref{pint:preproc}.The result is shown in Fig. \ref{remove_text_66}. Finally, this image is multiplied by the original one and the result is shown in Fig. \ref{remove_text_7}. This last image is the final result of the pre-processing, with the original digital levels inside the solar disk and with digital levels of zero (corresponding to the black color) outside of it. 

\begin{figure}[!hb]
	\subfloat[ \label{remove_text_0}]{%
		\includegraphics[width=0.5\textwidth]{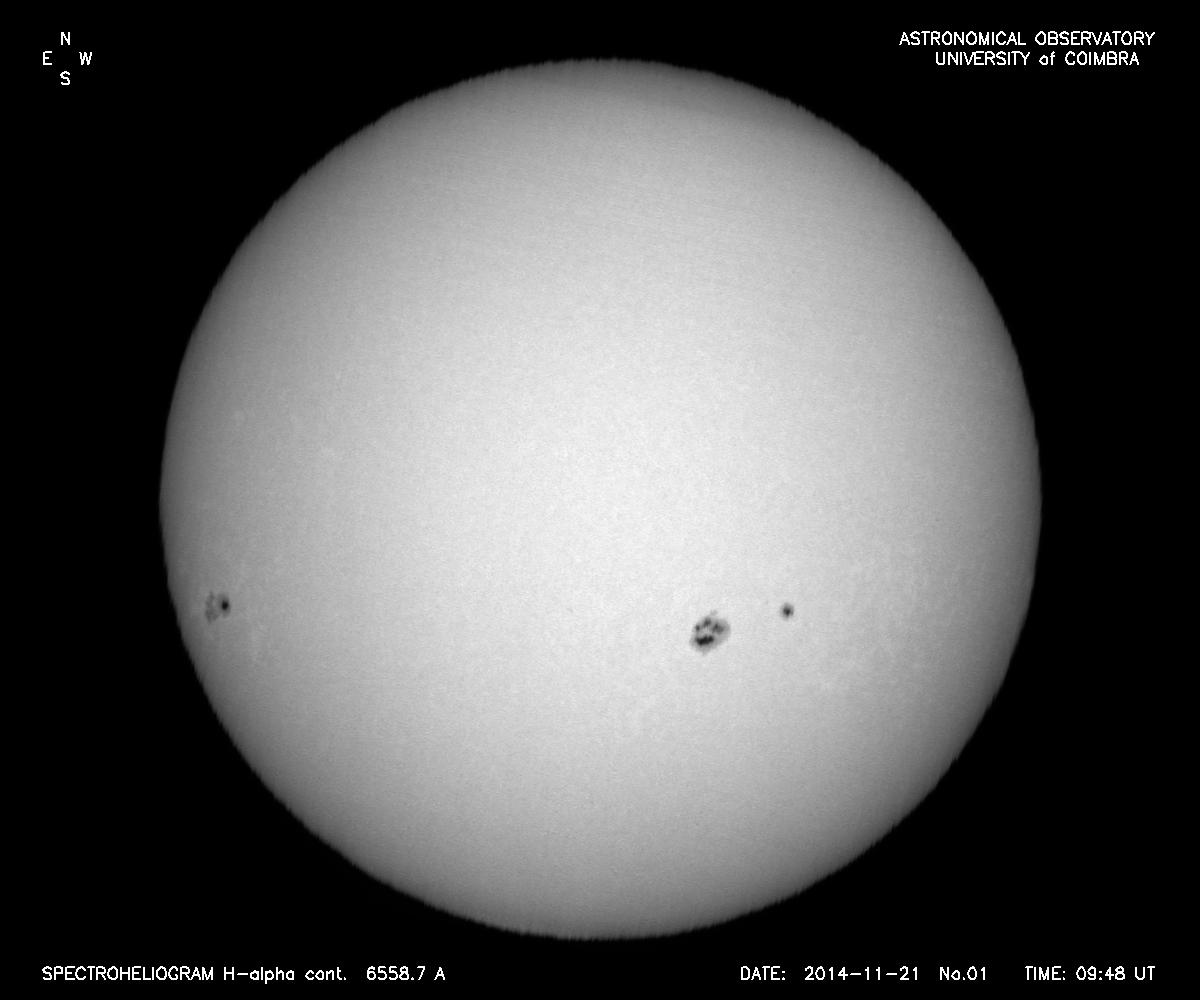}
	} \hfill
	\subfloat[\label{remove_text_1}]{%
		\includegraphics[width=0.5\textwidth]{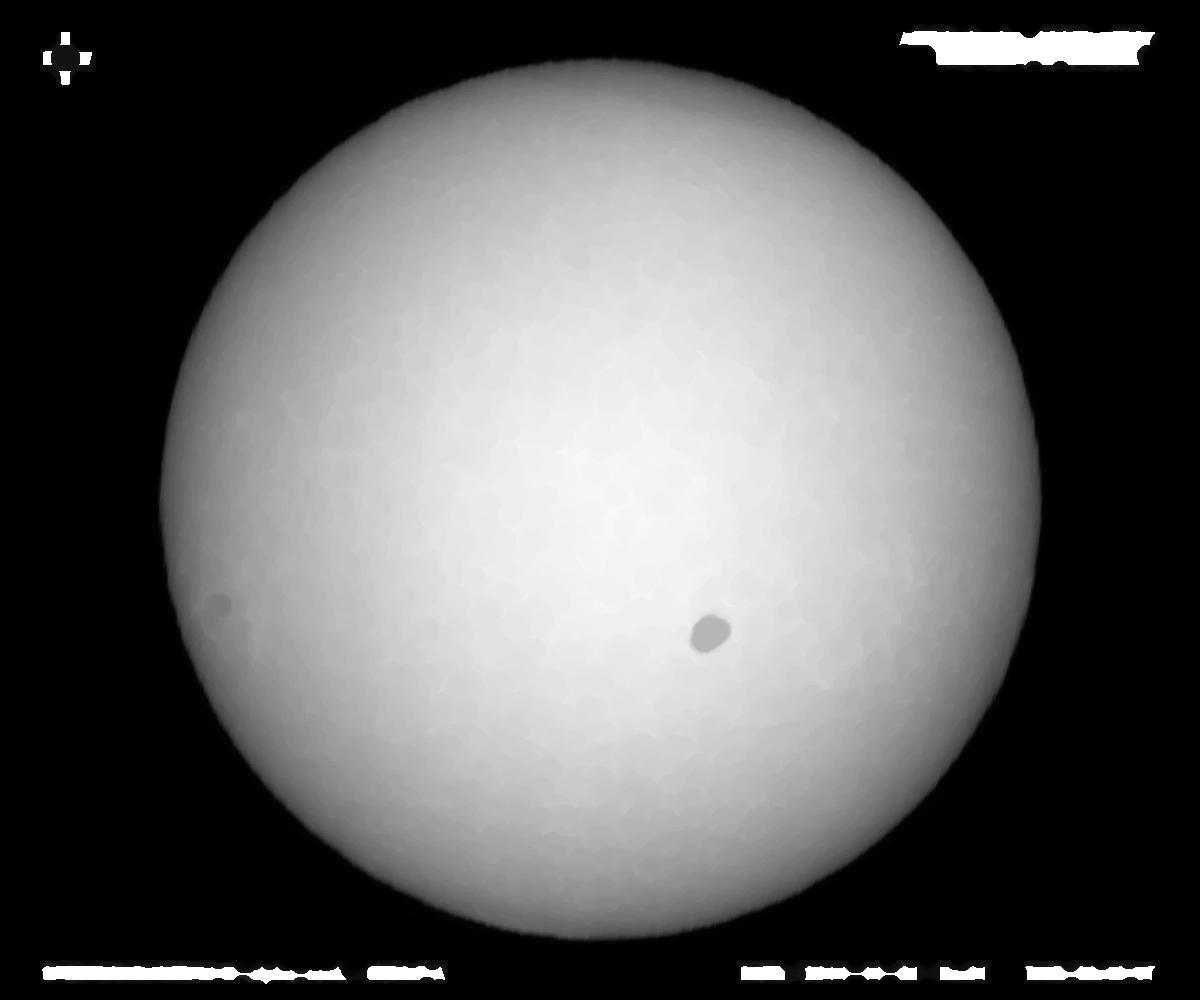}
	}
	\\ \hfill
	\subfloat[\label{remove_text_2}]{%
		\includegraphics[width=0.5\textwidth]{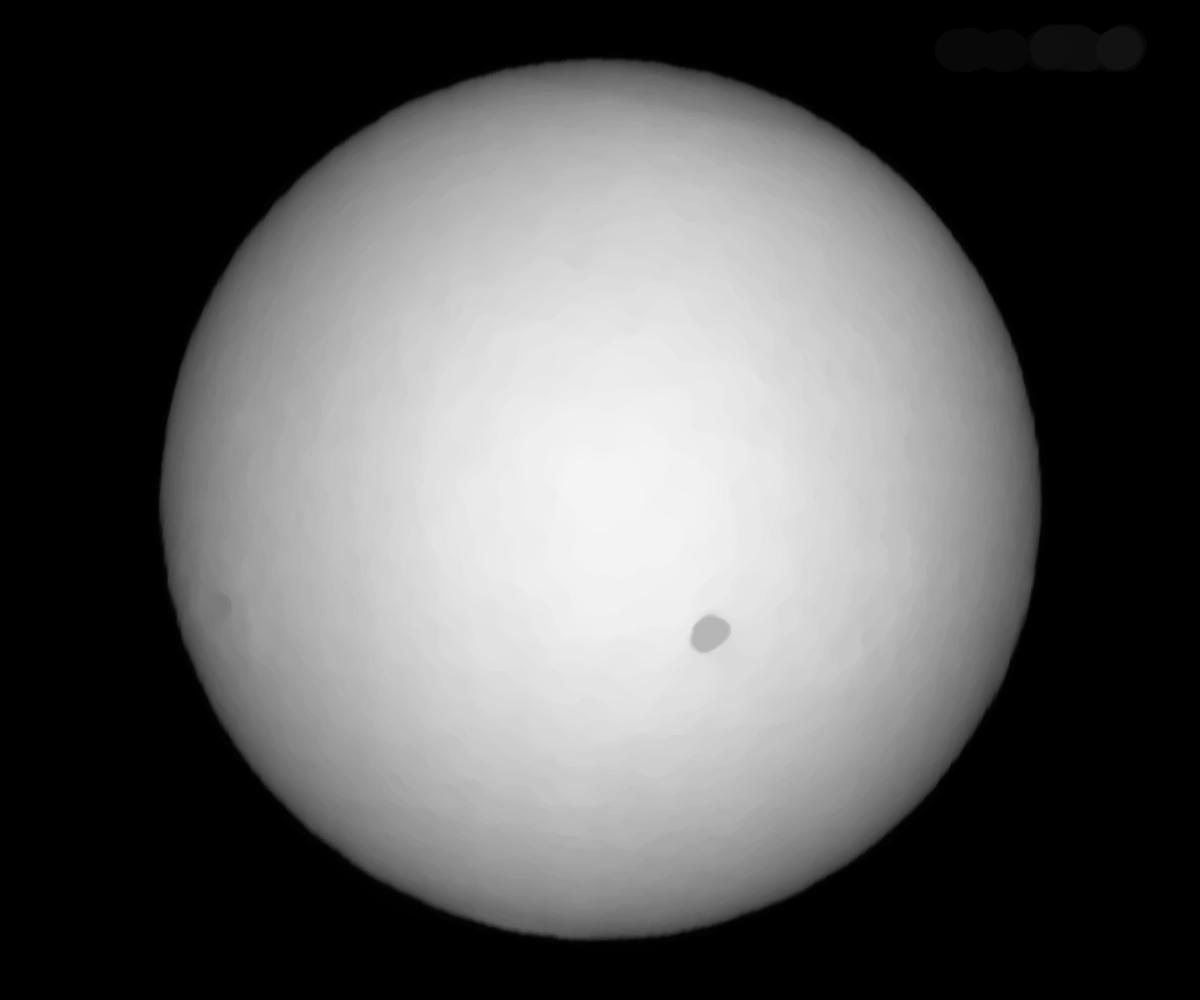}
	} 
	\hfill
	\subfloat[\label{remove_text_3}]{%
		\includegraphics[width=0.5\textwidth]{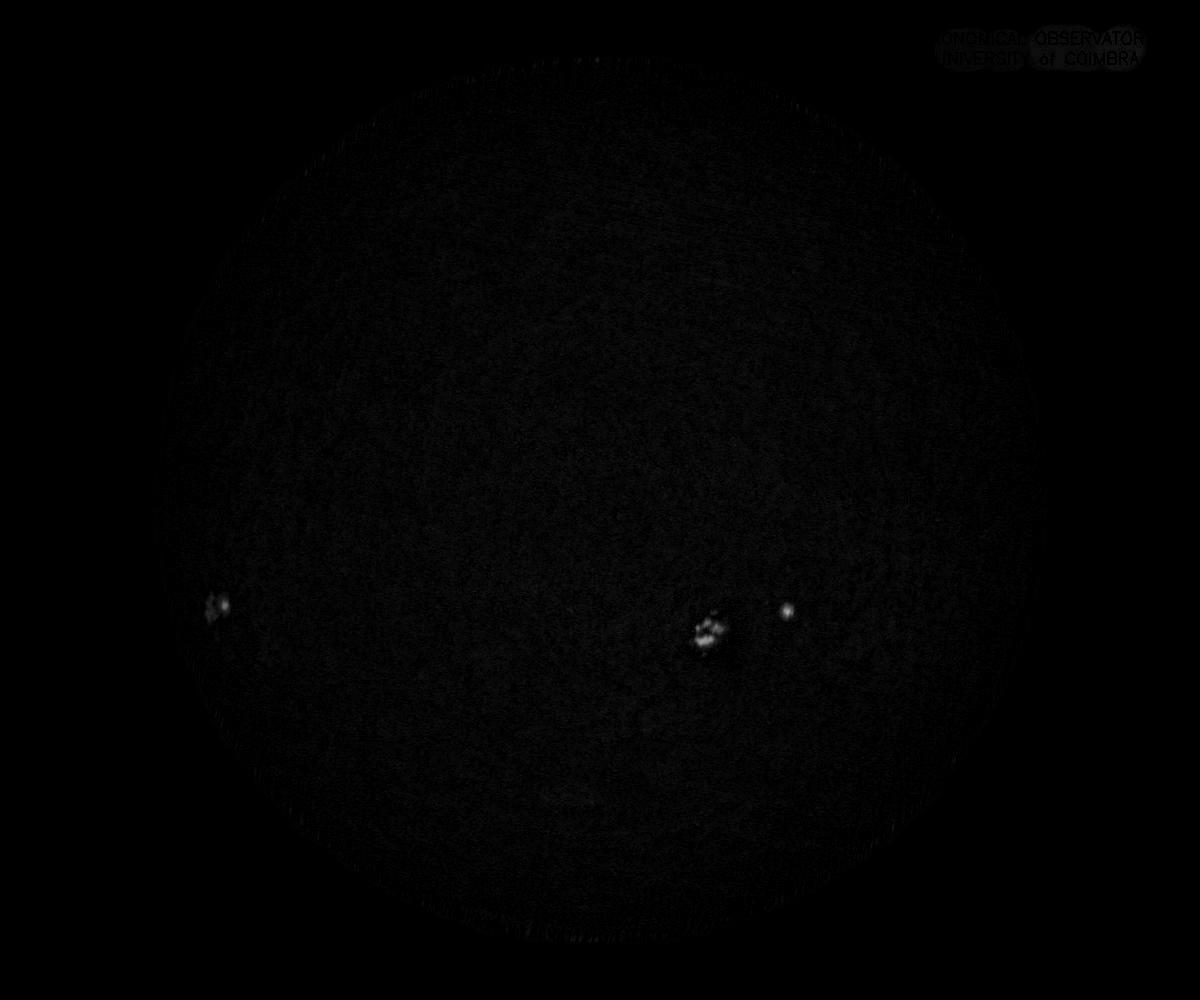}
	}
	\\	\hfill
	\subfloat[\label{remove_text_4}]{%
		\includegraphics[width=0.5\textwidth]{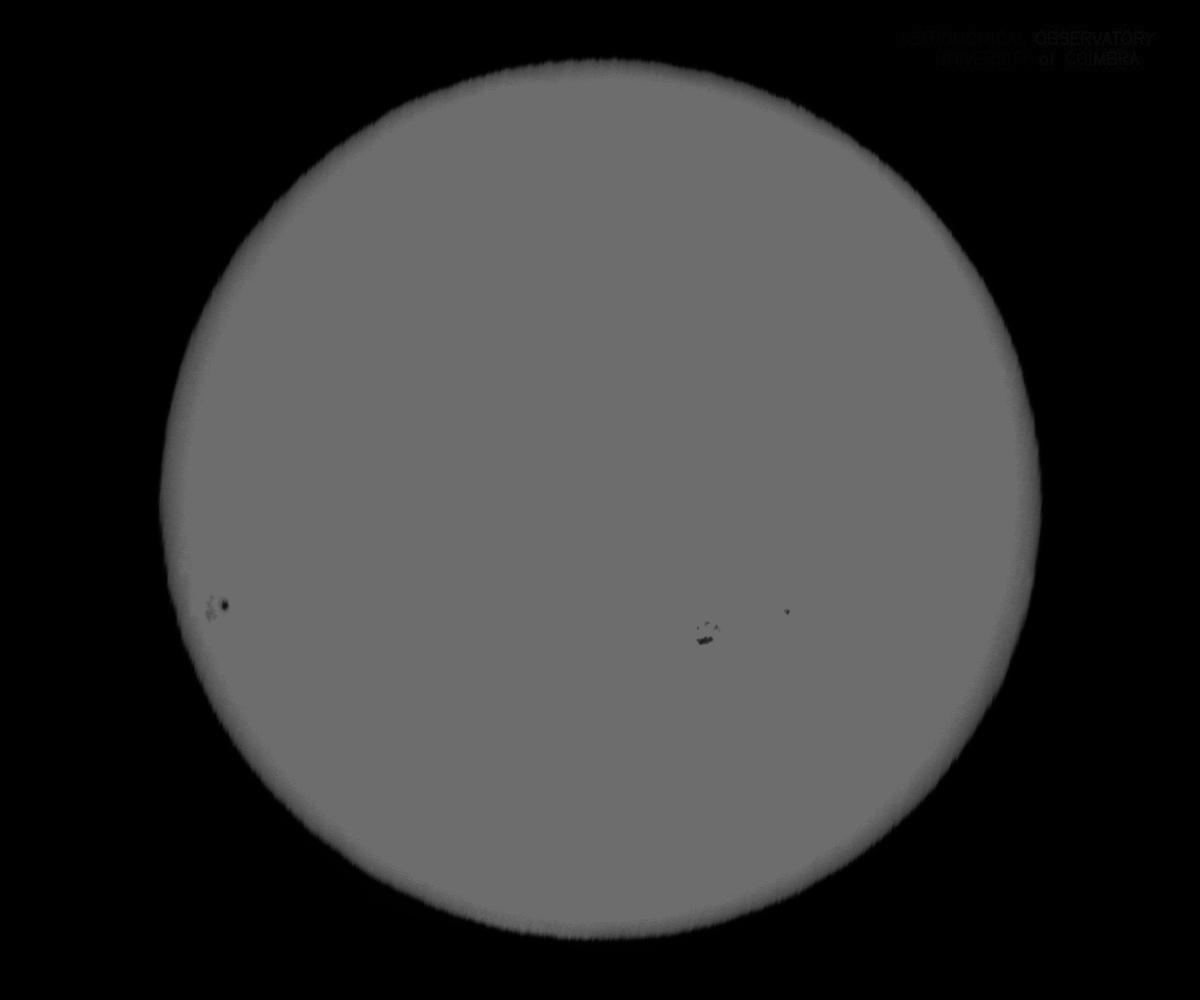}
	}
	\hfill
	\subfloat[\label{remove_text_5}]{%
		\includegraphics[width=0.5\textwidth]{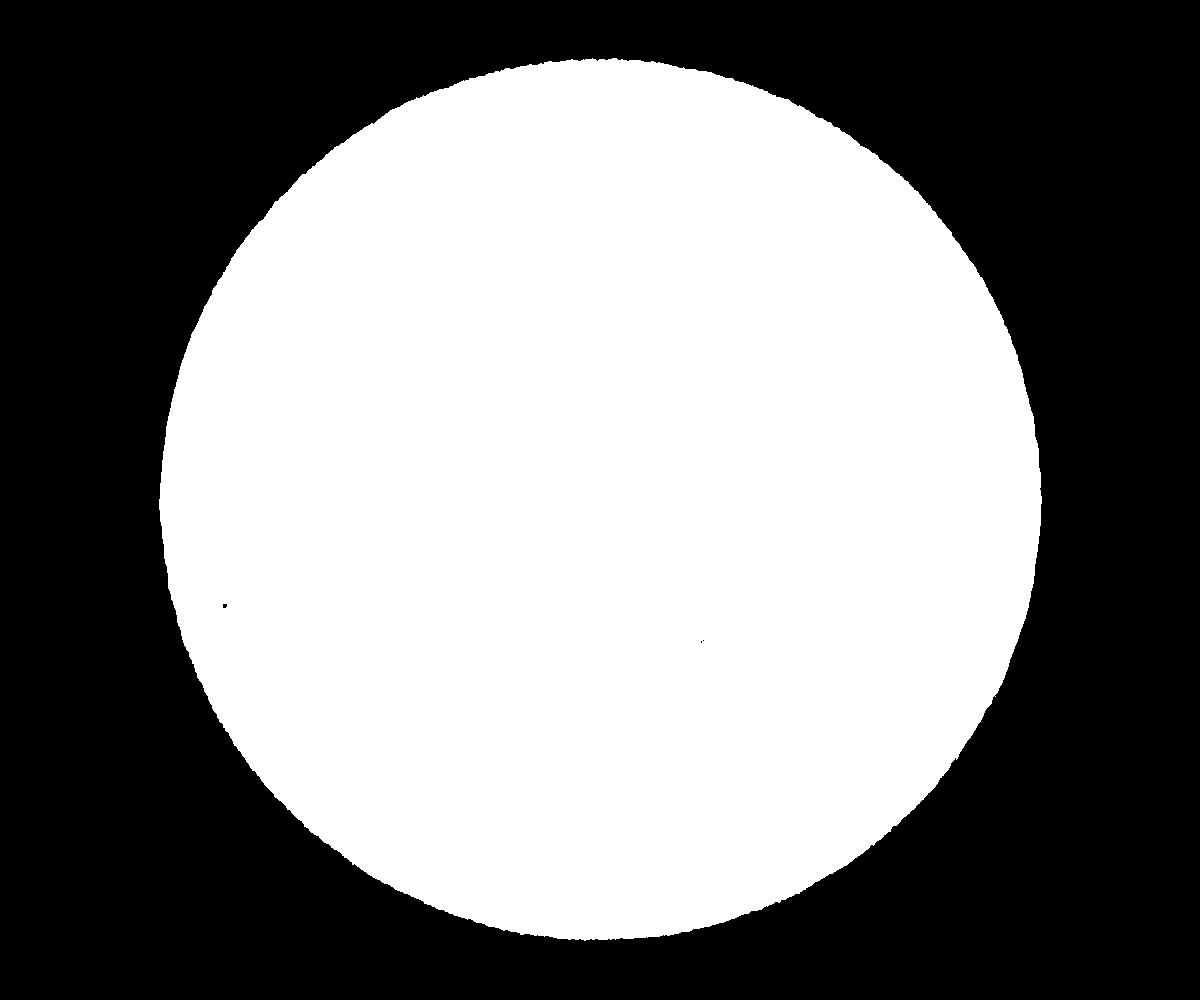}
	}\\
	\label{mm:preprocess}
	\phantomcaption
\end{figure}
\begin{figure}[!ht]\ContinuedFloat
	\subfloat[\label{remove_text_6}]{%
		\includegraphics[width=0.5\textwidth]{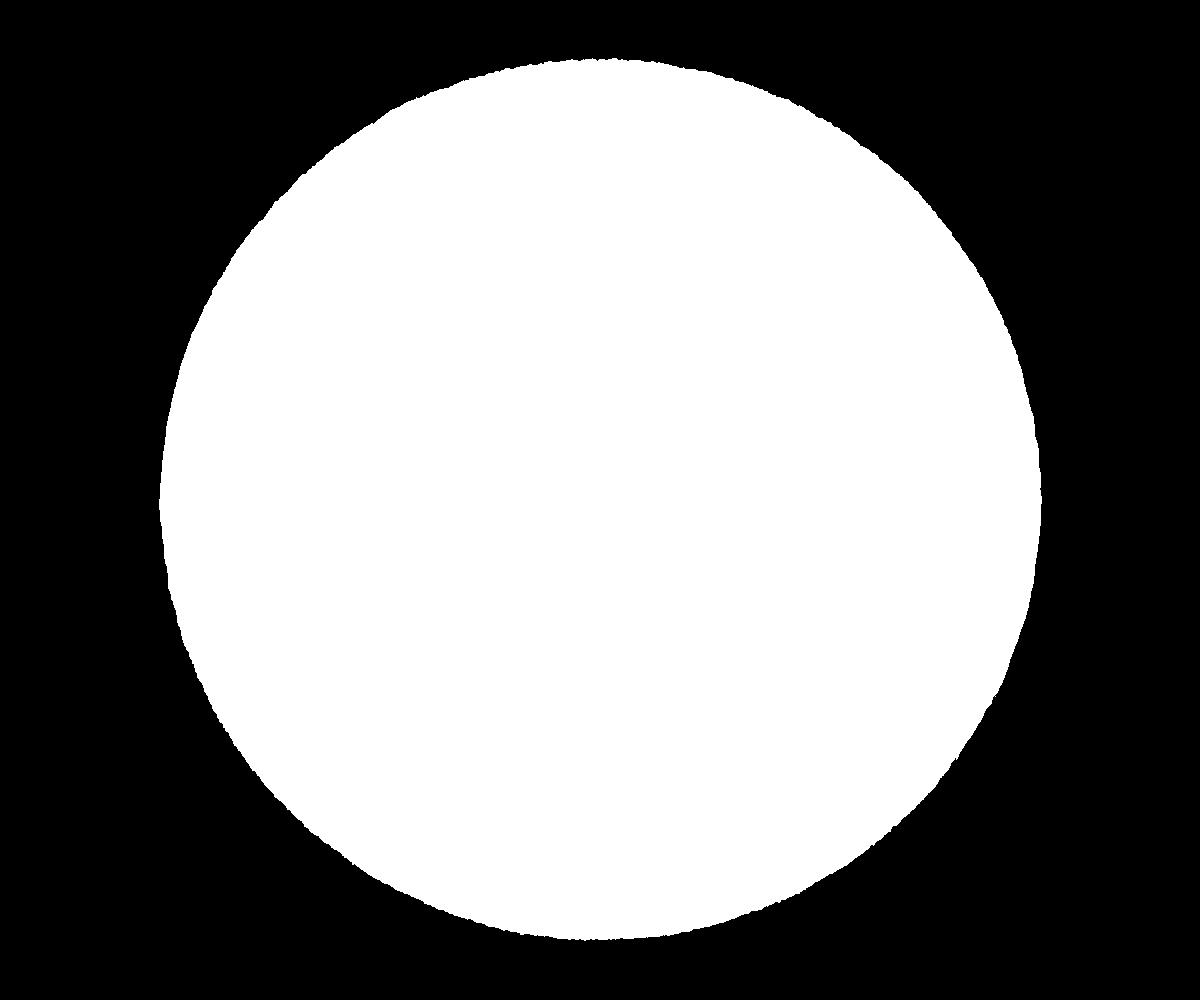}
	}	\hfill
	\subfloat[\label{remove_text_66}]{%
		\includegraphics[width=0.5\textwidth]{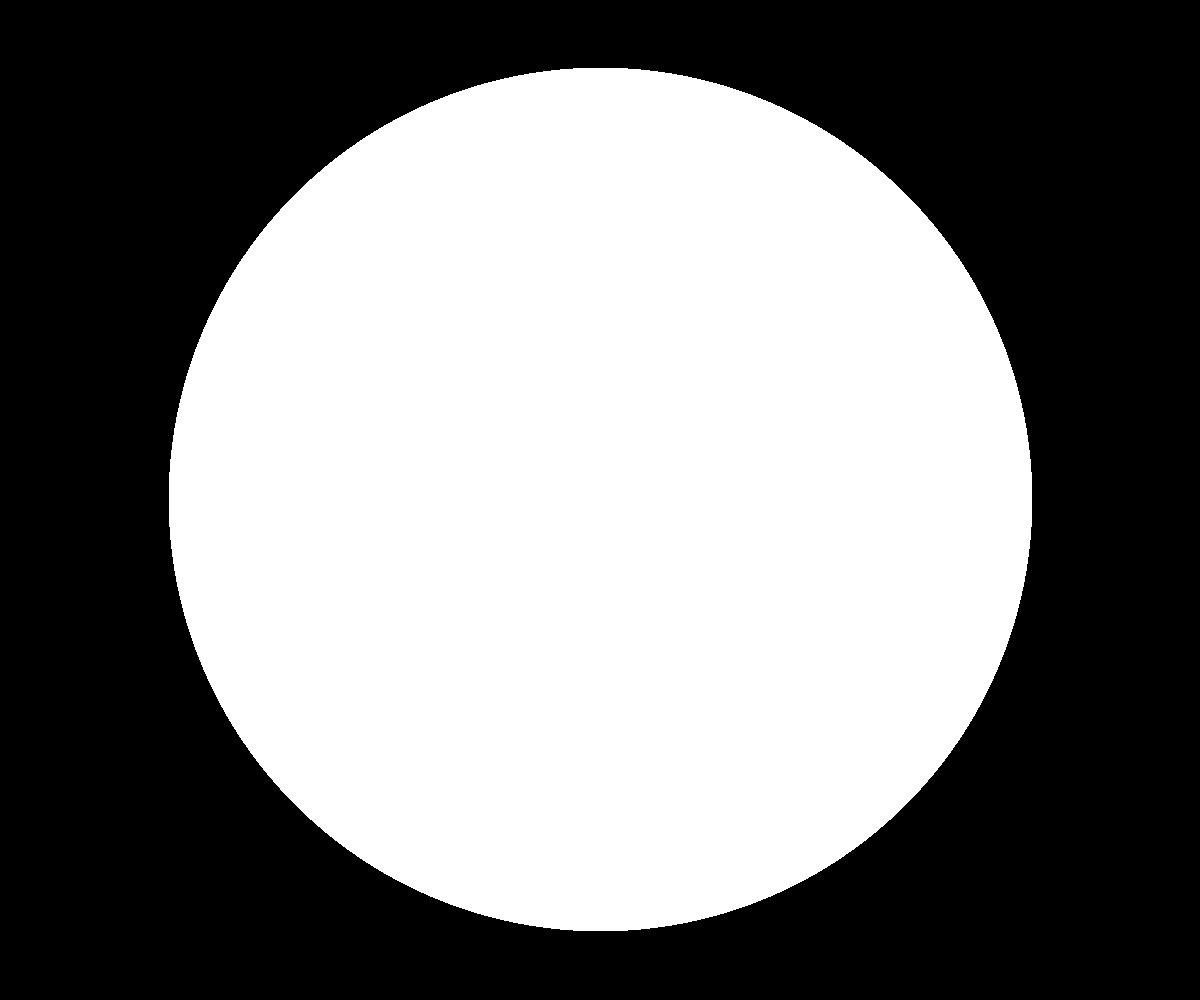}
	} \\\hfill

	\qquad\qquad\qquad\qquad\quad	
	\subfloat[\label{remove_text_7}]{%
	\includegraphics[width=0.5\textwidth]{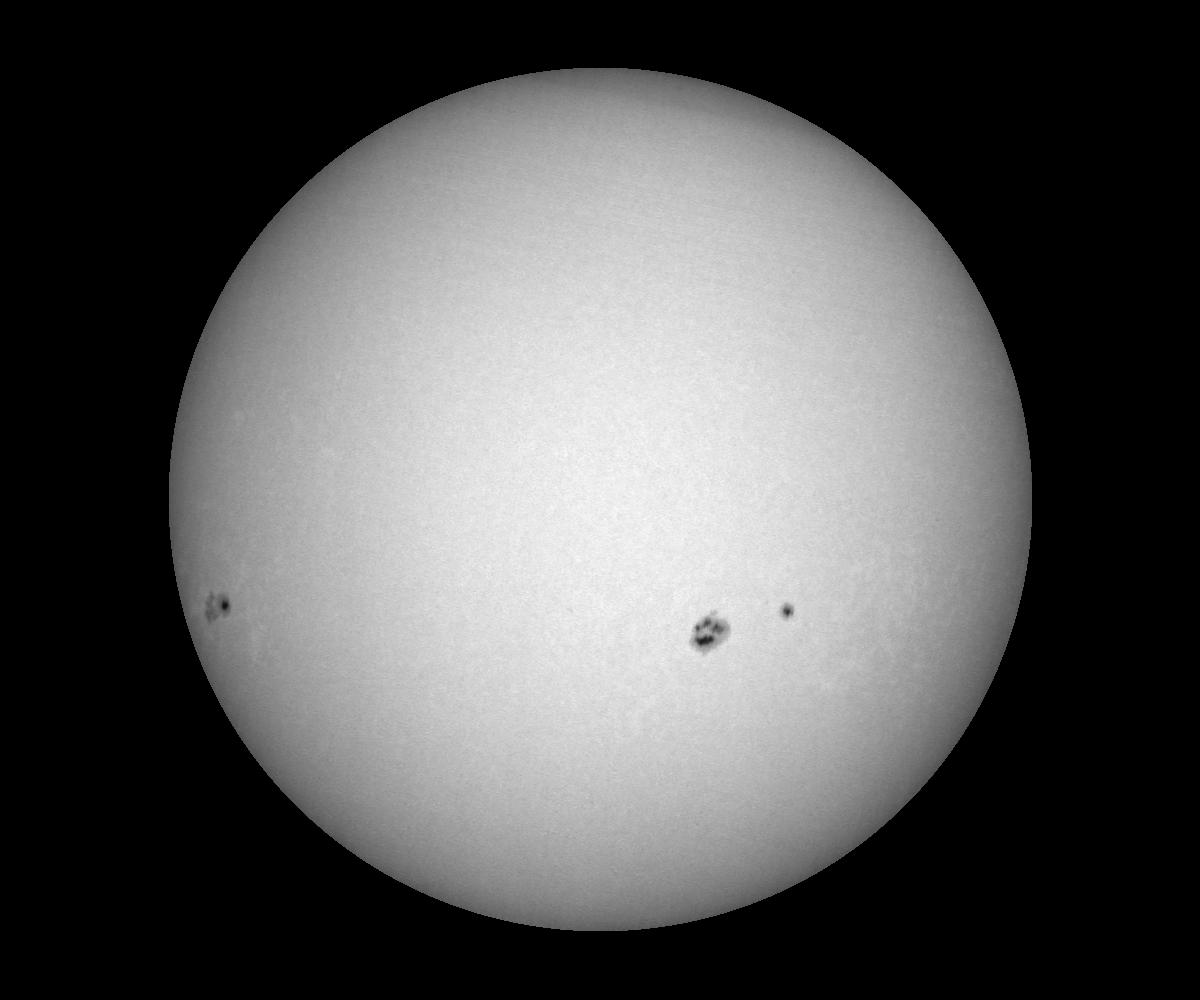}
	}\\	
	\caption{Pre-processing: (a) original image, that is, the mask; (b) \textit{close} of image (a); (c) \textit{open} of image (b); (d) image (c) subtracted by image (a), that is, the marker; (e) reconstruction of the marker (d) under the mask (a); (f) adaptative threshold of image (e), with values between 30 and 124; (g) filling the holes of image (e); (h) solar disk of (g) transformed into a perfect circle; (i) the final image of the pre-processing obtained by the multiplication of (h) by (a).}
	\label{mm:preprocess}
\end{figure}

\subsection{Morphological detection of sunspots}
\label{morph:detectsunspots}
\begin{figure}[]
	\subfloat[ \label{identification_1}]{%
		\includegraphics[width=0.5\textwidth]{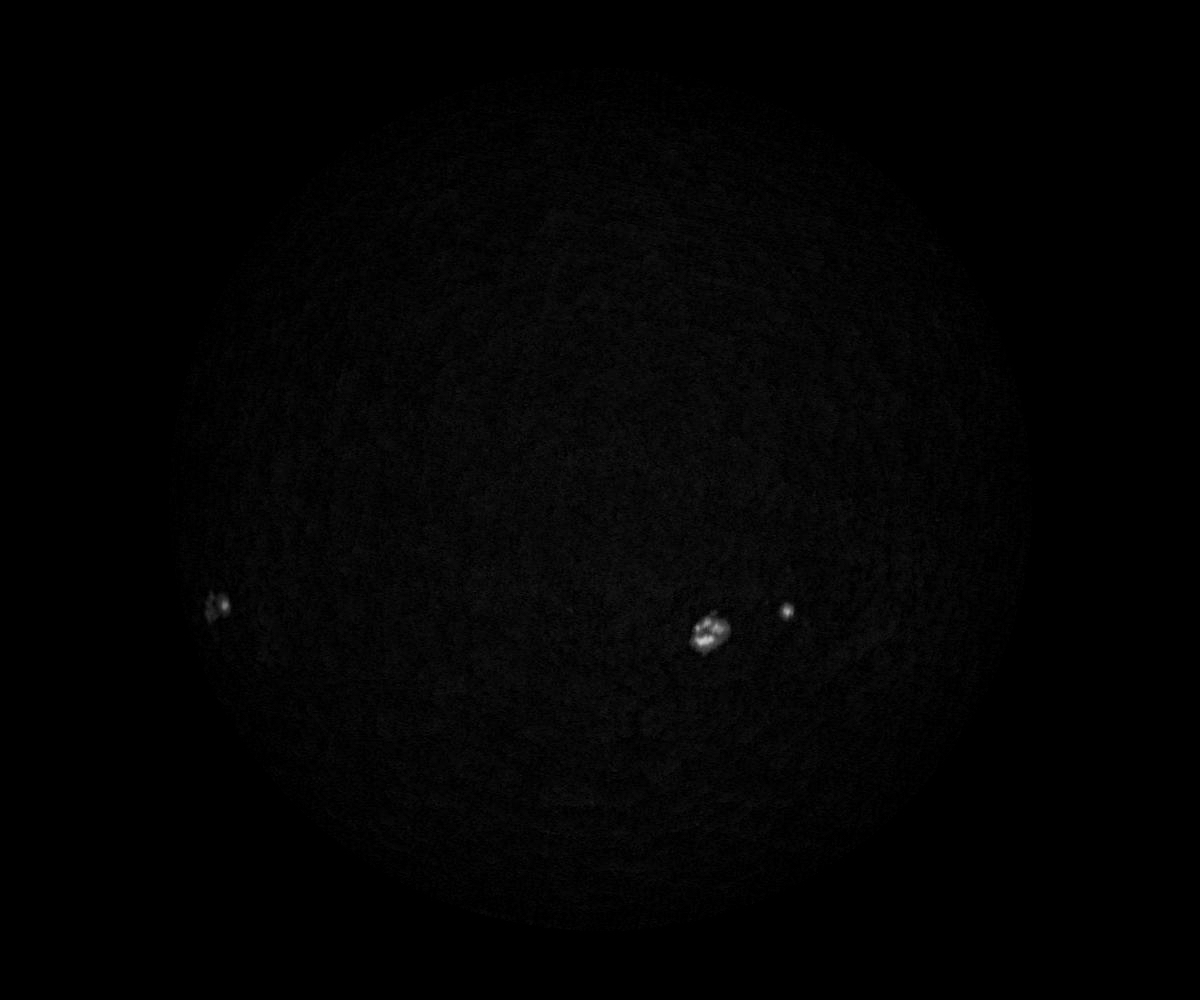}
	} \hfill
	\subfloat[\label{identification_2}]{%
		\includegraphics[width=0.5\textwidth]{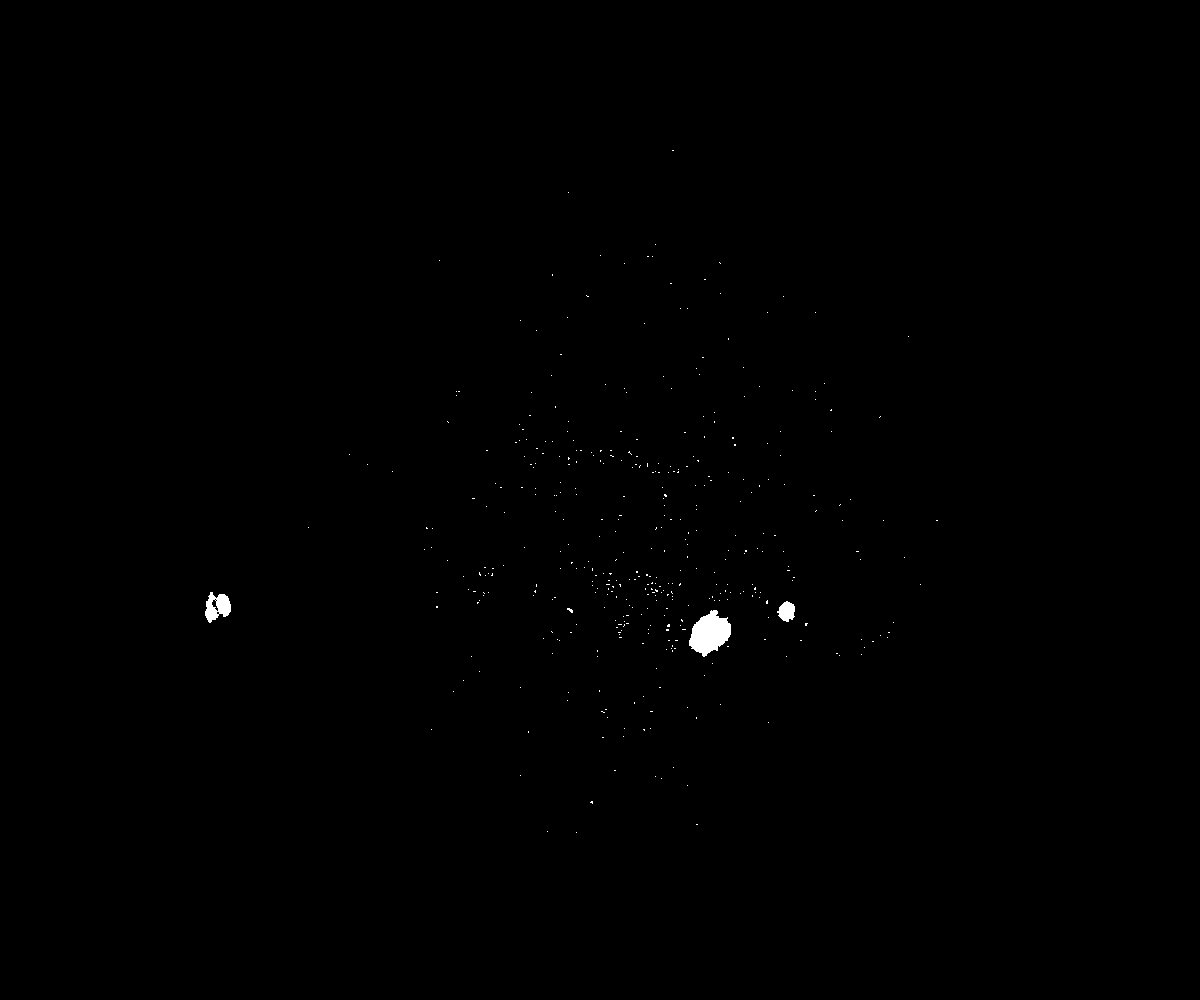}
	} \\ \hfill
	\subfloat[\label{identification_3}]{%
		\includegraphics[width=0.5\textwidth]{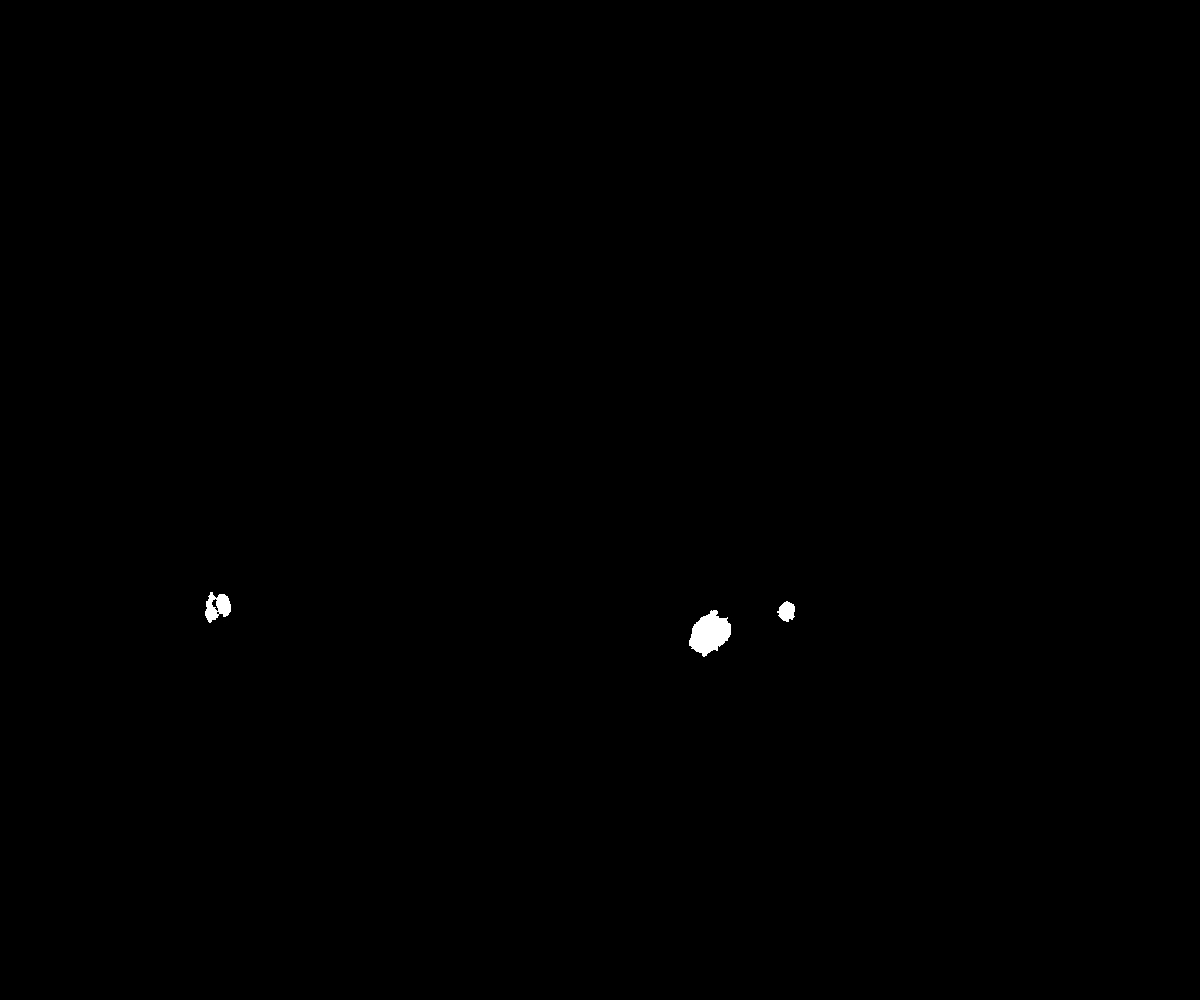}
	} \hfill
	\subfloat[ \label{identification_4}]{%
		\includegraphics[width=0.5\textwidth]{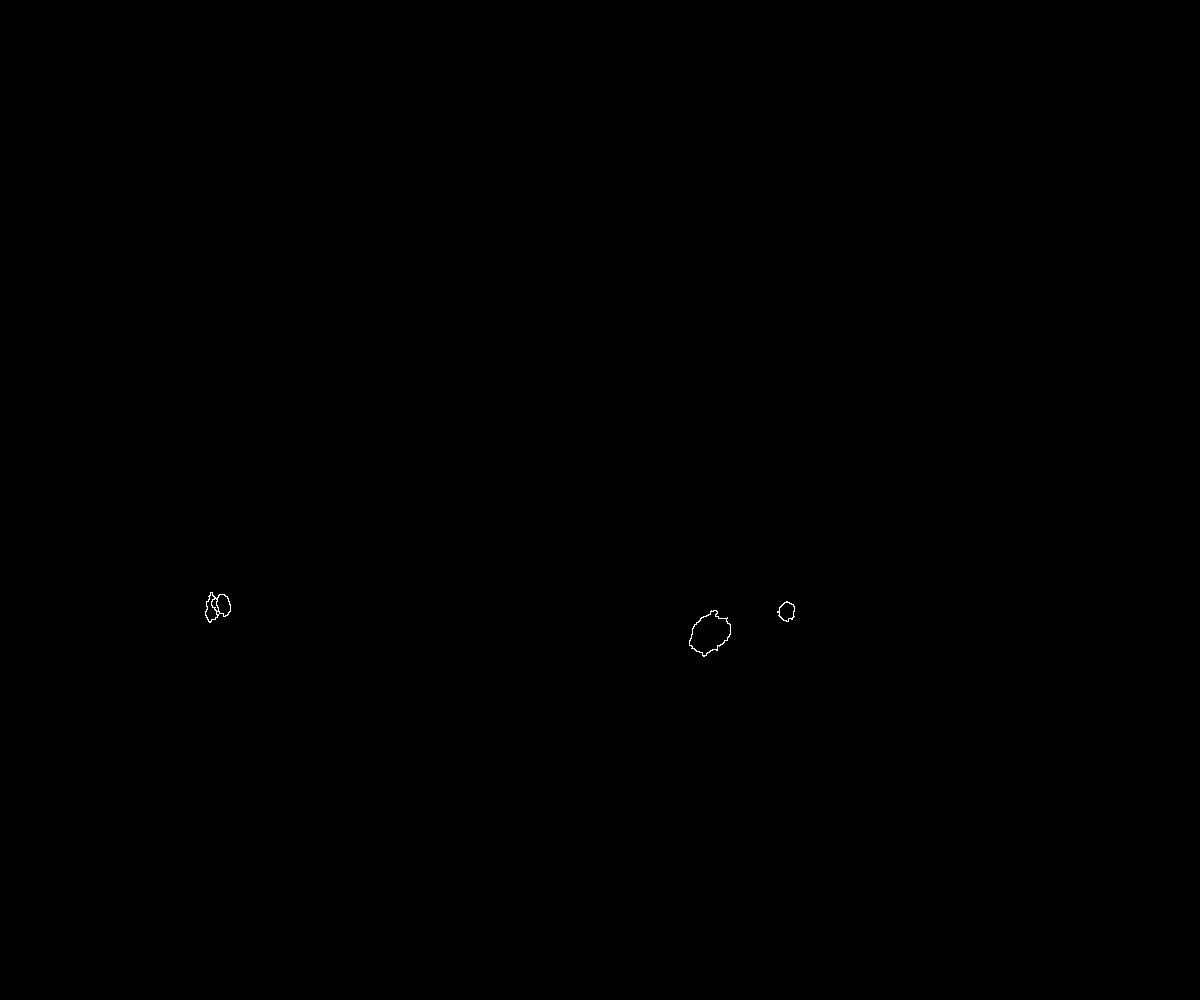}
	} \\ \hfill
	\subfloat[\label{identification_5}]{%
		\includegraphics[width=0.5\textwidth]{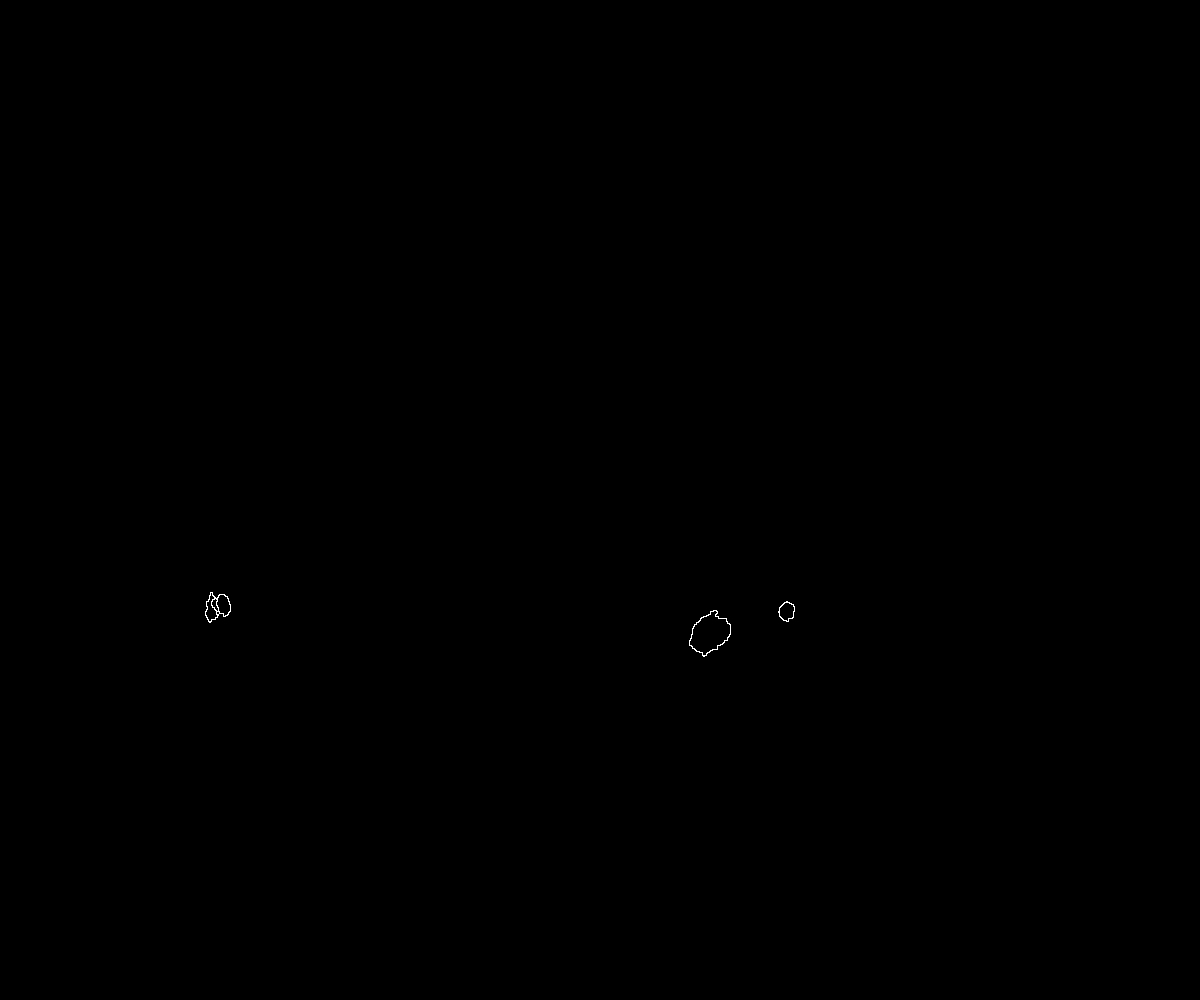}
	}\hfill
	\subfloat[\label{identification_6}]{%
		\includegraphics[width=0.5\textwidth]{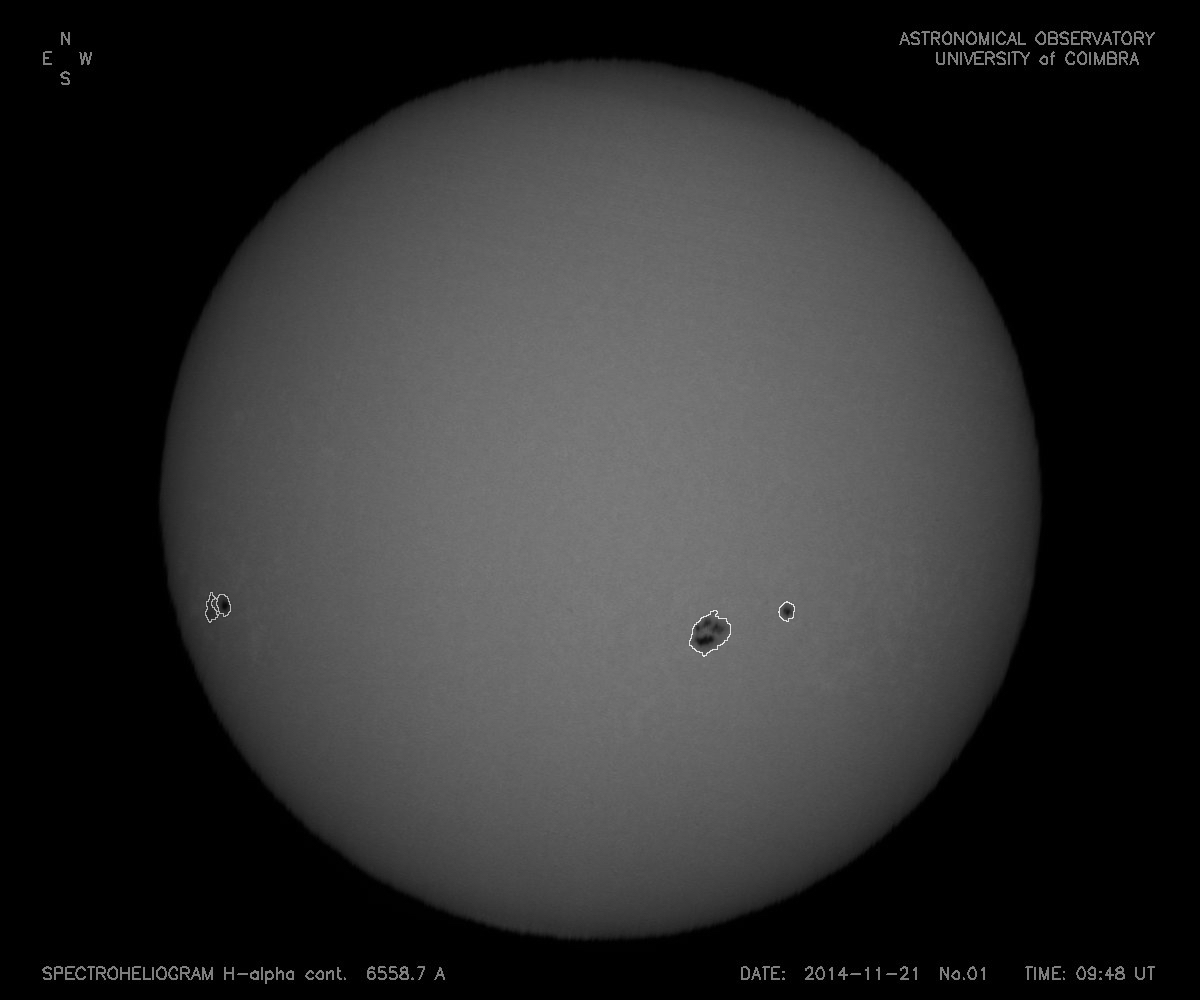}
	}\\
	
	\caption{Sunspots detection algorithm: (a) black top-hat applied to the image in Fig. \ref{remove_text_7}; (b) adaptative threshold of the image (a); (c) reconstruction of image (b) after being subjected to an erosion; (d) morphological gradient; (e) sunspots contours; (f) sunspots contours superimposed over the original image.}
	\label{preprocessamento}
	\end{figure}

The main goal of this algorithm, based on morphological transforms and designed by MM, is the automatic detection of sunspots in Coimbra’s spectroheliograms. The initial image fed into the algorithm is the final image of the pre-processing (Fig. \ref{remove_text_7}). To enhance the (possible) sunspots on the image, a \textit{black top-hat} transform was applied-which essentially extracts the small elements and details seen in the image. This transform consists of the difference between the closing by a disk of 20 pixels in diameter of the initial image (Fig. \ref{remove_text_7}) and that very same initial image. The result of this transform is the image in Fig. \ref{identification_1}. Then, the image resulted from the black top-hat transform is used to extract the contour of sunspots. For that, an adaptative threshold was applied with limits 20 and 255, originating the image shown in Fig. \ref{identification_2}. As one can see, not only the sunspots were identified but also a lot of noise. In order to eliminate it, an \textit{erosion} is applied, using a disk of diameter 1, eliminating the noisy specs, but since that also erodes the real features a \textit{reconstruction} must be applied for those features to recover their original shape. The result is shown in Fig. \ref{identification_3}. The sunspots are now correctly identified, and the next step is the extraction of its contours which is done through the morphological \textit{gradient} operation (Fig. \ref{identification_4}) followed by a \textit{thinning} operation (Fig. \ref{identification_5}). The gradient operation enhances the contours of the sunspots and the thinning operation allows to reduce that contour to one pixel only, and to remove pixels on the boundaries of sunspots, preserving the relation between structures and holes, which allows to obtain the sunspots skeletons. The final result of this stage of the algorithm is shown in Fig. \ref{identification_6}, where the sunspots’ contours were superimposed over the original image.

\subsection{Umbra-Penumbra Segmentation }

Another aim of the algorithm is the umbra-penumbra segmentation. Sunspots are, generally, constituted by umbra and penumbra, which leads to a bimodal distribution of grey levels within each spot. Nevertheless, there are sunspots constituted only by umbra and, in this case, it is assumed that the distribution of grey levels is unimodal. For this reason, a representative set of sunspots was selected, and their histograms analyzed and, for all practical purposes, if the difference between its maximum and minimum grey levels was greater than 20, the sunspot would have a bimodal distribution, otherwise the distribution would be considered unimodal. Furthermore, concerning the bimodal distributions, a threshold value t is automatically estimated between the two peaks of each histogram to segment umbra and penumbra, following: 
\begin{equation} \label{segmentationformula}
t=\frac{sunspot_{max} - sunspot_{min}}{2} + sunspot_{min},
\end{equation}
where $sunspot_{max} $ and $sunspot_{min}$ are, respectively, the maximum and the minimum values of the grey level inside the sunspot. 

The stage of umbra and penumbra segmentation then starts by labeling each sunspot so that each one could be treated separately (Fig. \ref{segmentation_0}). A sunspot in the image was chosen, as an example, to explain the segmentation implementation in detail. It is highlighted with a square around in Fig. \ref{segmentation_1}. To get the original grey levels of this sunspot, two operations are necessary: first, to isolate the sunspot, an adaptative threshold is applied to the image in Fig. \ref{segmentation_0} using the label number as limits, which results in the binarized image represented in Fig. \ref{segmentation_2}; after that, an intersection between that image and the original image is carried out resulting the image in Fig. \ref{segmentation_3}. Hereupon, the $sunspot_{max}$ and $sunspot_{min}$ are computed in order to determine the type of grey-scale distribution. In the case of the sunspot chosen as an example, the distribution was bimodal and therefore the value t was estimated following Eq. \ref{segmentationformula}. However, in the case of unimodal distribution, t was assumed to be 2. Thereafter, two adaptative thresholds were performed: the first one with limits $1$ and $t-1$, to segment the umbra, and the second one with limits t and 255 to segment the penumbra. The results of these operations are shown in Fig. \ref{segmentation_4} and Fig. \ref{segmentation_5}, respectively. This stage ends with the creation of four images: the first one resulting of the union of all the segmented umbras (Fig. \ref{segmentation_6}), the second one resulting of the union of all the segmented penumbras (Fig. \ref{segmentation_7}), the third one resulting of the sum of all umbras and penumbras together (Fig. \ref{segmentation_8}), and the last one resulting of the composite of the gradients of the umbras and the penumbras, superimposed over the original image (Fig. \ref{segmentation_9}). 

\begin{figure}
	\centering
	\subfloat[ \label{segmentation_0}]{%
		\includegraphics[width=0.5\textwidth]{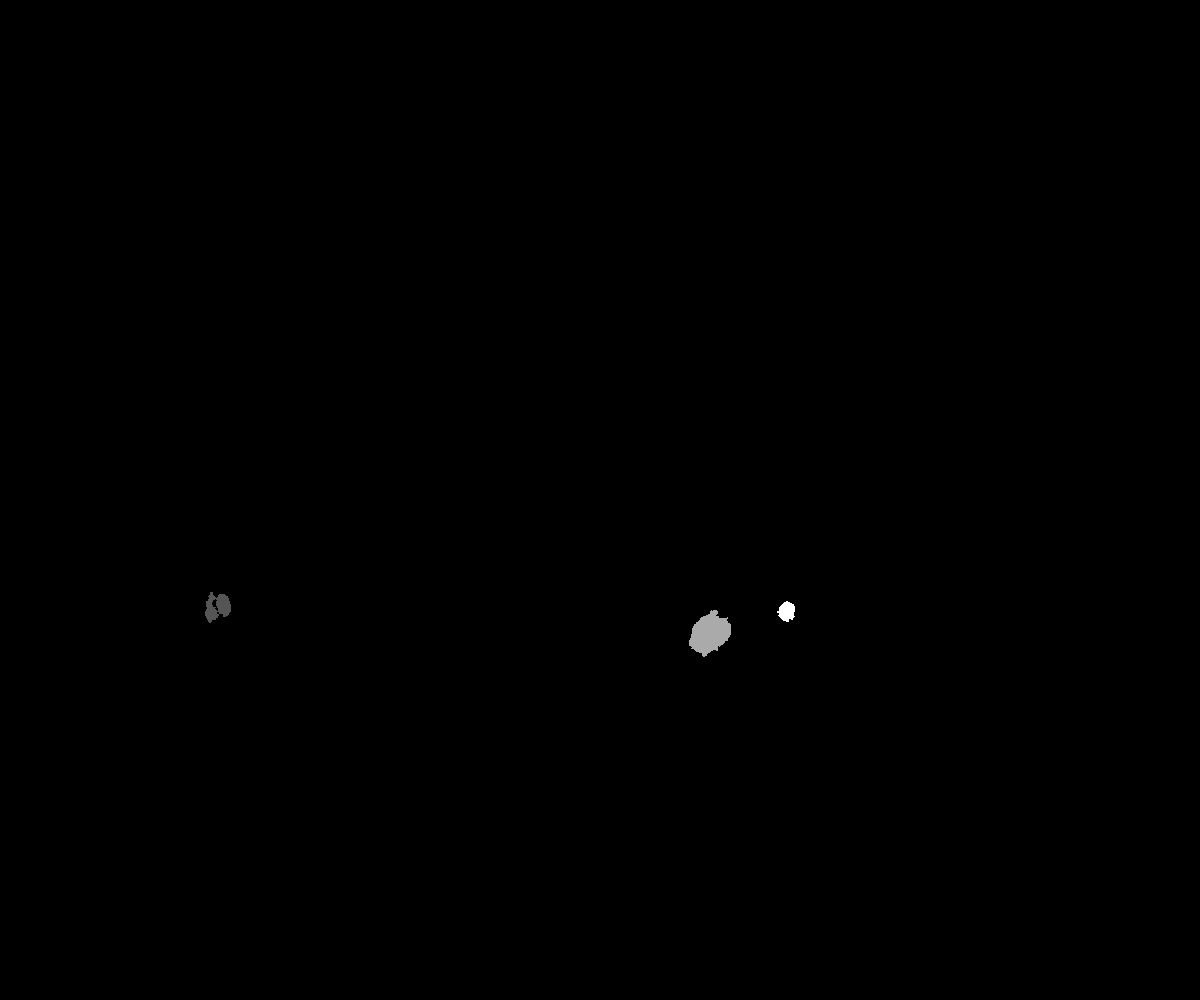}
	} \\ 
	\subfloat[ \label{segmentation_1}]{%
		\includegraphics[width=0.5\textwidth]{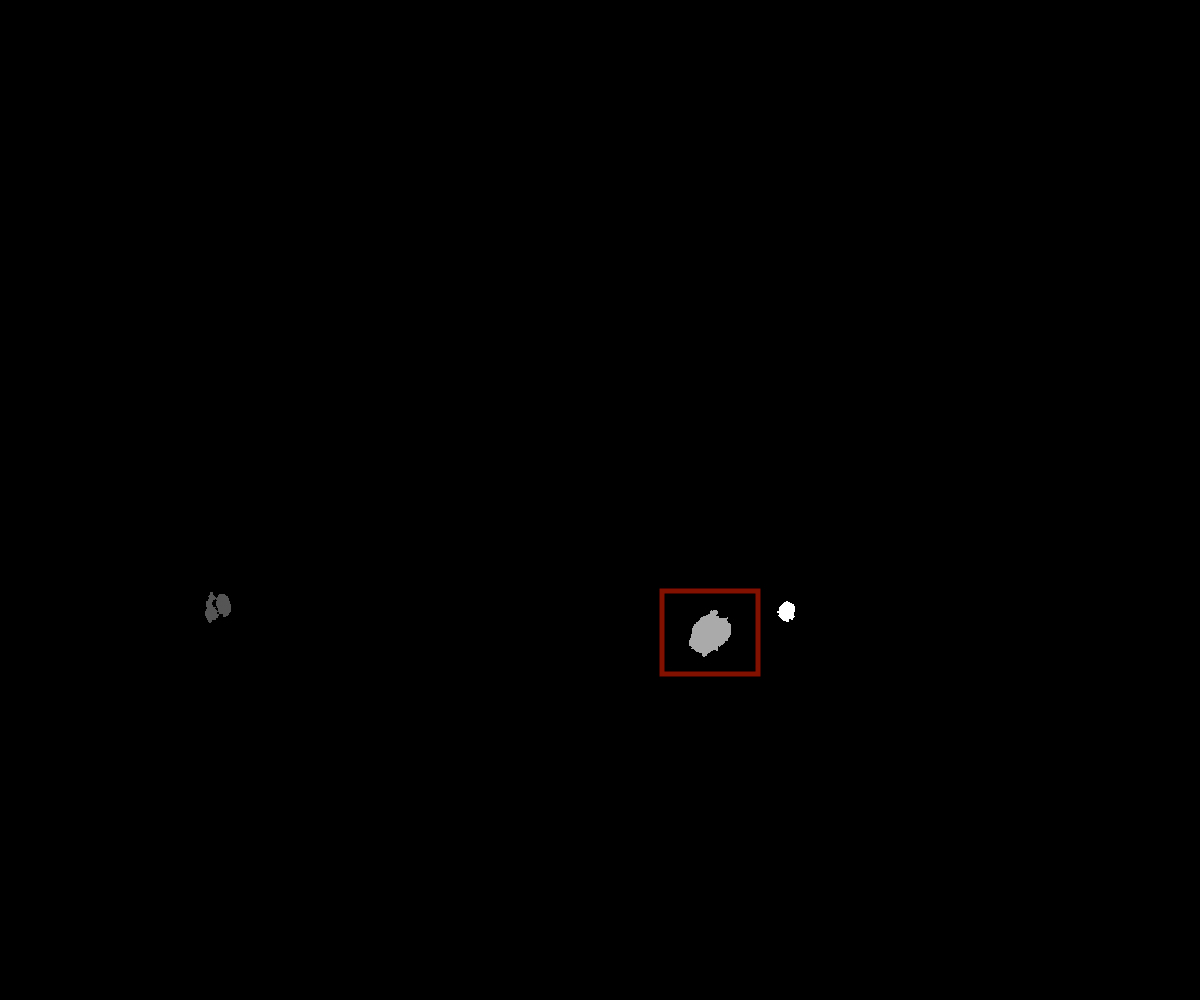}
	} \\ 
	\subfloat[\label{segmentation_2}]{%
		\includegraphics[width=0.5\textwidth]{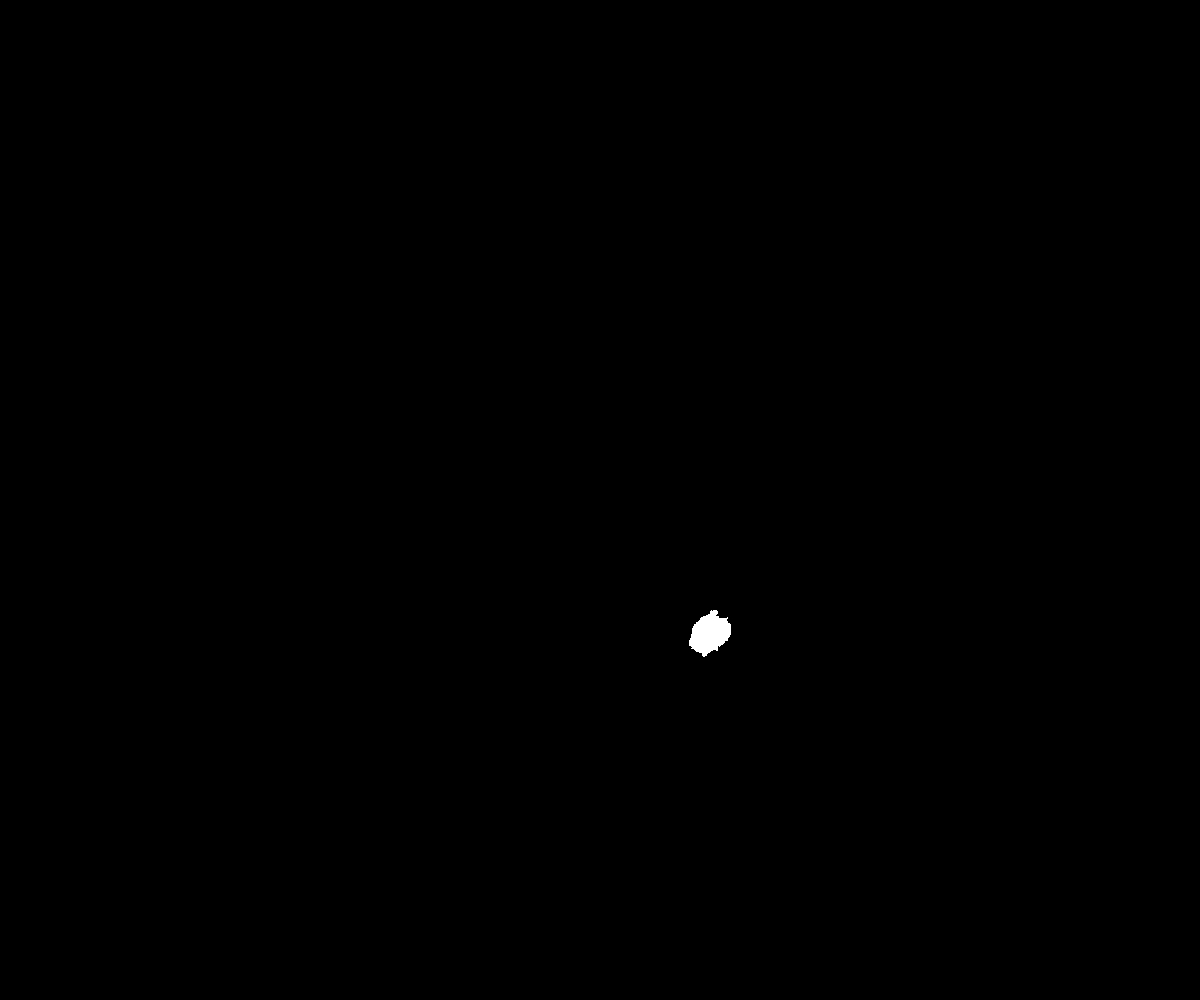}
	} \\
\phantomcaption
\end{figure}

\begin{figure}\ContinuedFloat
	\centering
	\subfloat[ \label{segmentation_3}]{%
		\includegraphics[width=0.5\textwidth]{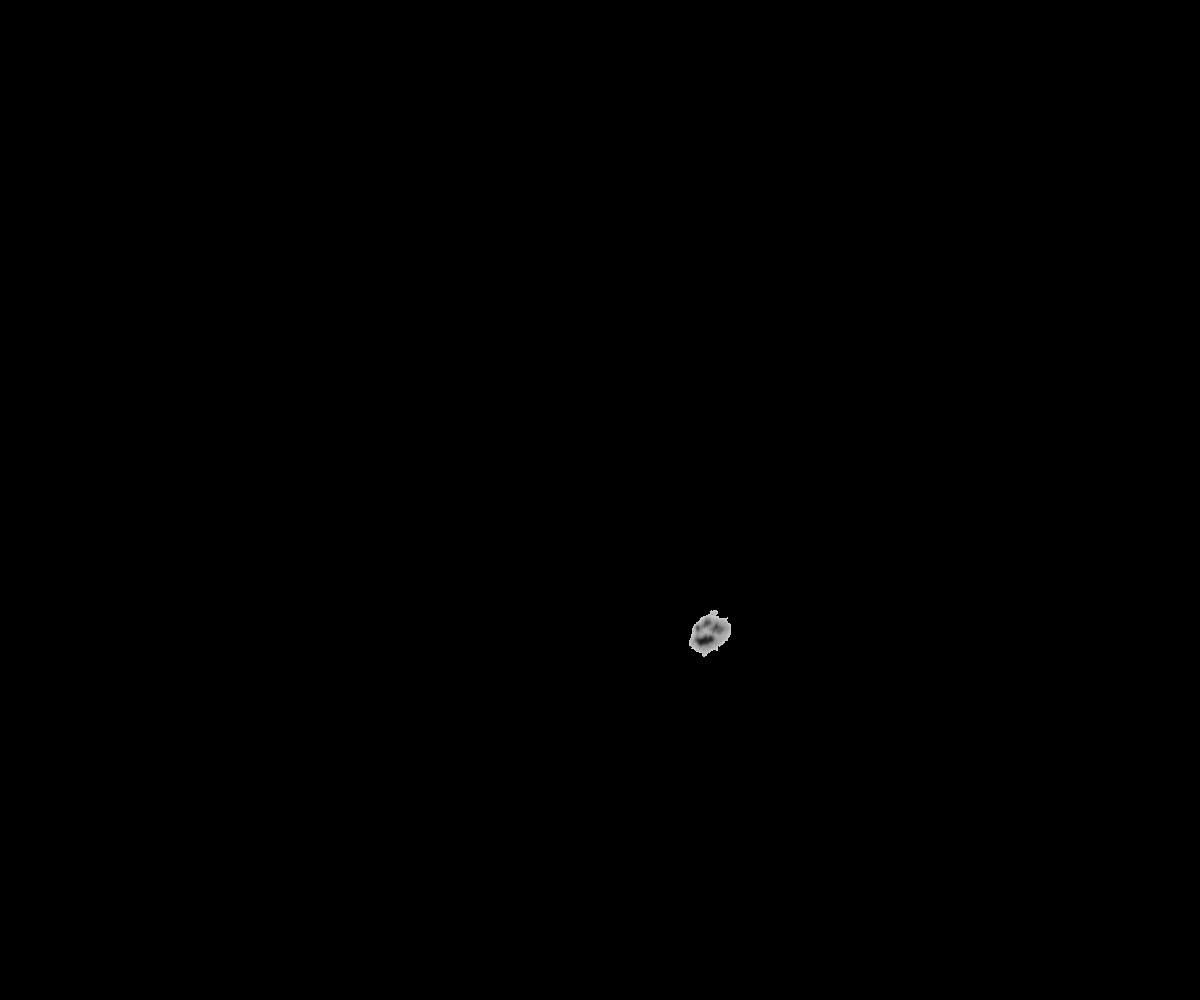}
	} \\
	\subfloat[\label{segmentation_4}]{%
		\includegraphics[width=0.5\textwidth]{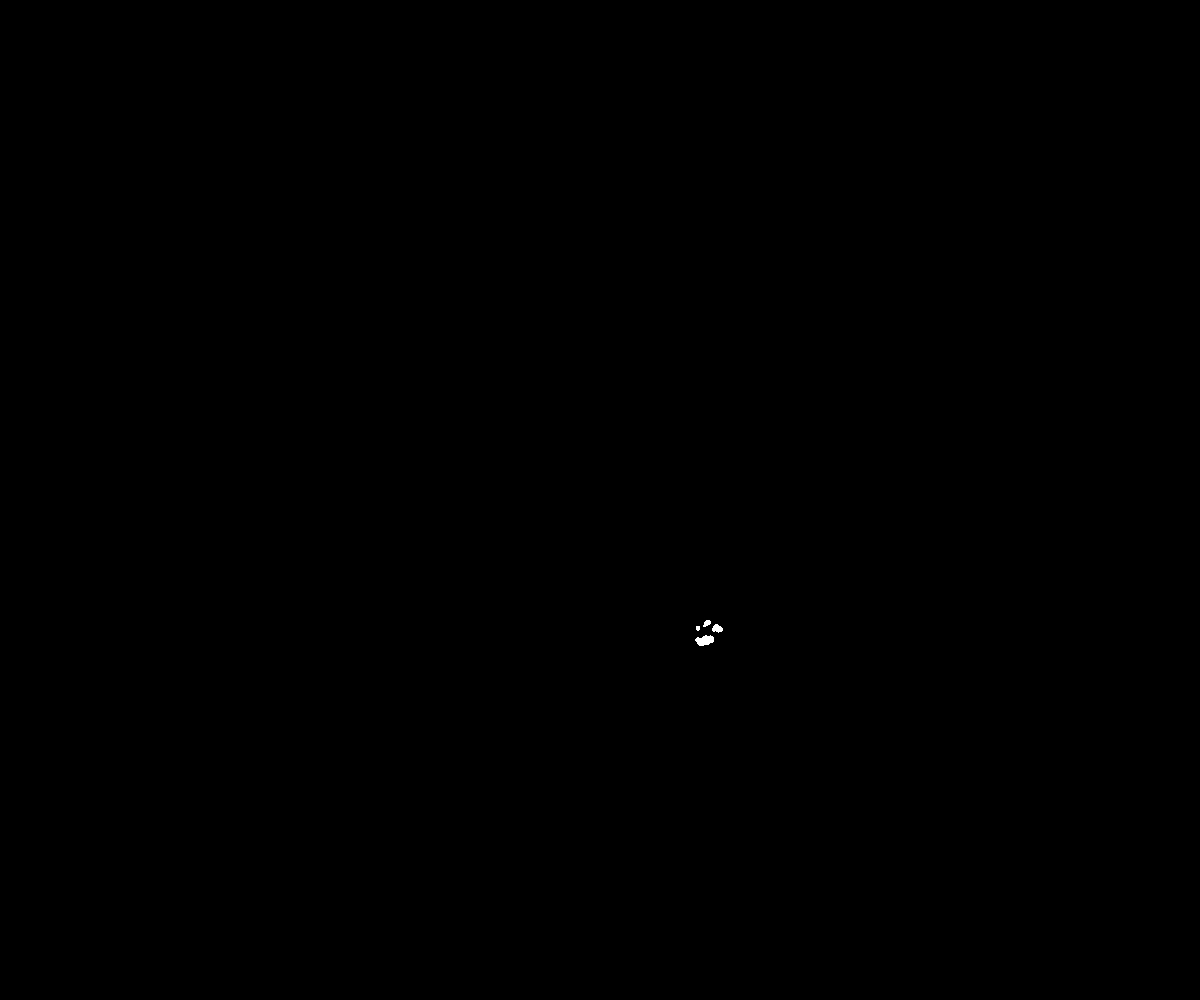}
	}\\
	\subfloat[\label{segmentation_5}]{%
		\includegraphics[width=0.5\textwidth]{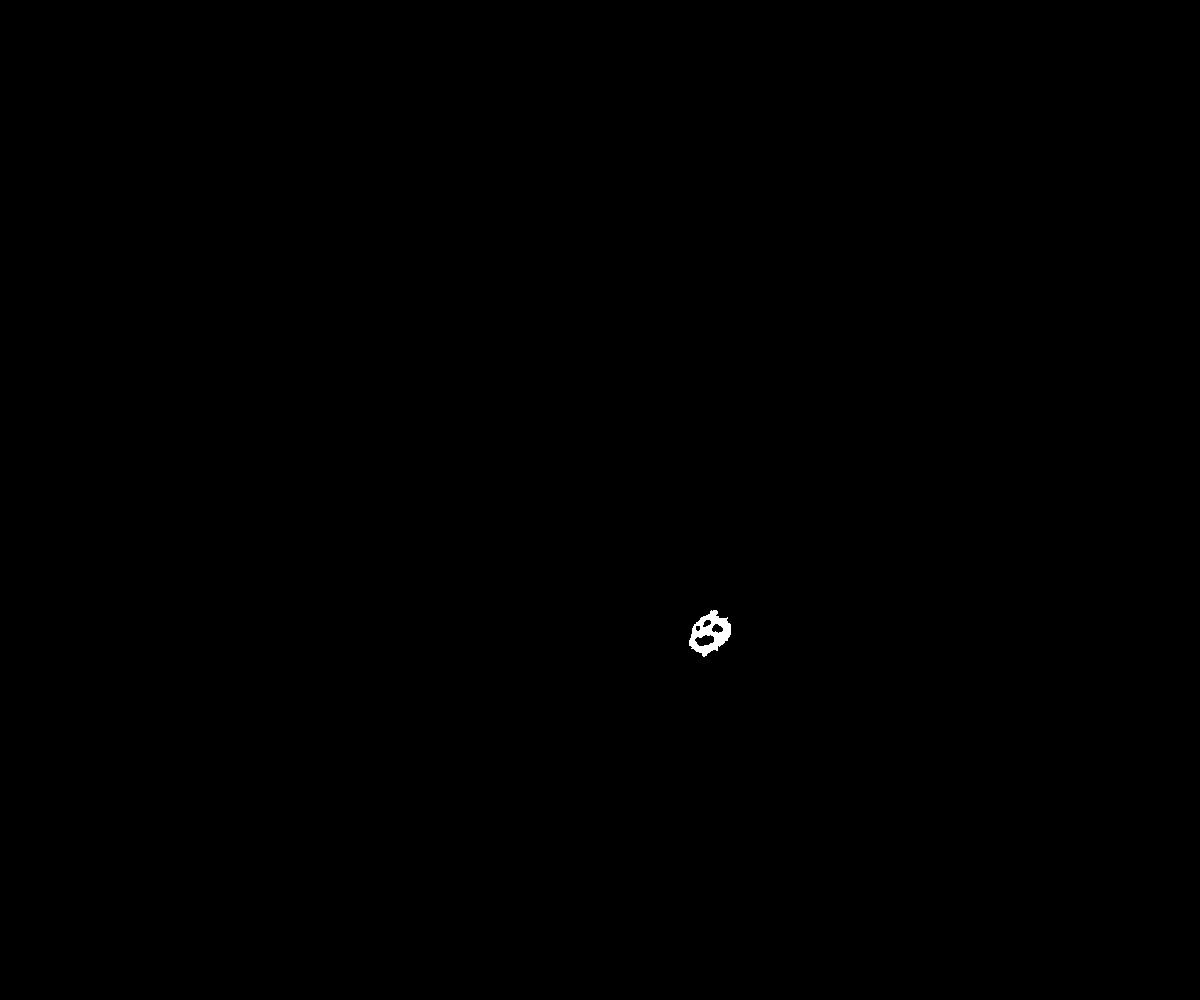}
	}\\
	\caption{ Segmentation algorithm: (a) Sunspots labeled; (b) sunspot used as an example high-lighted with a square around; (c) adaptative threshold of the image (a); (d) intersection of the image (c) and the original image; (e) umbra segmentation; (f) penumbra segmentation.}	
\end{figure}

\begin{figure}[!ht]
	\centering
	\subfloat[ \label{segmentation_6}]{%
		\includegraphics[width=0.5\textwidth]{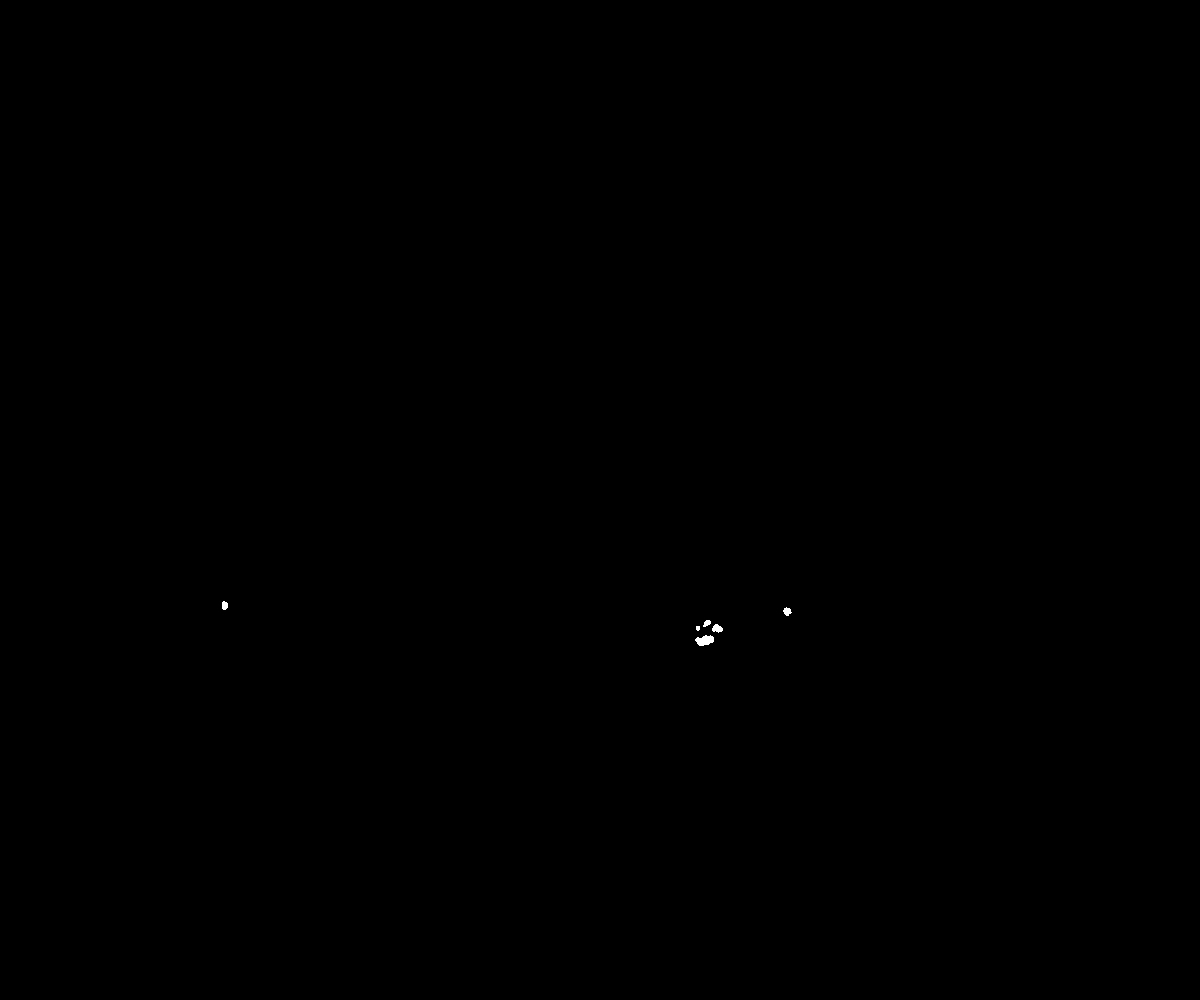}
	} 
	\subfloat[ \label{segmentation_7}]{%
		\includegraphics[width=0.5\textwidth]{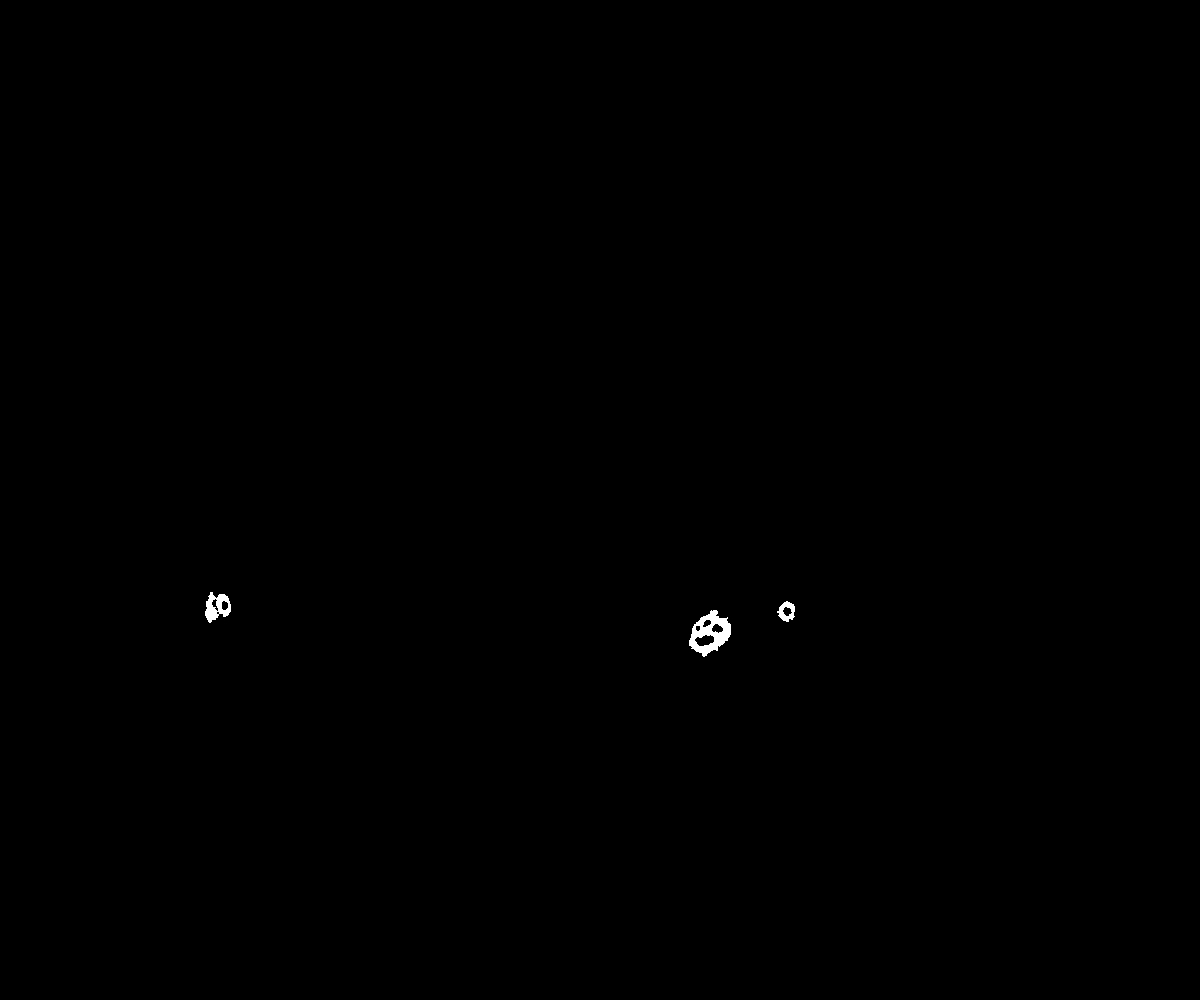}
	}\\
	\subfloat[ \label{segmentation_8}]{%
		\includegraphics[width=0.5\textwidth]{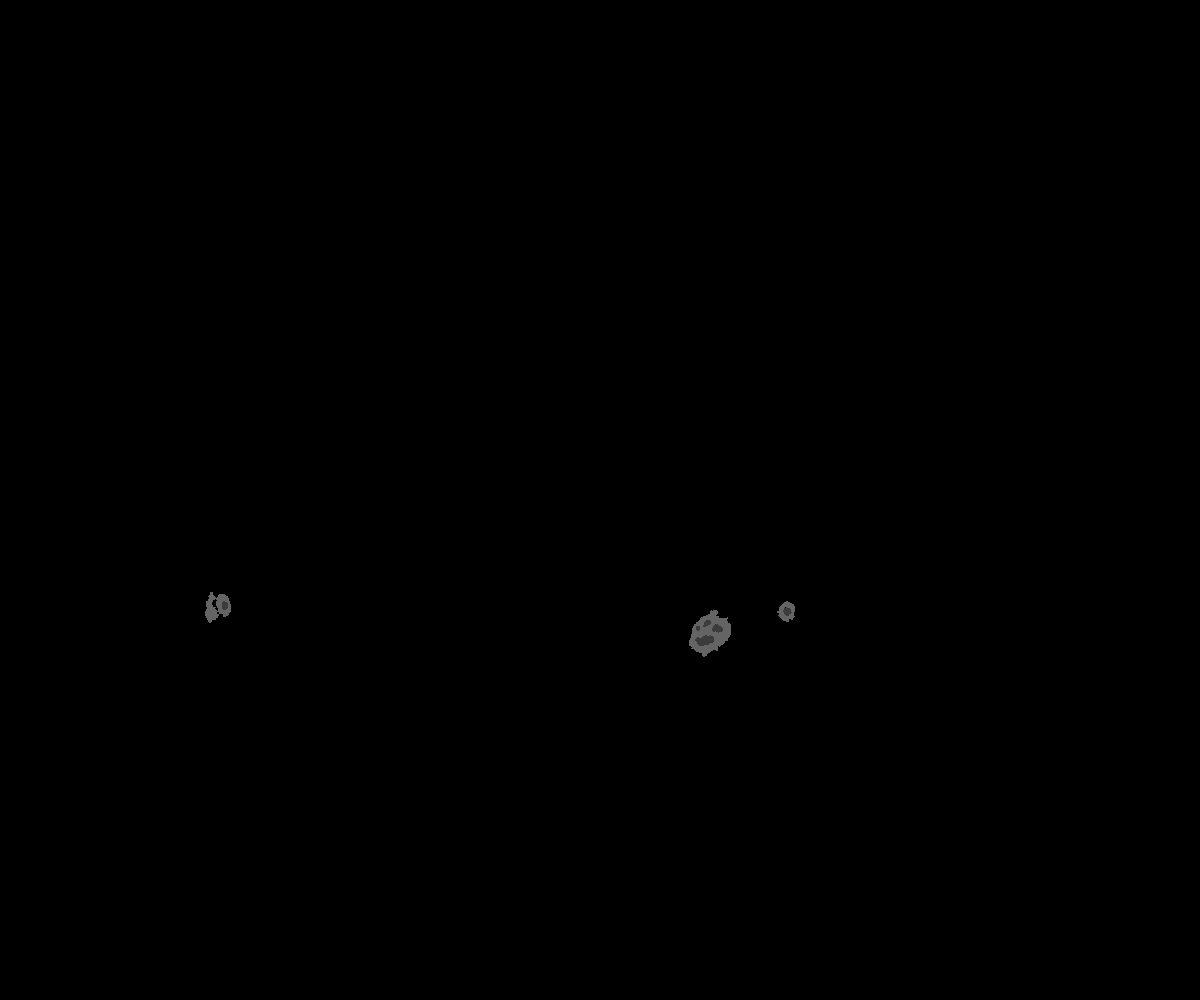}
	} 
	\subfloat[ \label{segmentation_9}]{%
		\includegraphics[width=0.5\textwidth]{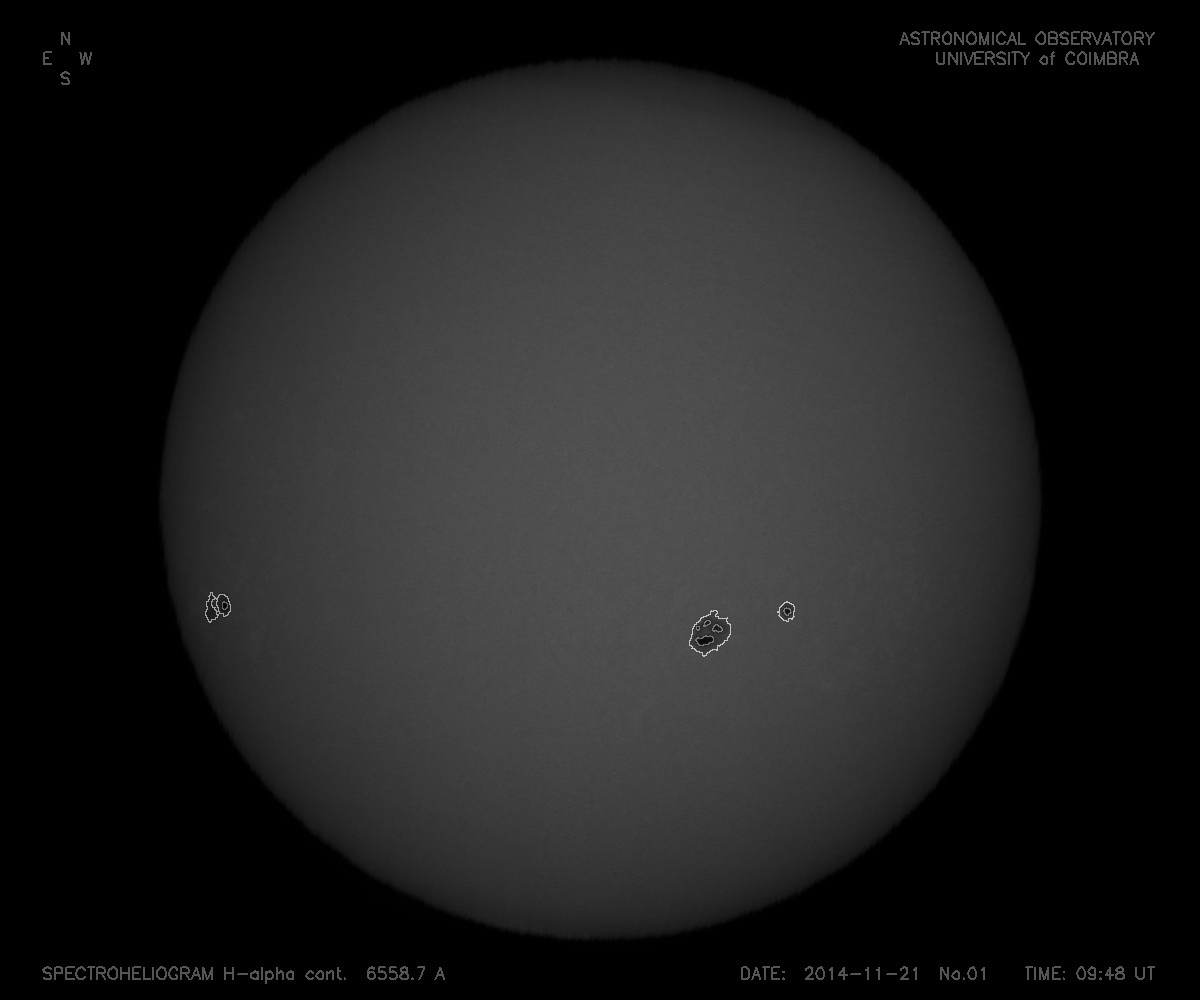} 
	}\\
	\caption{Segmentation results: (a) union of all umbras segmented; (b) union of all penumbras segmented; (c) sum of all umbras and penumbras; (d) umbras and penumbras contours superimposed over the original image.}	
\end{figure}

\section{Automatic detection of sunspots based on pixel intensity}
\label{pint}
This approach is based on the intensity of the digital level of the pixels, designated by PI. The method proposes a two-stage process: an intensity normalization of the original image, followed by a detection stage of sunspots and segmentation of umbra and penumbra. The intensity normalization aims to create a synthetic image, in which the solar disk presents homogenous digital levels and, simultaneously, the image background has a digital level of zero. The difference between this image and the original image, will be used to proceed to the detection stage. Also, like in the morphological approach, this method also needs a pre-processing step, which is explained in the next section.
\subsection{Pre-processing data}
\label{pint:preproc}
The firsts steps of pre-processing the image are similar to the morphological method (MM). As in previous method, the original image (Fig. \ref{pint_preproc_1}) is submitted to a close (Fig. \ref{pint_preproc_2}) followed by an opening (Fig. \ref{pint_preproc_3}), but by disk of size 20 in both operations.  The subtraction of the original image and the open image, whose result is shown in Fig. \ref{pint_preproc_4} is used as marker in the reconstruction, using the original image as a mask. The reconstruction operation is shown in Fig. \ref{pint_preproc_5}. From this point on, the steps of pre-processing are different: two thresholds will be applied to the reconstructed image (Fig. \ref{pint_preproc_5}) to obtain binarized solar disk. Through visual inspection the letters (the highest digital values) of the spectroheliograms occupy about 2\% of the total area of the image. Thus, the 2\% higher values will be assigned to zero. Furthermore, the radius of the solar disk is of the order of 450 pixels, that is, the solar disk fills an area about 53\% of the image. Therefore, the 45\% lowest value will be set as the lower threshold. If the value is 0, then the threshold will be replaced by 13. So, all the values of the image of Fig. \ref{pint_preproc_5} lower than lower threshold, will be assigned to zero. The image of the Fig. \ref{pint_preproc_6} shows the result of the threshold, and as it can be observed, there are still some text residue left on the image. These pixels have digital values comparable to the ones on the solar disk. If the percentil is increased to calculate the upper threshold, the pixels corresponding to the solar disk can be removed. Also, if those pixels are on the edge of the solar disk, this would affect the correct detection of them. To remove it completely, an open was applied, as shown in Fig. \ref{pint_preproc_7}. After the opening operation some images can presented some holes inside the solar disk, which can be filled by a hole fill. The result is presented on the image of Fig. \ref{pint_preproc_8}, the binary image with the digital level of 1 for the all pixels of the solar disk, and 0 for the background. The final image of pre-processing (Fig. \ref{pint_preproc_9}) is obtained by multiplying the binary image (Fig. \ref{pint_preproc_8}) and the original one (Fig. \ref{pint_preproc_1}), which allows us to obtain the solar disk without labels and with a homogeneous background.
\begin{figure}[!hb]
	\subfloat[ \label{pint_preproc_1}]{%
		\includegraphics[width=0.5\textwidth]{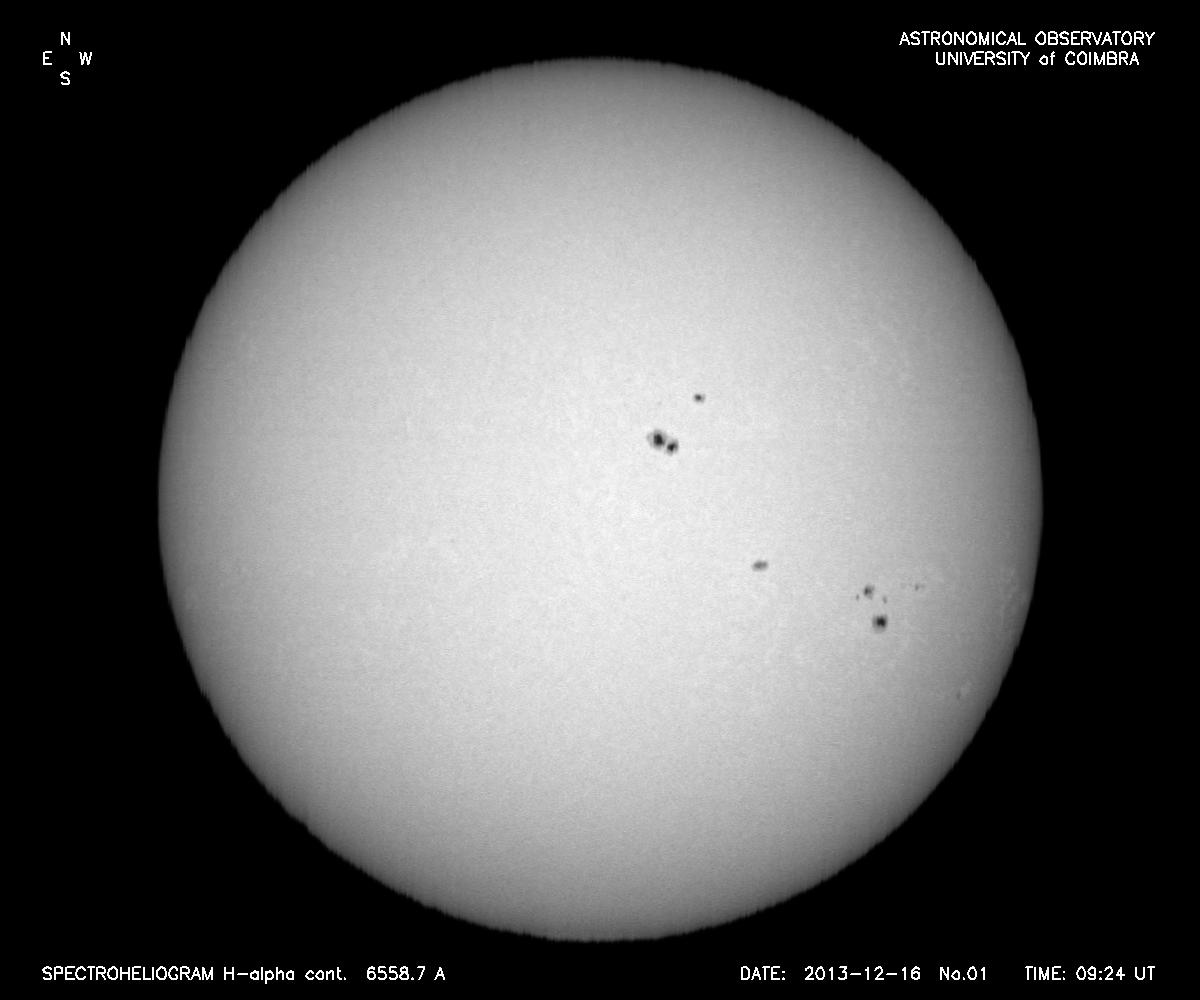}
	} \hfill
	\subfloat[\label{pint_preproc_2}]{%
		\includegraphics[width=0.5\textwidth]{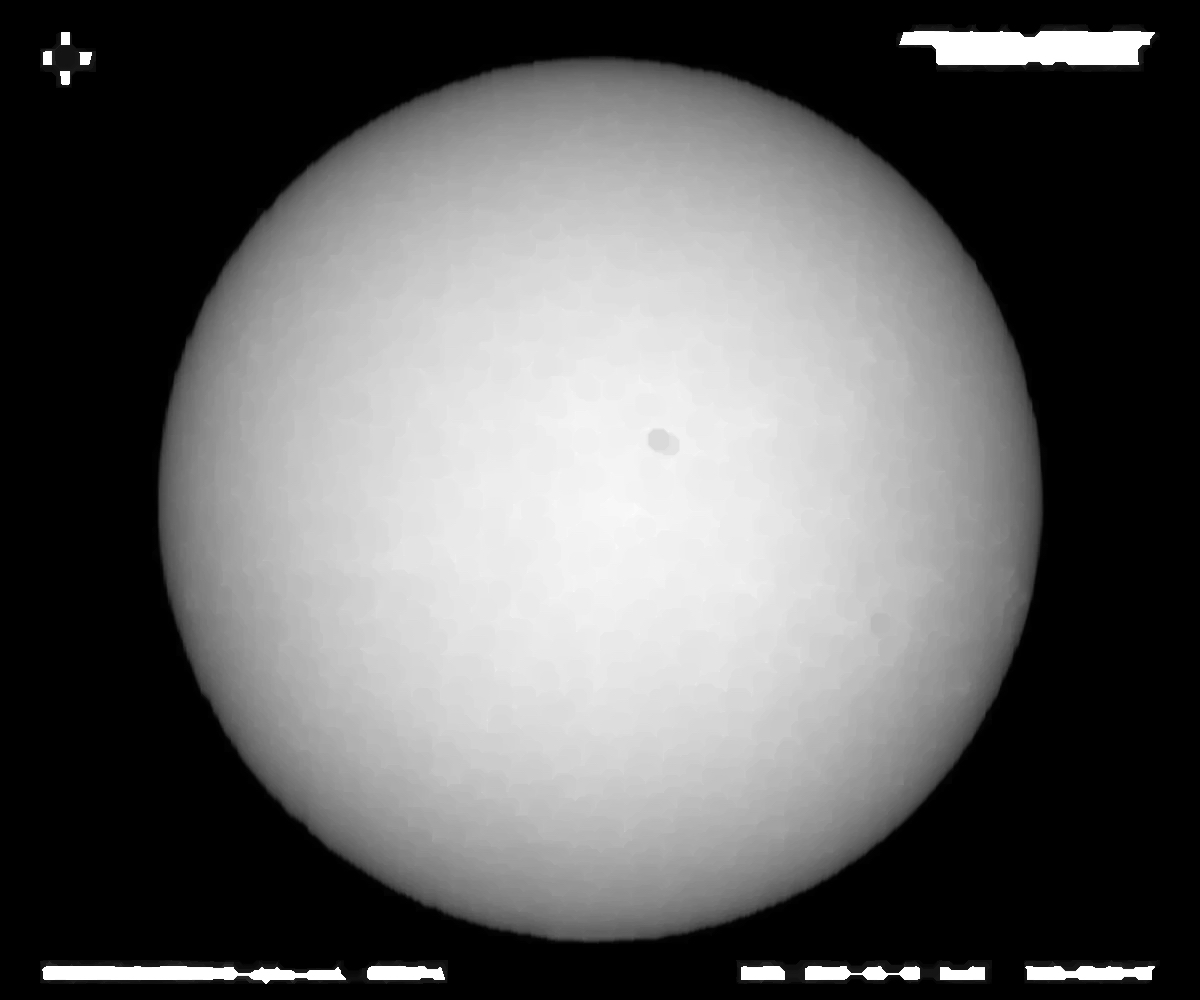}
	}\\ \hfill
	\subfloat[\label{pint_preproc_3}]{%
		\includegraphics[width=0.5\textwidth]{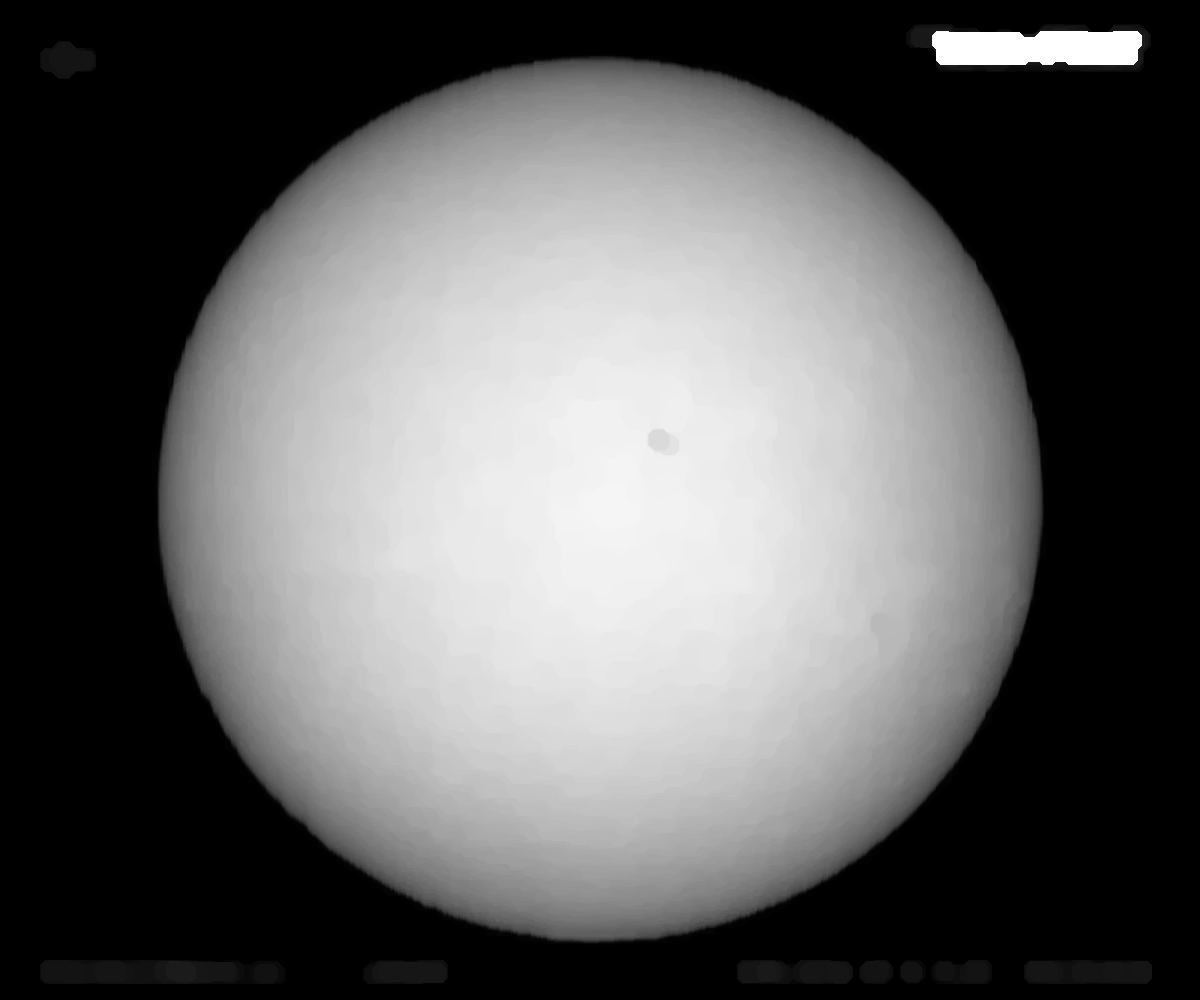}
	} 
	\hfill
	\subfloat[\label{pint_preproc_4}]{%
		\includegraphics[width=0.5\textwidth]{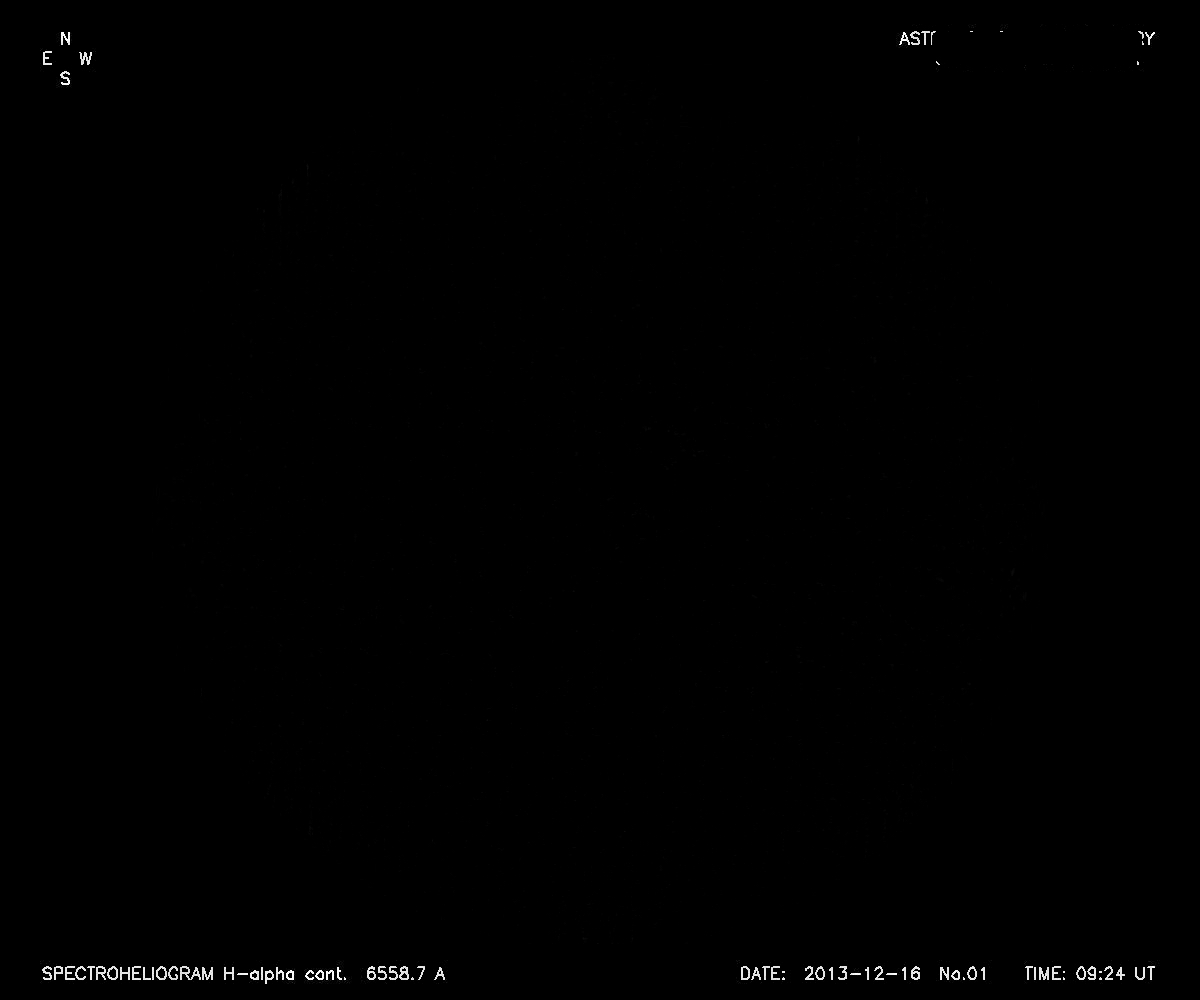}
	}	\\ \hfill
	\subfloat[\label{pint_preproc_5}]{%
		\includegraphics[width=0.5\textwidth]{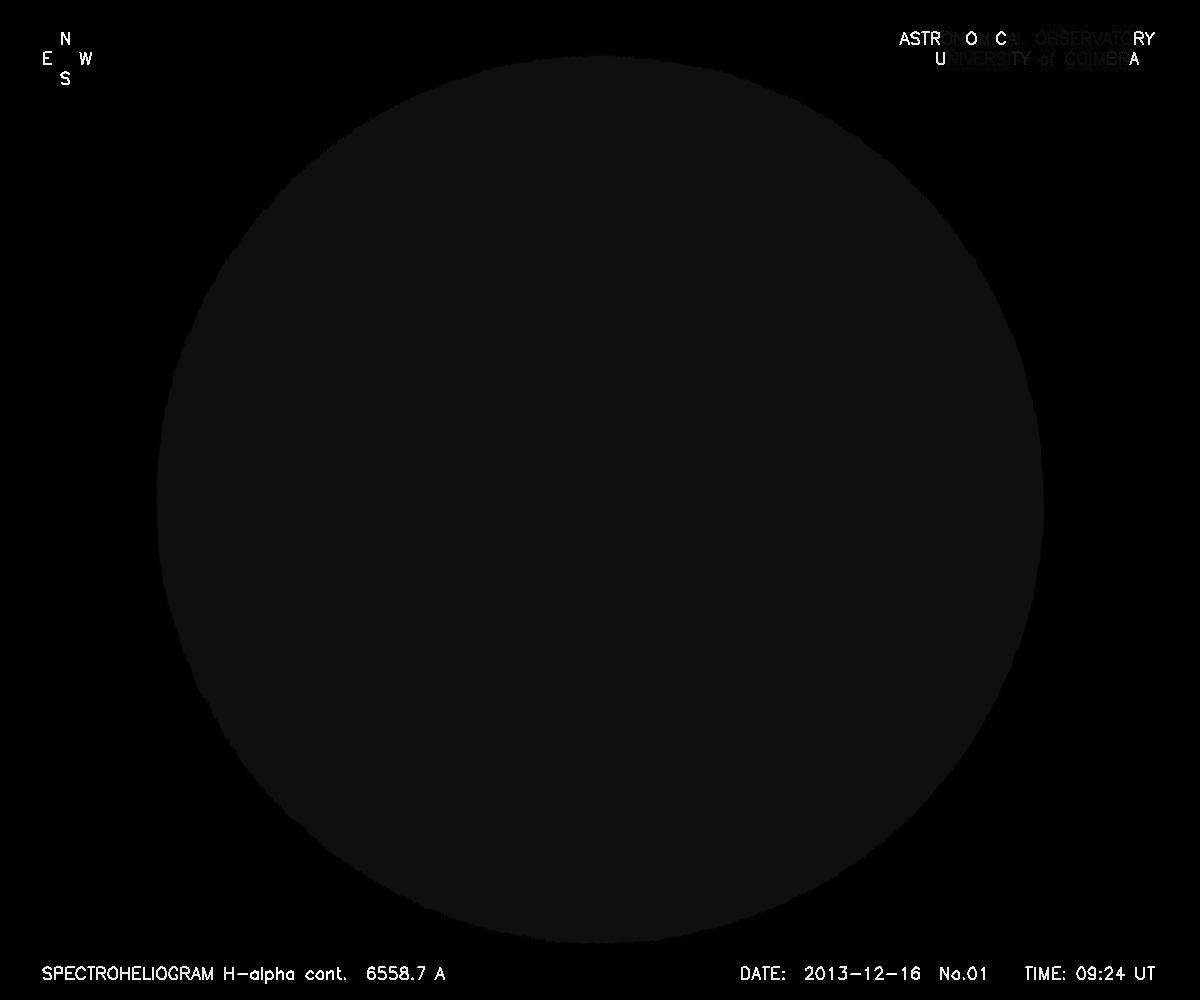}
	}
	\hfill
	\subfloat[\label{pint_preproc_6}]{%
		\includegraphics[width=0.5\textwidth]{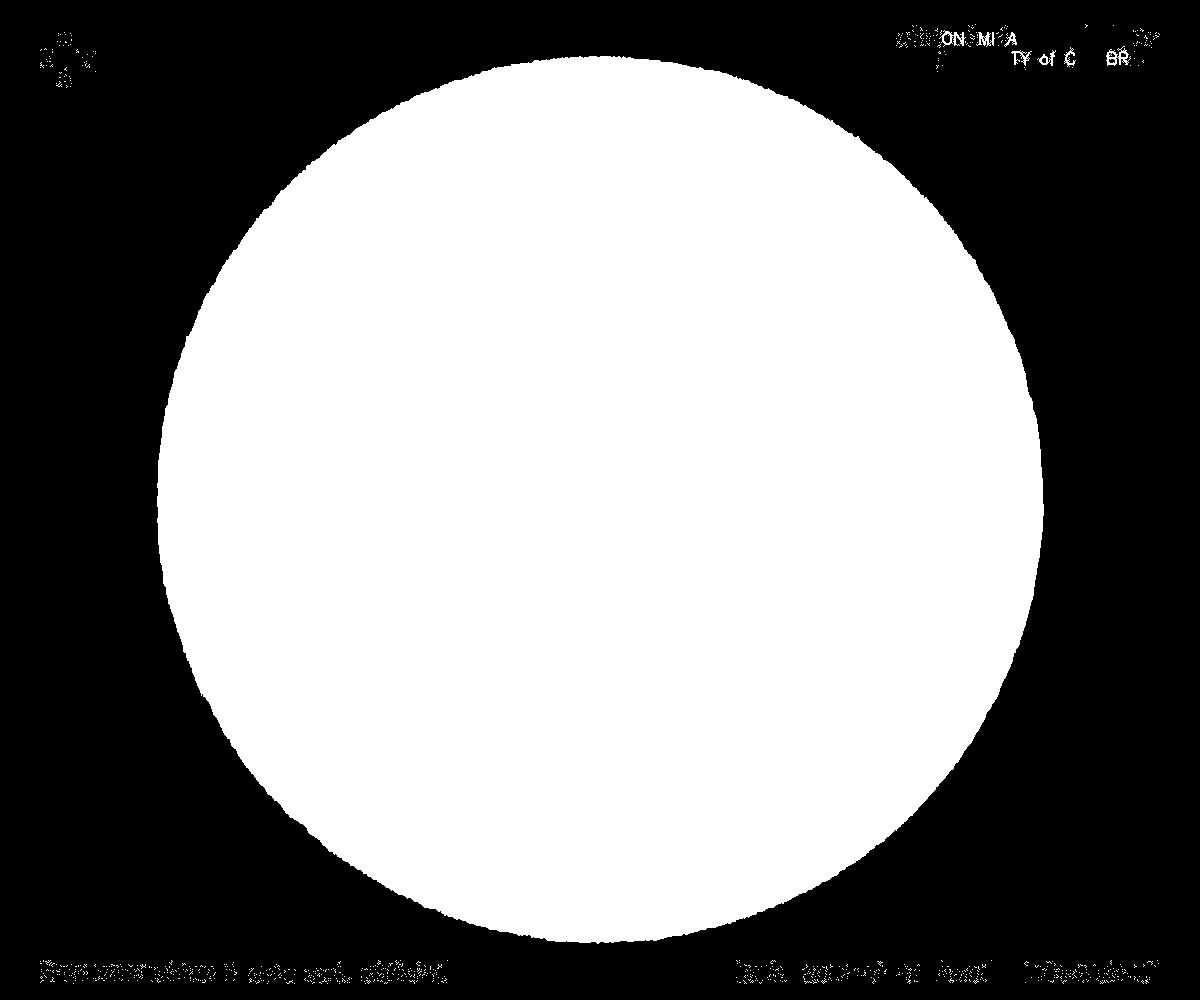}
	} 	\hfill
\phantomcaption
\end{figure}

\begin{figure}[!ht]\ContinuedFloat
		\subfloat[\label{pint_preproc_7}]{%
		\includegraphics[width=0.5\textwidth]{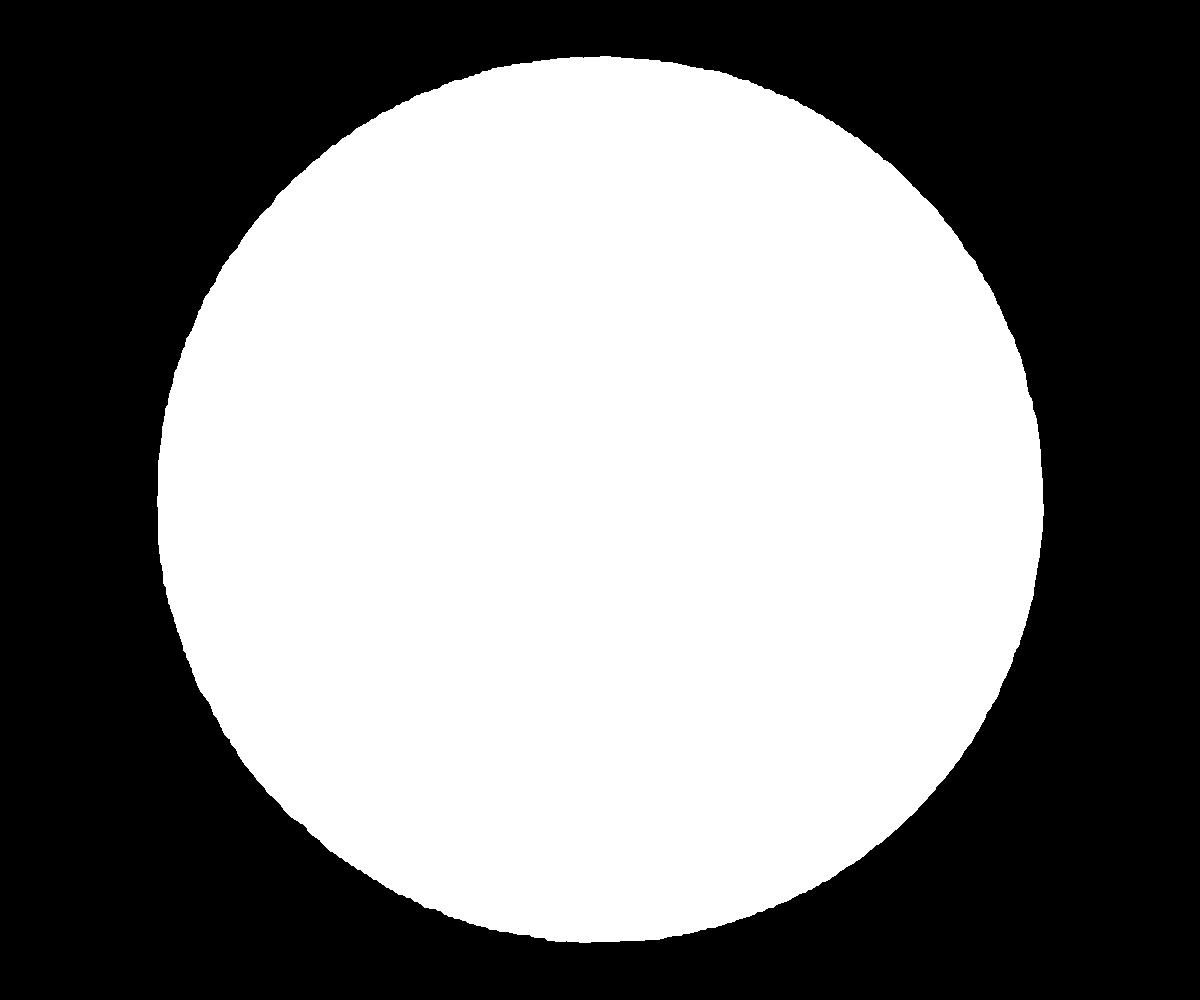}
	}	\hfill
	\subfloat[\label{pint_preproc_8}]{%
		\includegraphics[width=0.5\textwidth]{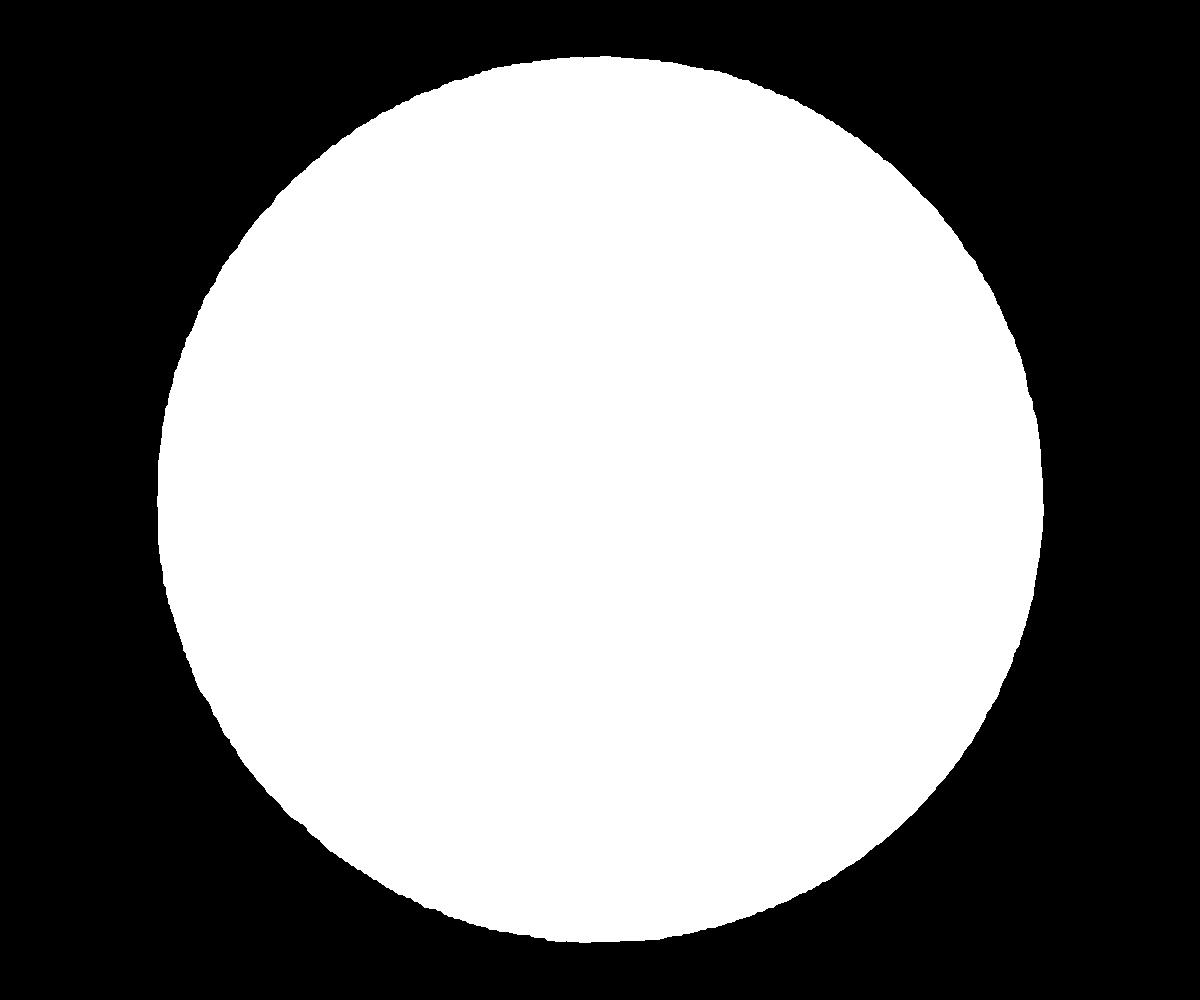}
	}\\  \\  \hfill
	
	\qquad\qquad\qquad\qquad\quad	
	\subfloat[\label{pint_preproc_9}]{%
		\includegraphics[width=0.5\textwidth]{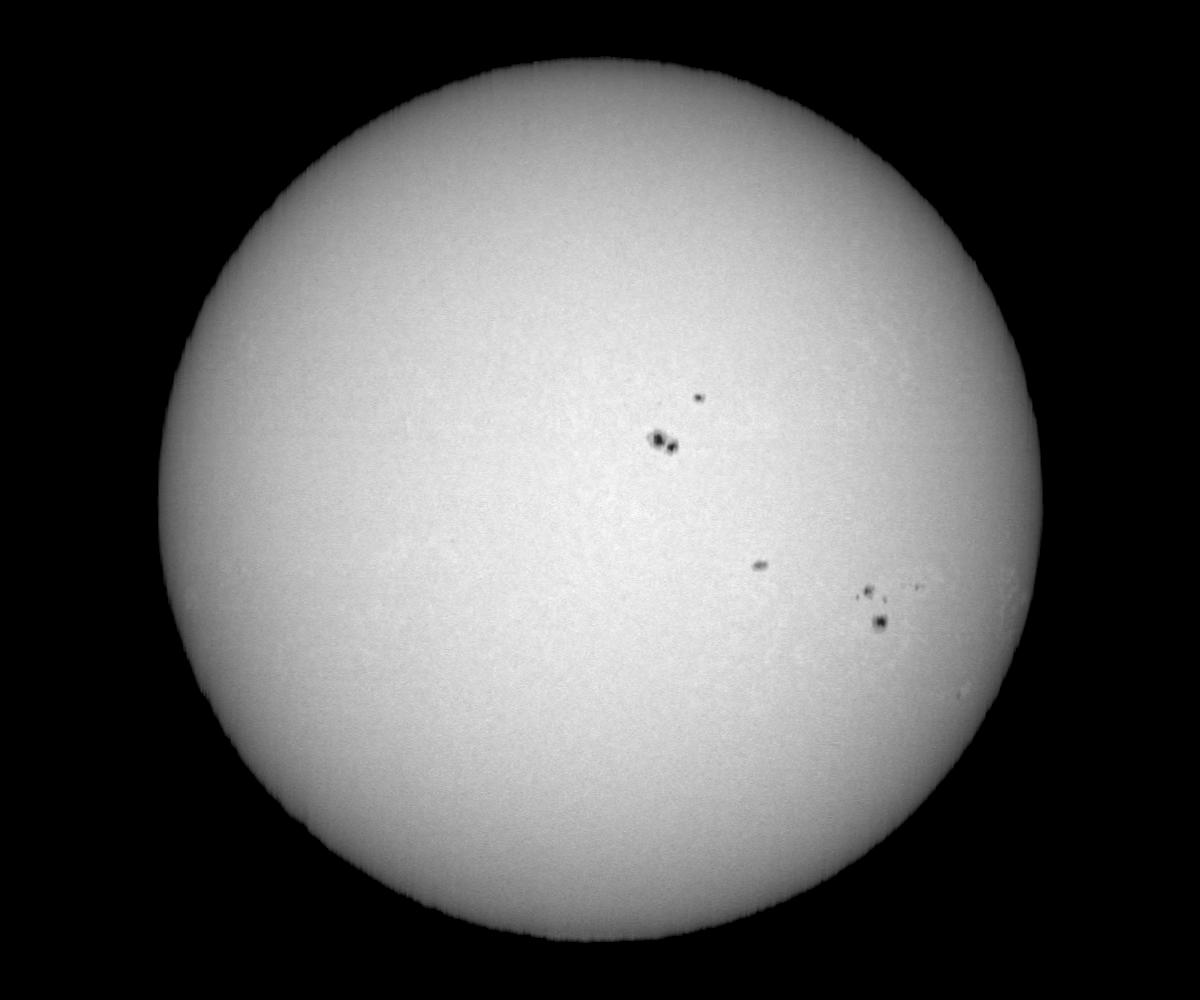}
	}\\
	
	\caption{Pre-processing: (a) original image; (b) close image (a); (c) open of the image (b); (d) subtraction of the image (c) by (a);  (e) reconstruction reconstruction of the image marker (d) under the image mask (a); (f)  threshold of the image (e); (g) open of the image (f); (h) hole fill of the image (g); (i) final image of the pre-processing obtained by the multiplication of (h) by (a).}
	\label{pint:preprocess}
\end{figure}

\begin{figure}[!ht]
	\centering
	\subfloat[ \label{pint_solardisk_1}]{%
		\includegraphics[width=0.8\textwidth]{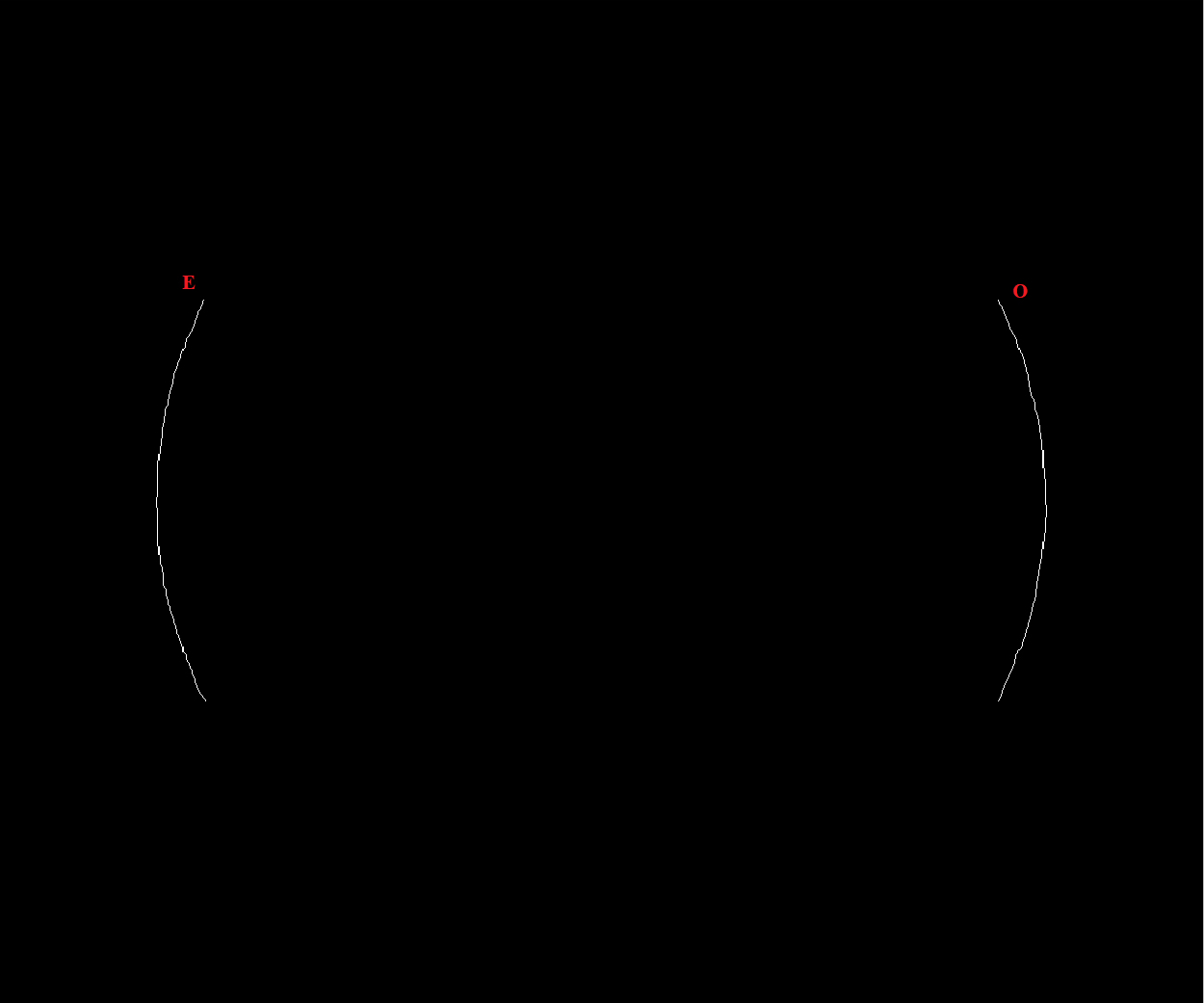}
	} \\
	\subfloat[\label{pint_solardisk_2}]{%
		\includegraphics[width=0.8\textwidth]{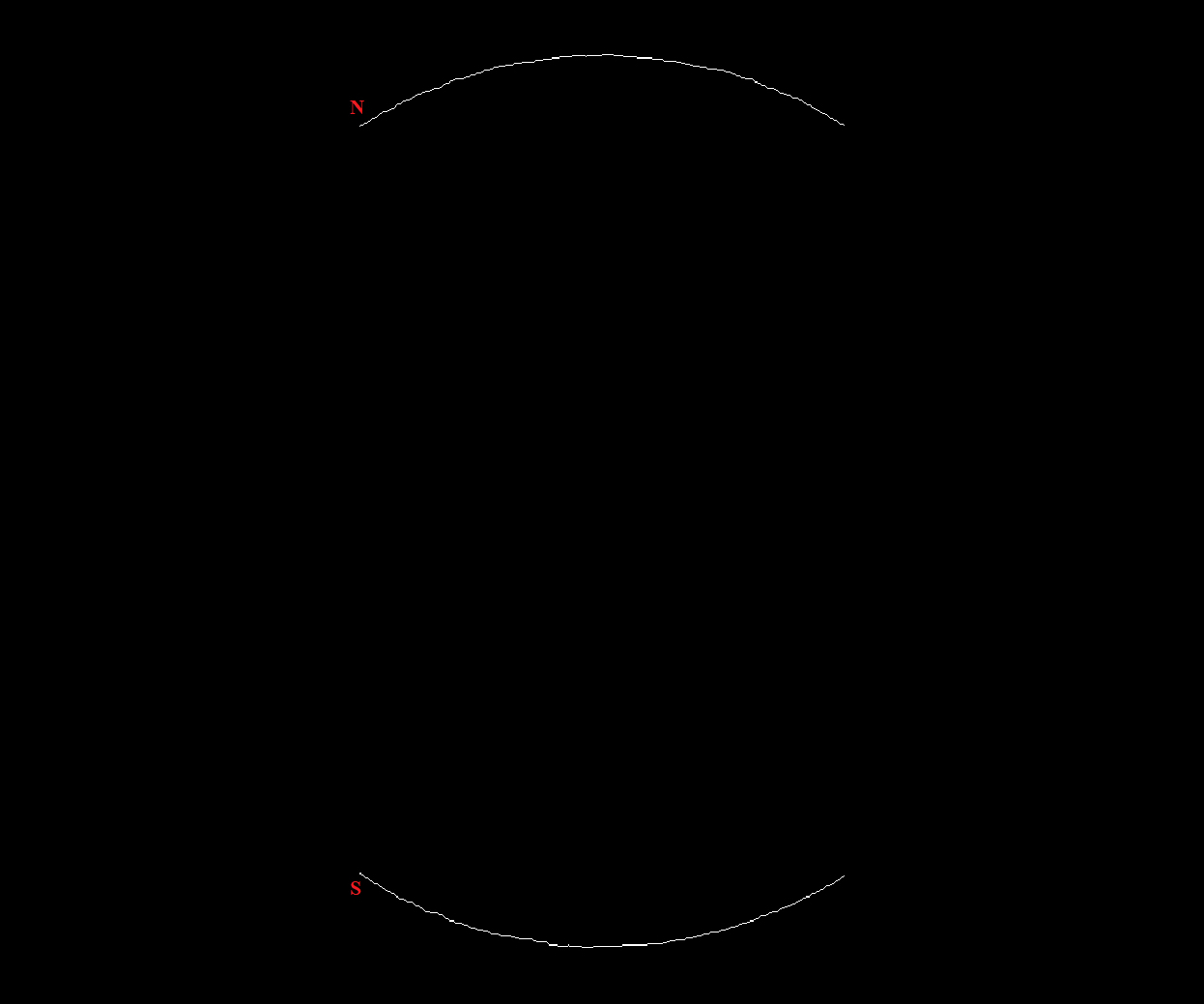}
	}\\	
	\caption{The solar disk contours: (a) rows and (b) columns.}
	\label{pint:solardisk}
\end{figure}

The calculation of the center and the solar disk radius is a crucial step to obtain the limb darkening profile, which in turn is fundamental for the algorithm based on the intensities. Following \citet{denker1999}, to determinate the x-coordinate of the center is necessary to calculate the first ($x_{FIRST}$) and last pixels ($x_{LAST}$), with values higher than zero, of 20\% of the rows above the geometrical center of the image, and of 20\% of the rows below the geometrical center. The y-coordinate of the center is calculated analogously. This concept is illustrated on Fig. \ref{pint_solardisk_1} for rows and Fig. \ref{pint_solardisk_2} for the columns. Due to the circular shape of the solar disk the pixels of the outermost solar disk of the center will induce several errors due to their geometrical obliquity with the rows of the image. 

The horizontal and vertical coordinates of the center of the disk are given by the equations \ref{h_center} and \ref{v_center}, respectively, 

\begin{equation}\label{h_center}
Horizontal = \frac{1}{1000 \cdot 0.4} \cdot \sum_{i=300}^{700} \Bigg({x_{FIRST}}_{i} + \frac{{x_{LAST}}_{i}-{x_{FIRST}}_{i}}{2} \Bigg),
\end{equation}

\begin{equation}\label{v_center}
Vertical = \frac{1}{1200 \cdot 0.4} \cdot \sum_{i=360}^{840} \Bigg({y_{FIRST}}_{i} + \frac{{y_{LAST}}_{i}-{y_{FIRST}}_{i}}{2} \Bigg),
\end{equation}

where the limits of the sum in \ref{h_center} are obtained by,

\begin{align} \label{horlimits}
\begin{cases}
\textit{upper limit} = \frac{n_{rows}}{2} + n_{rows} \cdot 0.2 &\\
& ,n_{rows} = 1000\\
\textit{lower limit} =  \frac{n_{rows}}{2} - n_{rows} \cdot 0.2 
\end{cases}
\end{align}

where the limits of the sum in \ref{v_center} are obtained by,

\begin{align} \label{verlimits}
\begin{cases}
\textit{upper limit} = \frac{n_{columns}}{2} + n_{columns} \cdot 0.2 &\\
& ,n_{columns} = 1200\\
\textit{lower limit} =  \frac{n_{columns}}{2} - n_{columns} \cdot 0.2 
\end{cases}
\end{align}

The $n_{rows}$ and $n_{colums}$ are the number of the rows and columns, respectively.

\begin{figure}[!ht]
	\centering
	\includegraphics[width=0.8\textwidth]{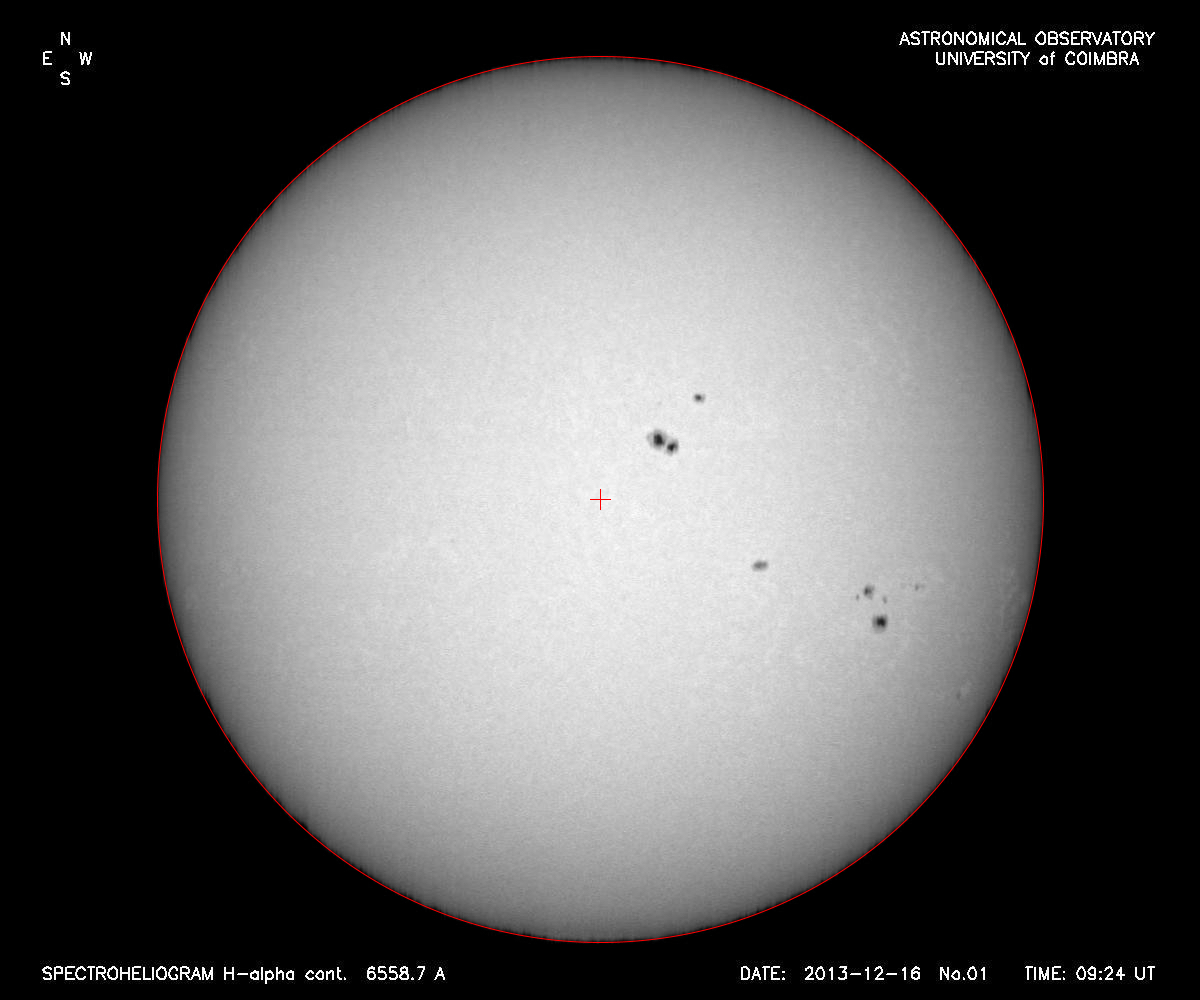}
	\caption{Center of the solar disk and the circle superimposed to the original image.}
	\label{pint_center}
\end{figure}

Finding the center of the solar disk allows to determinate the solar radius, which can be obtained from the average between the distances of every pixel used in the calculation of the center, and the center itself (Fig. \ref{pint_solardisk_1} and \ref{pint_solardisk_2}). If the solar radius is zero, that means that the lower threshold applied in the pre-processing phase is too high. This can be solved by repeating all the steps (from the two thresholds). In this case, the lower threshold must be 1\% smaller than the previous, until the solar disk radius being higher than zero. Fig. \ref{pint_center} shows an example of a correctly identified center of the solar disk. 

\subsection{Intensity normalization}
\label{pint:intnorm}

\begin{figure}[!ht]
	\centering
	\includegraphics[width=0.6\textwidth]{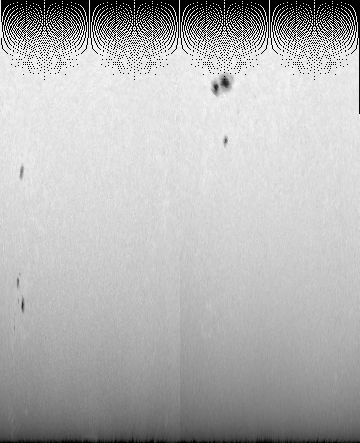}
	\caption{Image of Fig. \ref{pint_preproc_9} converted to polar coordinates.}
	\label{pintpolar}
\end{figure}

The central regions of the solar disk are, normally, brighter than the contour of the disk due the effect of limb darkening. The intensity normalization aims to eliminate this phenomenon. Due the fact that limb darkening has a radial effect, it is preferable to work on images in polar coordinates. The input image for the cartesian to polar coordinate conversion, is the final image of the pre-processing step (Fig. \ref{pint_preproc_9}). The output of the conversion is a rectangular image where the number of rows is equal to the rounded value of the solar disk radius, i.e., each row represents the radial distance. The number of columns is $360$ and each column corresponds to the angle from 0 to 359 degrees. Fig. \ref{pintpolar} shows the result of the coordinates conversion.

\begin{figure}[!ht]
	\centering
	\subfloat[ \label{limbprofile}]{%
		\includegraphics[width=0.7\textwidth]{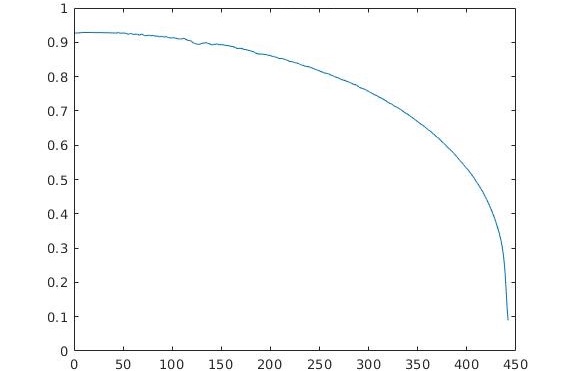}
	} \\
	\subfloat[\label{synthetic}]{%
		\includegraphics[width=0.7\textwidth]{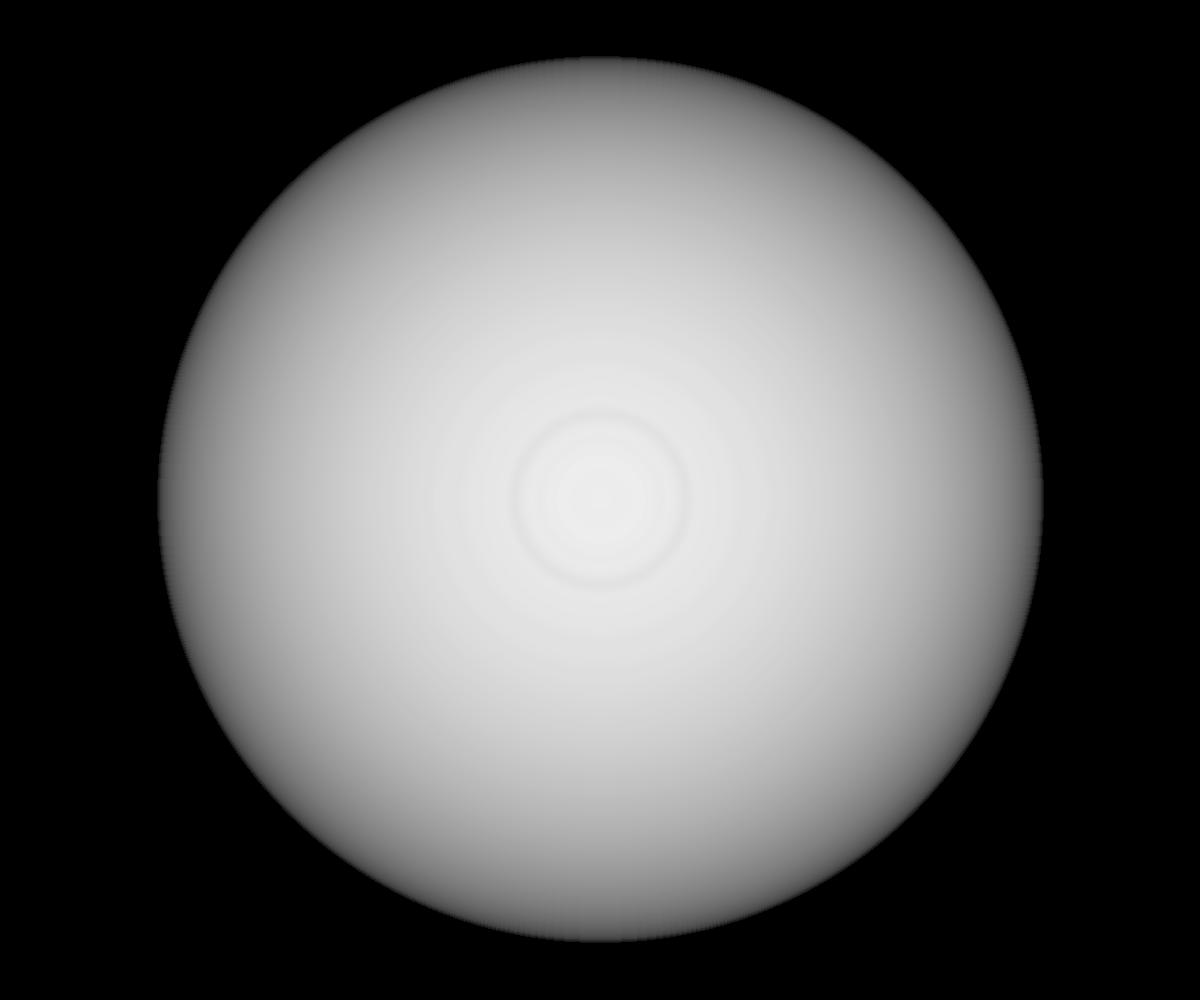}
	}\\	
	\caption{Limb darkening process: (a) limb darkening profile; (b) synthetic image of solar disk.}
	\label{pint:limbdarkning}
\end{figure}

The image of Fig. \ref{pintpolar} allows to calculate the mean value of each row, represented by a vector, obtaining the limb darkening profile from the average value for each radial position (Fig. \ref{limbprofile}). Any eventual sunspots present at small radial distances would strongly affect the average value of the profile, since that value is computed from a low number of pixels from the input image. Therefore, the 10\% inner radial position will be replaced by a median of higher radial values. This process is given by the equation \ref{radial}, where $m$ is equal to 10\% of the number of rows and the radial positions. 

\begin{equation}\label{radial}
r_i = \frac{\sum\limits_{n=i}^{m} r_n}{m-i+1}
\end{equation}

The limb darkening profile allows to create a synthetic solar image. This image has the same dimensions as the original image. The values of the limb darkening profile are assigned to each pixel, according their distance from the solar disk center (Fig. \ref{synthetic}). 

\begin{figure}[!ht]
	\centering
	\includegraphics[width=0.8\textwidth]{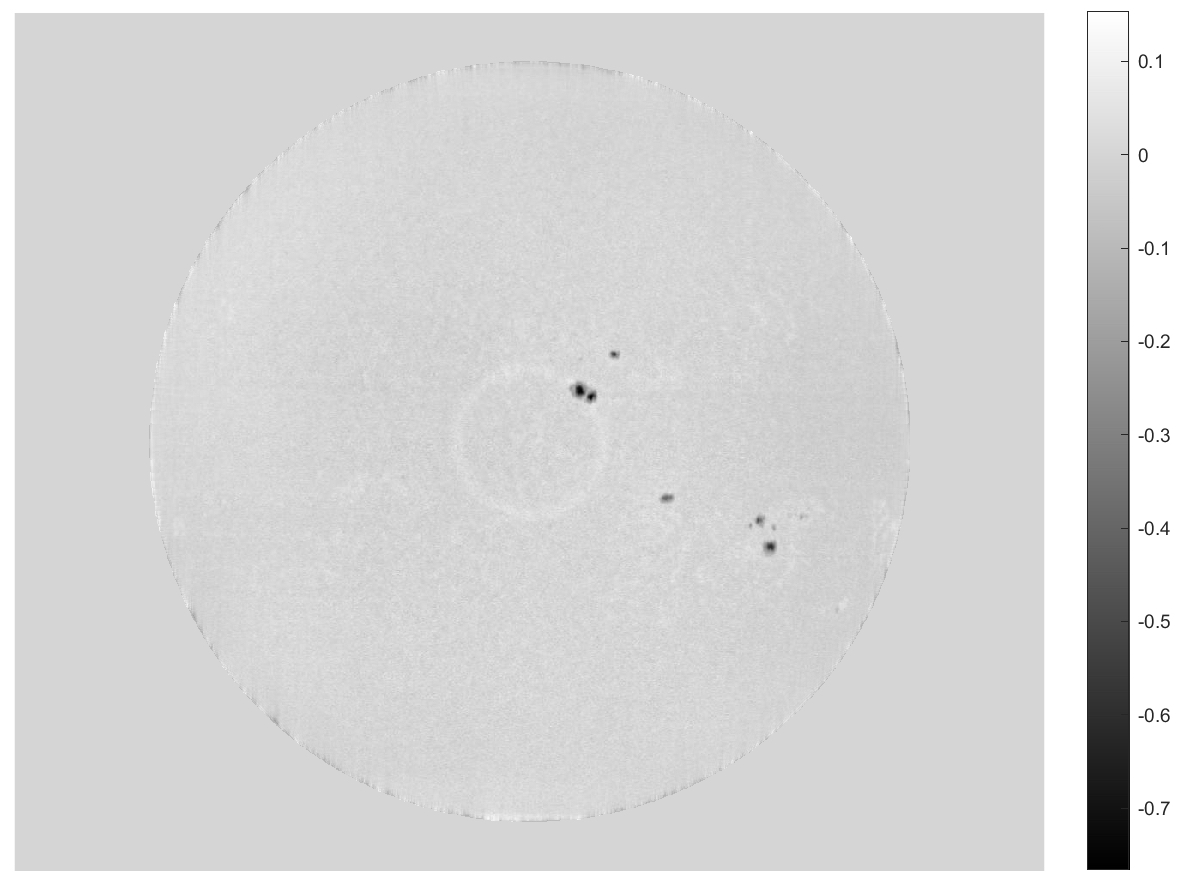}
	\caption{Final image of intensity normalization.}
	\label{pintnormalization}
\end{figure}

The final step of the intensity normalization consists in subtracting the synthetic image of Fig. \ref{synthetic} to the final image of the pre-processing step (Fig. \ref{pint_preproc_9}). The result can be observed on Fig. \ref{pintnormalization}. This resulting image will be used to proceed to the automatic detection of the approach based on pixel intensity.

\subsection{Detection of sunspots}
\label{pint:detsunspots}

The automatic detection allows to segment the umbra and penumbra of sunspots simultaneously. The detection starts with the application of two different thresholds, each one applied to the penumbra and umbra. The lower value threshold is applied to detect the penumbra, since all values between the lower and higher thresholds are penumbra. The higher threshold value determines the umbra. The best threshold values are based on a calibration process. Several values were applied, and the results compared with a set of 10 ground truth images, in order to choose the values with the best results. This comparison is made based on the true positive rate (TPR) and the true negative rate (TNR), which can be expressed by,

\begin{equation}\label{TPR}
TPR = \frac{TP}{TP+FN},
\end{equation}

\begin{equation}
TNR = \frac{TN}{FP+TN},
\end{equation}

where TP, is the number of true positives, TN the number of true negatives, FP the number of false positives and FN the number of false negatives. True positive (TP) means that a pixel in the output image detected as part of a sunspot is a sunspot pixel in the ground-truth image, and false positive (FP) means a pixel detected as part of a sunspot in the output image that is not a sunspot pixel in the ground-truth image. The false negative (FN) is a pixel not detected as part of a sunspot in the output image that is actually a sunspot pixel in the ground-truth image.
Additionally, two metrics were also used for the determination of the best thresholds: Quality Index (Q) and Overall Accuracy (OA). The following expression give us the Q,

\begin{equation}\label{Q}
Q = \frac{TP}{TP+FP+FN}.
\end{equation}

The Overall Accuracy (OA), is the ratio of pixels that were correctly classified to all the classified pixels. OA is used to evaluate the umbra penumbra differentiation, and is obtained by the following expression,

\begin{equation}\label{OA}
OA (\%) = \frac{UU + PP}{UU+UP+PU+PP} *100,
\end{equation}

where UU is the number of umbra pixels that are detected as umbra; PP is the number of penumbra pixels that are detected as penumbra; UP is the number of umbra pixels detected as penumbra pixels, and PU is the  number of penumbra pixels detected as umbra pixels. The metrics obtained are summarized on table \ref{table:evaluationthresh}.

\begin{table}
	\centering
\begin{tabular}{|m{1.6cm}|m{1.6cm}|m{1.4cm}|m{1.5cm}|m{1.3cm}|m{1.3cm}|}
		\hline 
		Lower Threshold (\%)  & Upper Threshold (\%)   &  TPR (\%) & TNR (\%) & Q(\%) &  OA(\%) \\
		\hline
		6.00 & 30.00 & 89.60 & 99.99 & 83.00 & 91.82 \\
		6.30 & 28.00 & 88.90 & 99.99 & 83.28 & 90.38 \\
		7.00 & 25.00 & 85.12 & 99.99 & 80.12 & 88.08 \\
		\hline
	\end{tabular}
	\caption{Evaluation performance to determinate the threshold values based on 10 images.}
	\label{table:evaluationthresh}
\end{table}

\begin{figure}
	\centering
	\subfloat[ \label{noise_1}]{%
		\includegraphics[width=0.5\textwidth]{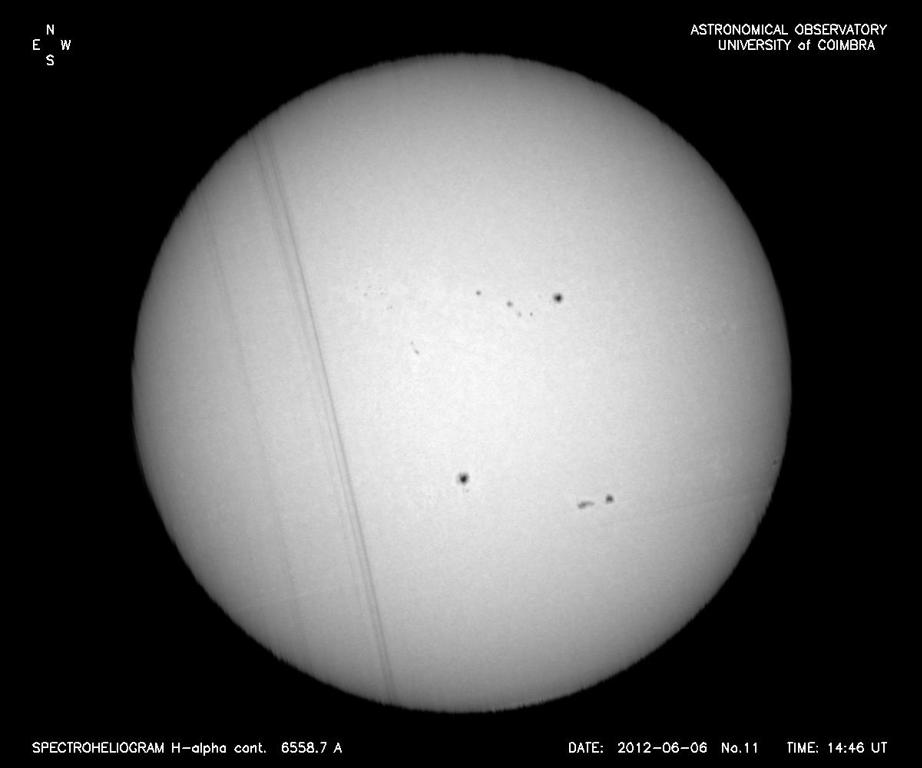}
	} 
	\subfloat[ \label{noise_2}]{%
		\includegraphics[width=0.5\textwidth]{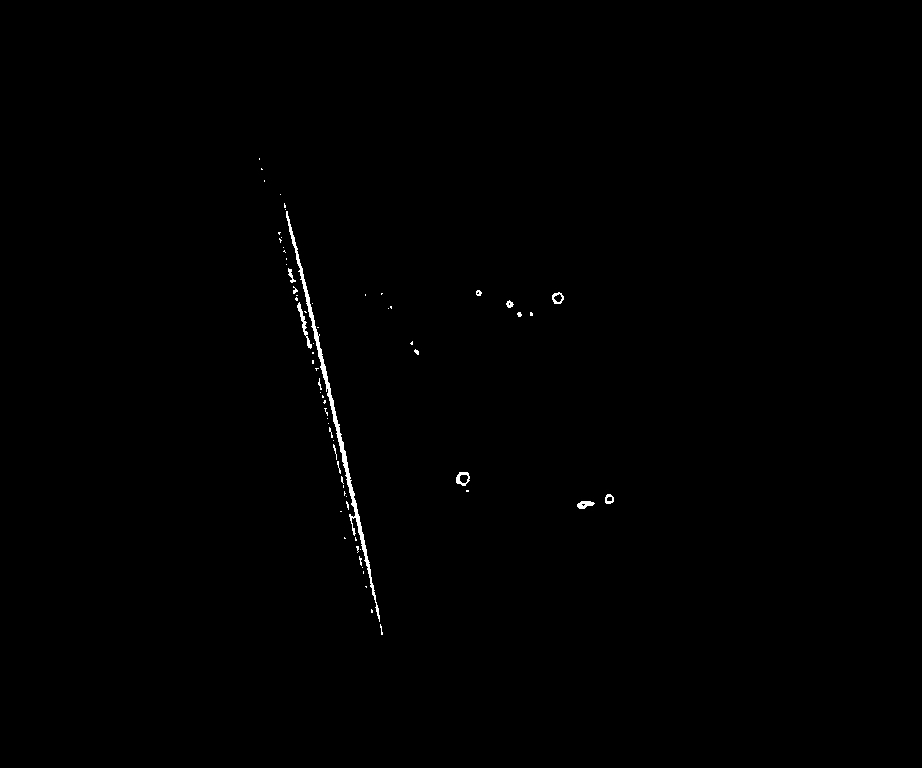}
	} \\ \hfill
	\subfloat[\label{noise_3}]{%
		\includegraphics[width=0.5\textwidth]{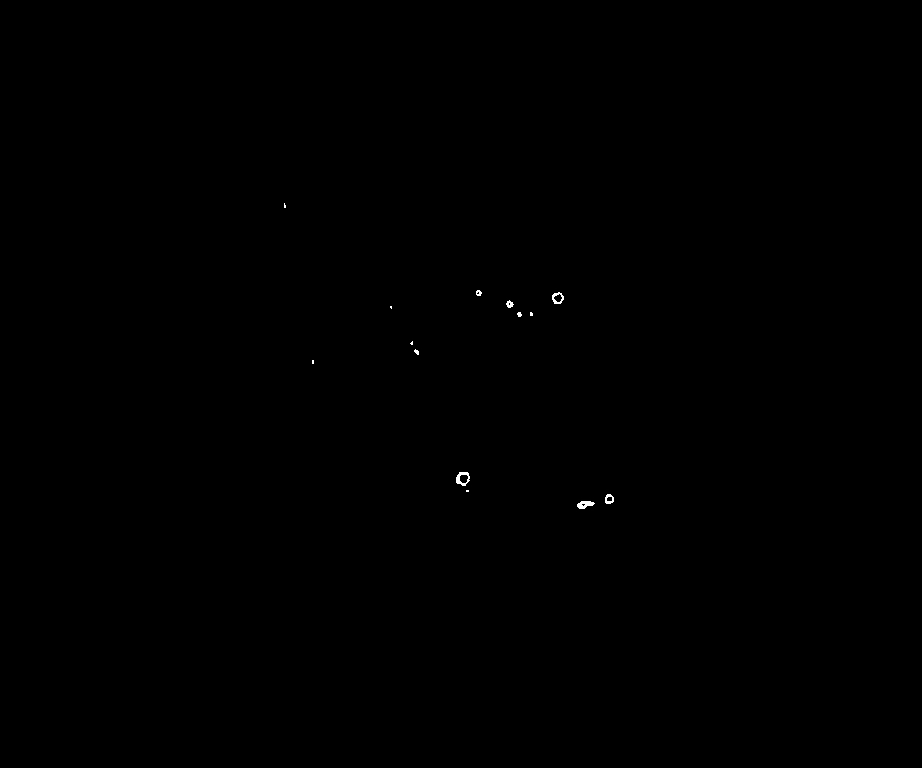}
	}
	\subfloat[ \label{noise_4}]{%
		\includegraphics[width=0.5\textwidth]{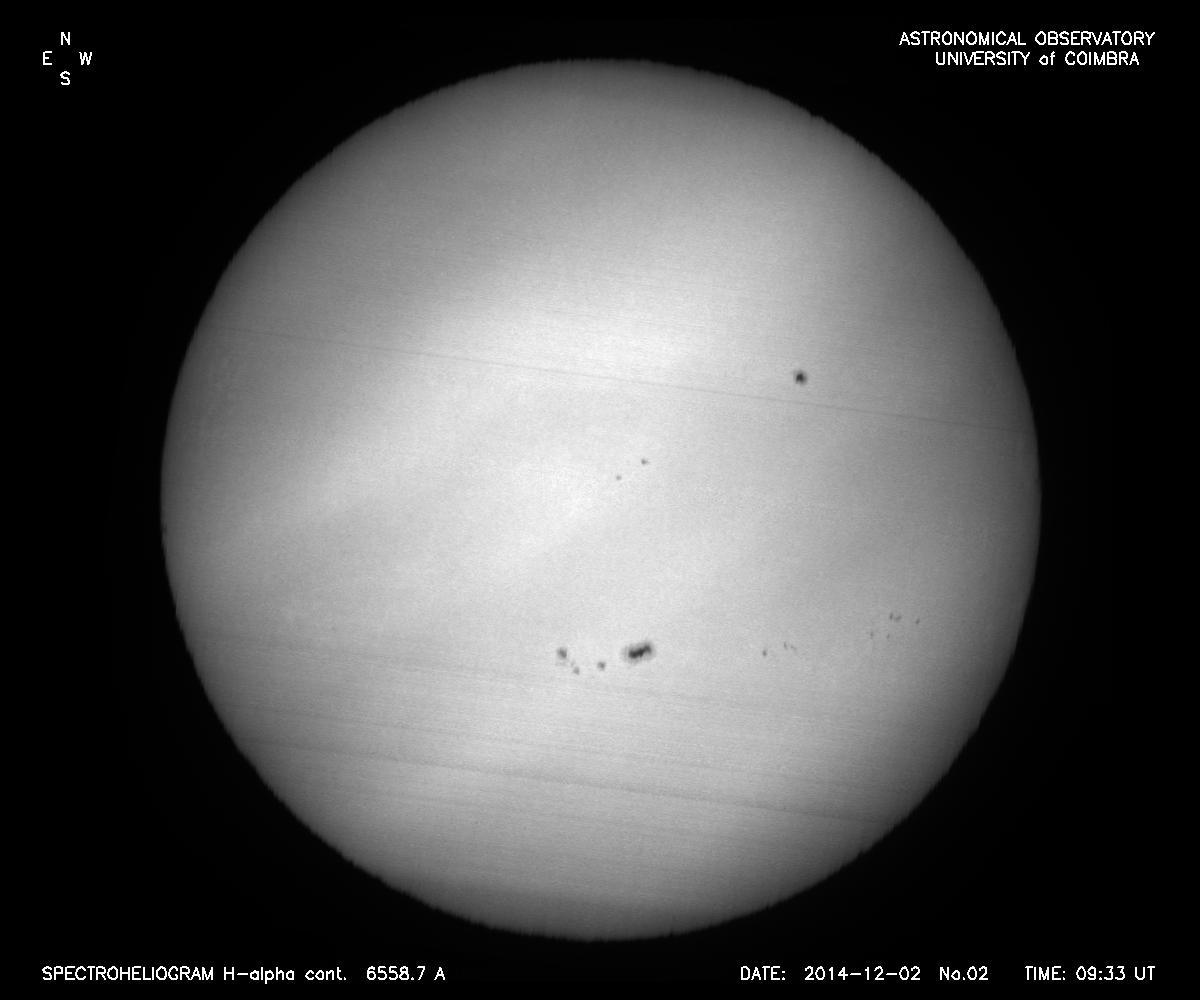}
	} \\ \hfill
	\subfloat[\label{noise_5}]{%
		\includegraphics[width=0.5\textwidth]{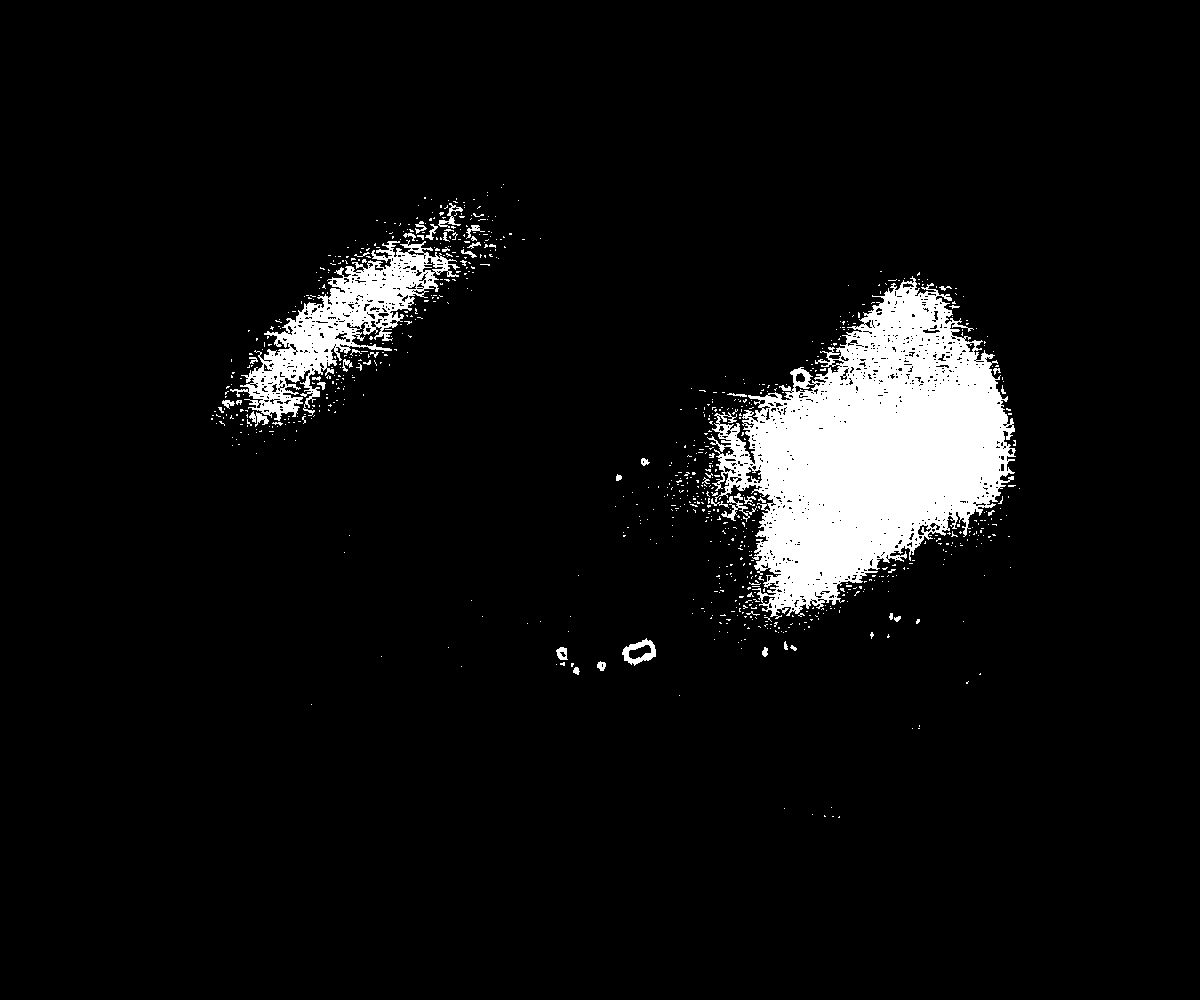}
	}
	\subfloat[\label{noise_6}]{%
		\includegraphics[width=0.5\textwidth]{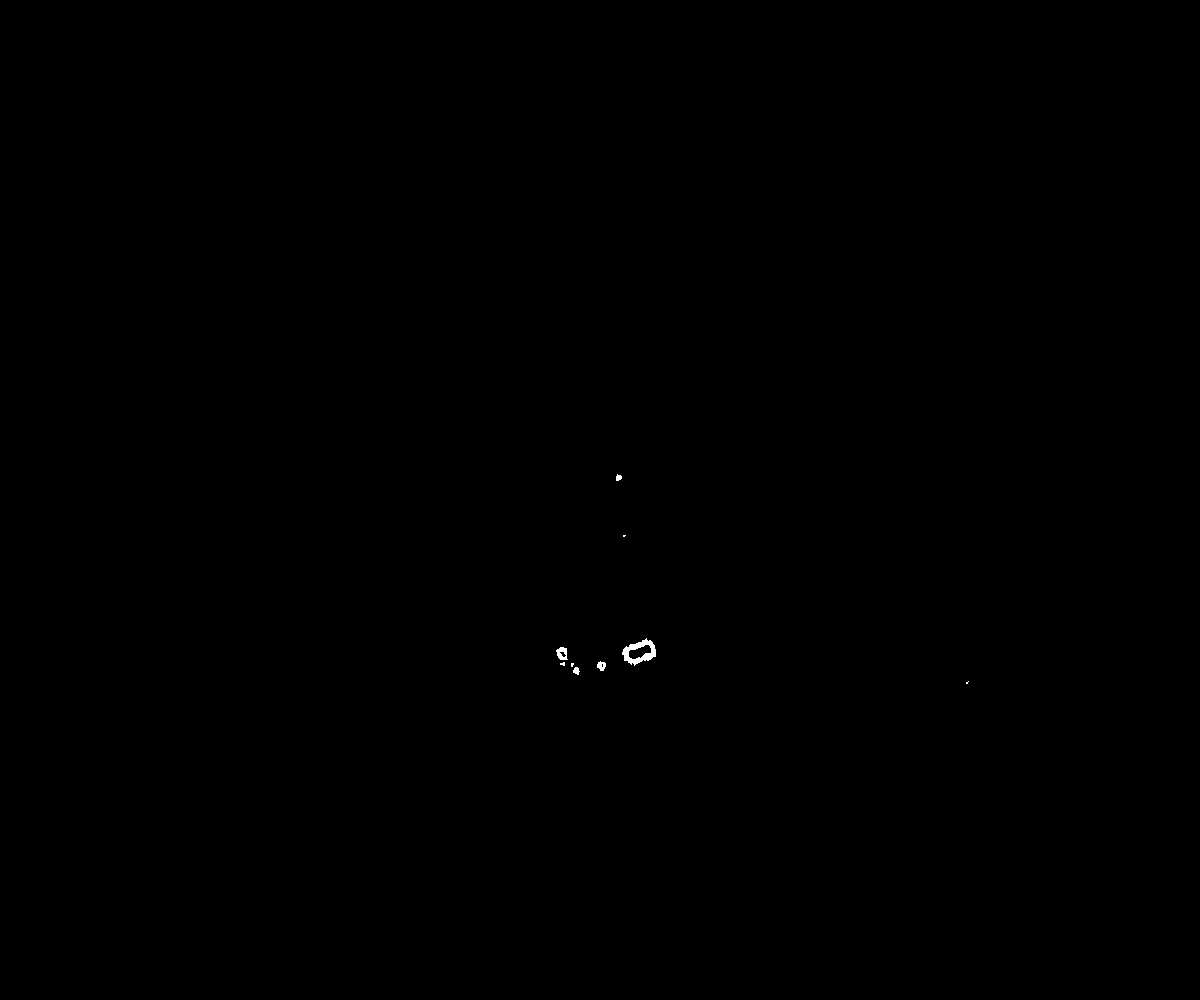}
	}
	\caption{Examples of the atmospheric effects and noise on spectroheliograms: (a) original image of 6/6/2012; (b) threshold of the image (a); (c) close of the image (b) followed by a segmentation; (d) original image of 1/12/2014; (e) threshold of the image (d); (f) close of the image (e) followed by a segmentation.}
	\label{pint:noise}	
\end{figure}

Based on the values of table \ref{table:evaluationthresh}, the lower and upper thresholds chosen are 6\% of the maximum value of the grayscale and 30\%, respectively. The threshold of the penumbra is not high enough to clear all the pixels that do not contain solar features. A higher value would discard pixels of the penumbra, so a filtering process is required to eliminate as much as possible the consequent false positives. The atmospheric effects and noise on the images acquired from the ground can increase the number of the false positives. Fig. \ref{pint:noise} shows examples of atmospheric effects and noise that can be observed on spectroheliograms. Fig. \ref{noise_1} shows some streaks. After the application of the two thresholds a morphological close was applied, followed by segmentation process based on the connectivity of the pixels (Fig. \ref{noise_2}). Then, all the segments of Fig. \ref{noise_2} with a height of at least four times the width of the segment will be rejected (Fig. \ref{noise_3}). Concerning strong atmospheric effects, like large clouds (Fig. \ref{noise_4}), the same steps are applied, but now we need to use empirical information on the largest sunspots registered until today. On April $8^{th}$, 1947, the largest sunspot was recorded with an area of 6100 millionths of the solar surface (www.petermeadows.com/html/area.html), which represents 0.61\% of the solar disk. Based on this value, all clouds (Fig. \ref{noise_5}) with more than 1\% of the total pixels of the solar disk are considered darker areas caused by clouds and consequently rejected (Fig. \ref{noise_6}).  

\begin{figure}[!ht]
	\centering
	\includegraphics[width=0.8\textwidth]{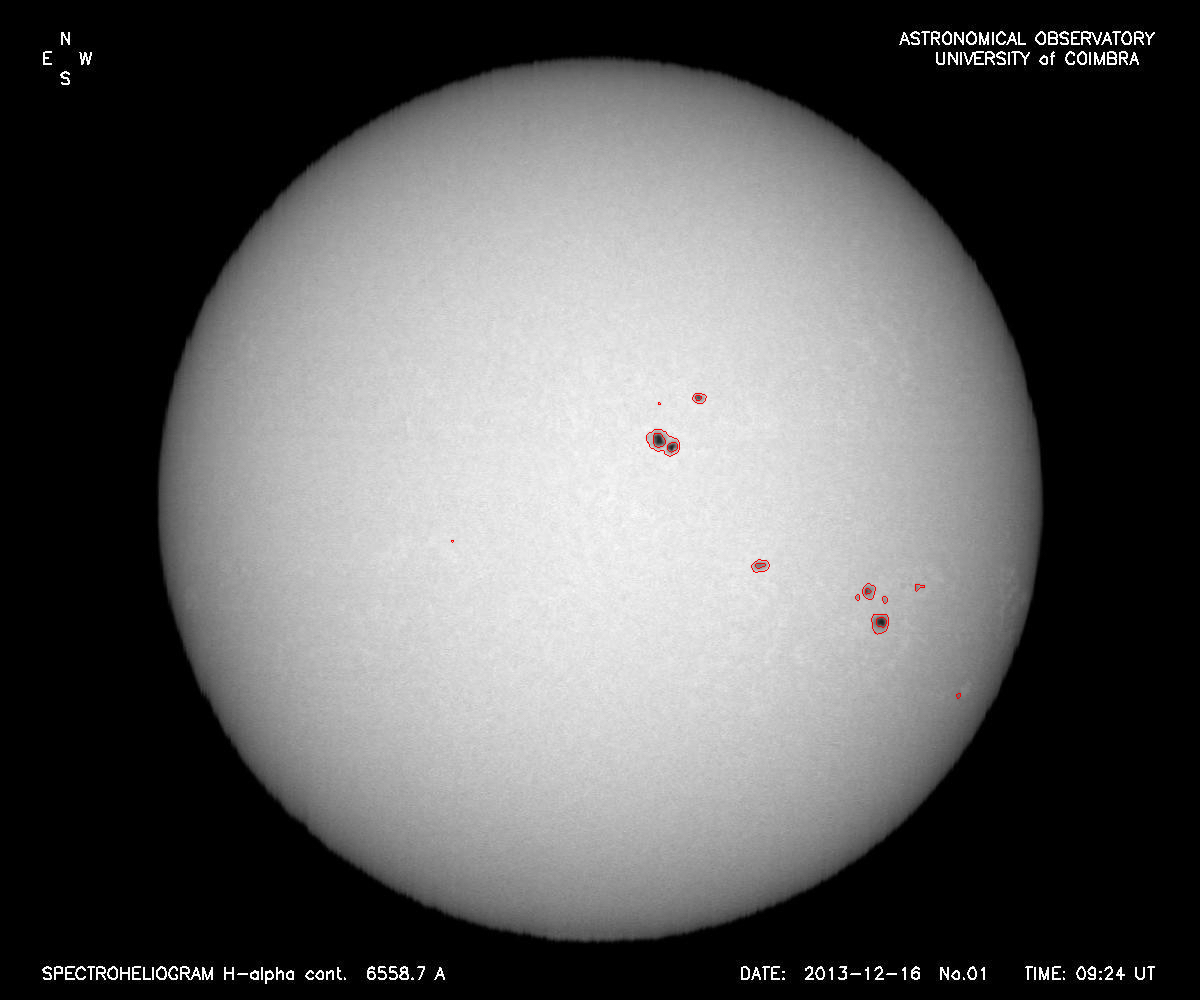}
	\caption{Umbras and penumbras detected by the intensity levels approach, superimposed to the original image.}
	\label{pint:umbraspenumbras}
\end{figure}

An example of the automatic detection of umbras and penumbras based on the intensity levels approach is shown on the Fig. \ref{pint:umbraspenumbras}. 

\section{Data analysis and discussion}
\label{analysisdiscussion}

The automatic algorithms developed to detect sunspots were applied to a representative set of 144 Coimbra’s spectroheliograms of the solar cycle 24. These images were compared with the correspondent ground-truth images built by a solar observer expert. Examples of the resulting images from the two methods are presented in this section, with a reduced contrast for better printed visualization. 
\subsection{Comparison between the two methods}
To evaluate the performance of both methods, two distinct evaluations stages were considered: one for sunspots detection quality, and another for the ability to separate umbra from penumbra. The metrics used in this evaluation rely on pixel-based comparisons. Therefore, each pixel of each output image needed to be classified a sunspot or non-sunspot pixel. The following metrics were used:

\begin{equation}\label{precision} 
Precision = \frac{TP}{TP + FP} 
\end{equation}

\begin{equation} \label{recall}
Recall  = TPR 
\end{equation}

\begin{equation} \label{f1score}
F-score = \frac{2 \ast (Precision \ast Recall)}{Precision + Recall}
\end{equation}

Precision measures the proportion of pixels that are actually positives out of all the pixels that are detected as positives. The recall, also known as sensitivity, or true positive rate (TPR), gives the information about the proportion of pixels that are detected as positives and are actually positives relatively to the universe of all pixels that are, in fact, positives (see equation \ref{TPR}). While the precision allows to evaluate the cost of having false positives (FP) in large number, the recall allows to select which is the best model when the number false negatives (FN) is high. The F-score represents a trade-off between the two previous metrics. This set of metrics was chosen because it is universally used in evaluation of binary detection algorithms in most diverse areas, for example in  \citet{Shuang}, \citet{Hong}, \citet{Ehsan}.
Table \ref{evaluation} shows the results of the evaluation between the two methods. 

\begin{table}
	\centering
	\begin{tabular}{|c|c|c|c|c|}
		\hline
		& Precision (\%)  &  Recall (\%) & F-score (\%) & OA(\%)\\
		\hline
		MM           & 81.33 & 79.42 &78.98 & 86.25\\
		PI            & 84.32 & 73.57 & 77.28 & 87.89\\
		\hline
	\end{tabular}
	\caption{Performance comparison between the methods.}
	\label{evaluation}
\end{table} 

\begin{figure}[!ht]
	\centering
	\subfloat[ \label{goodmmoriginal1}]{%
		\includegraphics[width=0.5\textwidth]{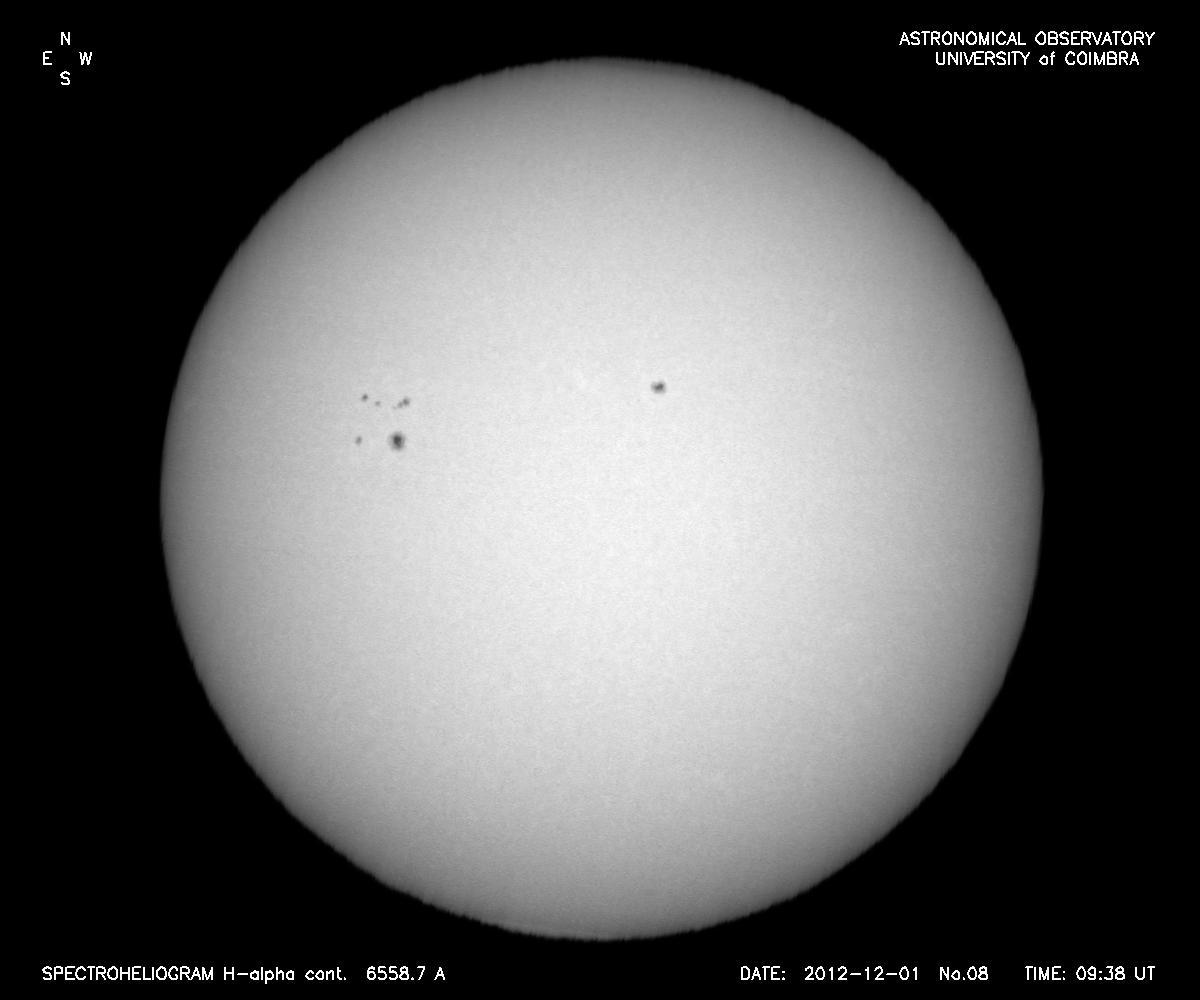}
	}
	\subfloat[\label{goodmm1}]{%
		\includegraphics[width=0.5\textwidth]{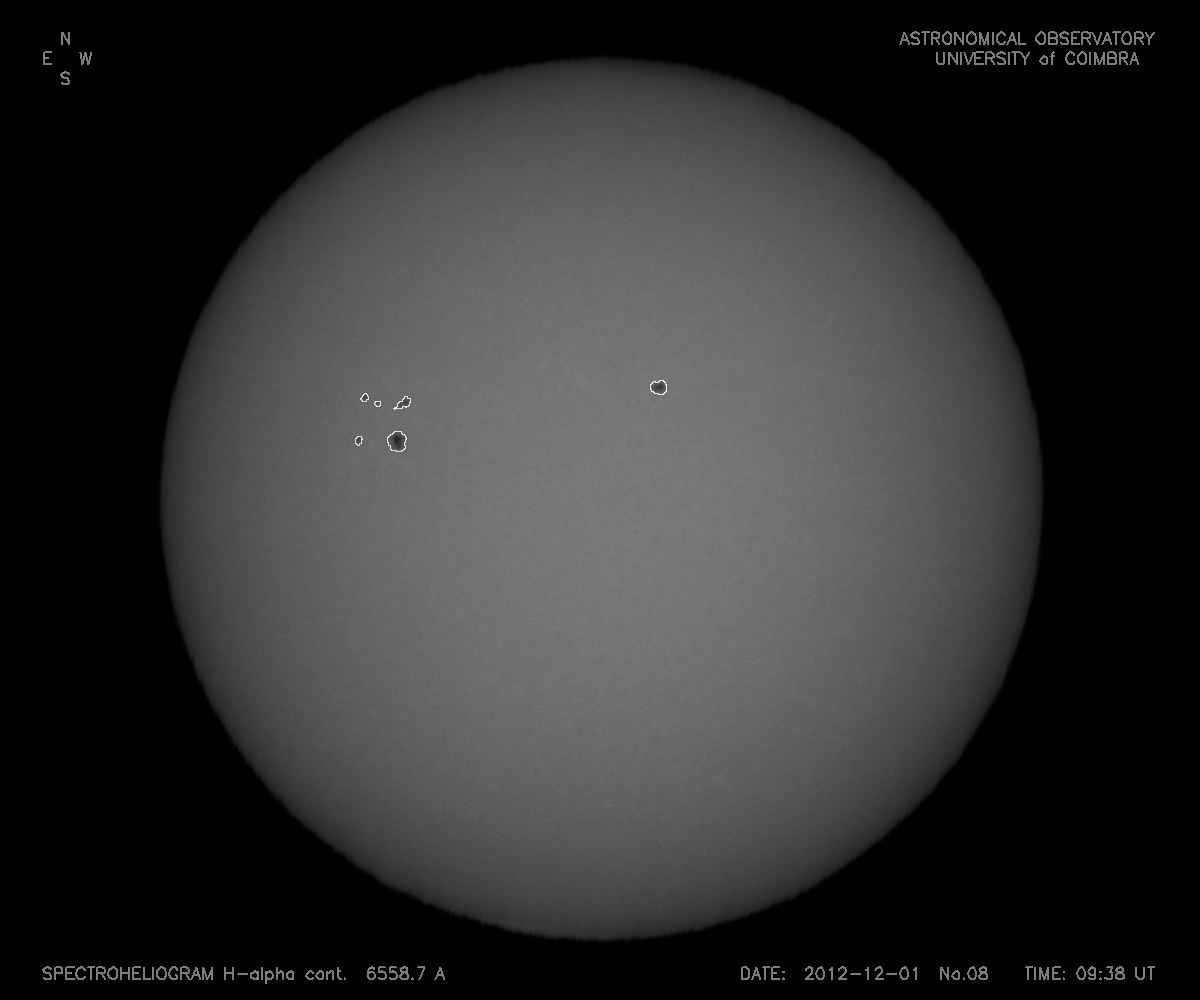}
	} \\
	\subfloat[\label{goodmmoriginal2}]{%
		\includegraphics[width=0.5\textwidth]{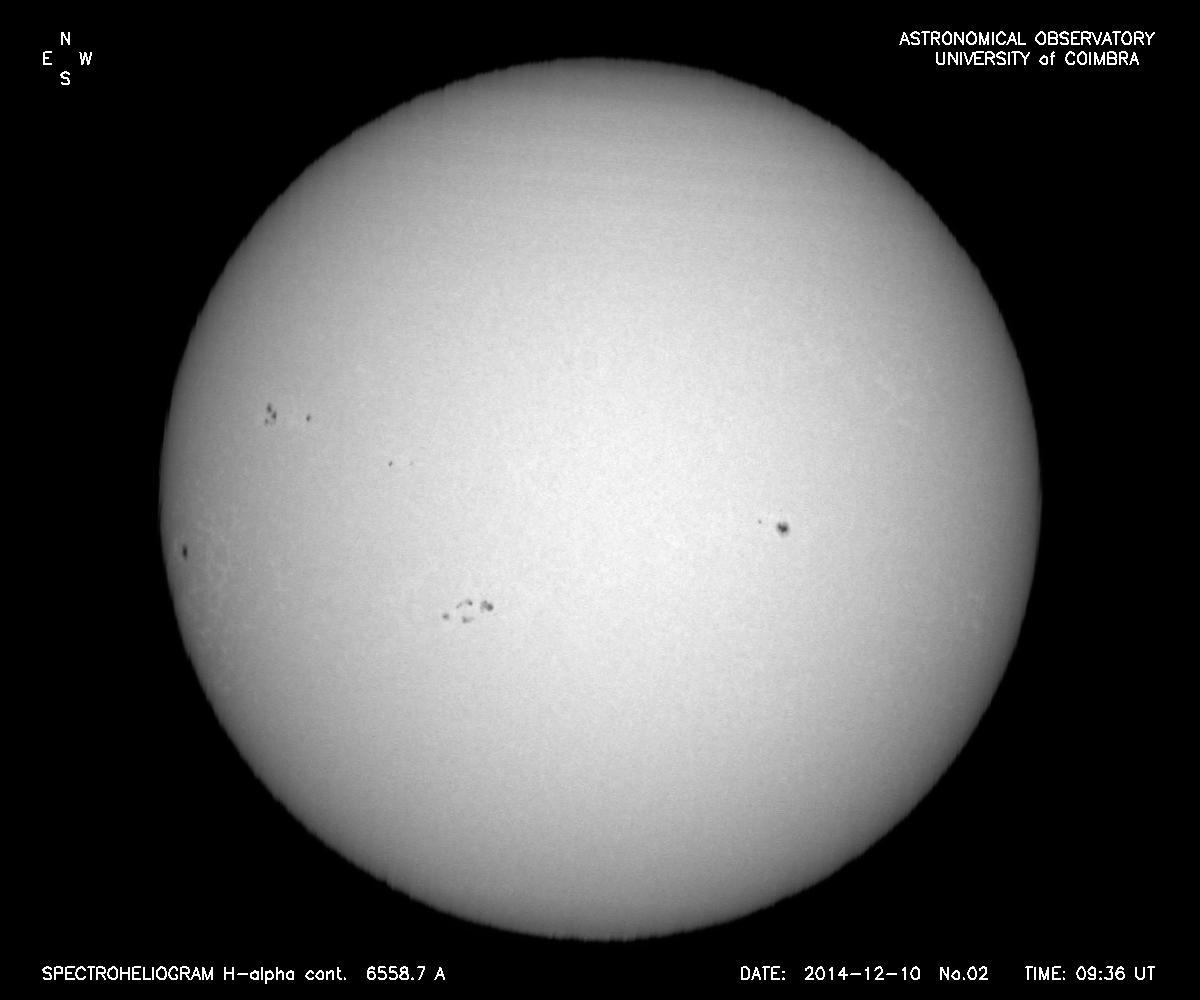}
	}
	\subfloat[\label{goodmm2}]{%
		\includegraphics[width=0.5\textwidth]{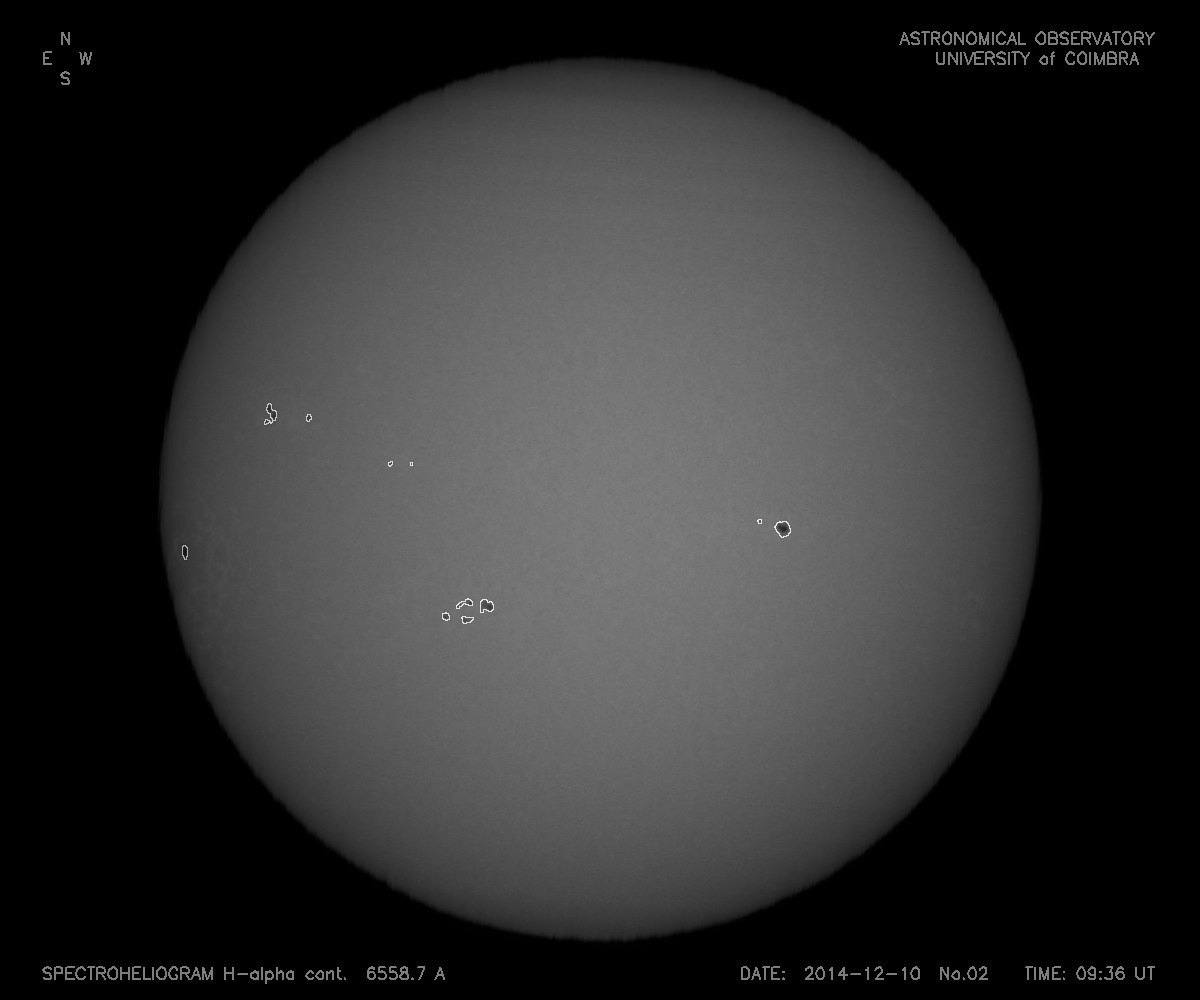}
	}
	\caption{Examples of morphological detection algorithm: (a) original image of 1/12/2012; (b) result for the image of 1/12/2012; (c) original image of 10/12/2014; (d) result for the image of 10/12/2014.}
	\label{gooddetectionmm}
	
\end{figure}

\begin{figure}[!ht]
	\centering
	\subfloat[ \label{goodpioriginal1}]{%
		\includegraphics[width=0.5\textwidth]{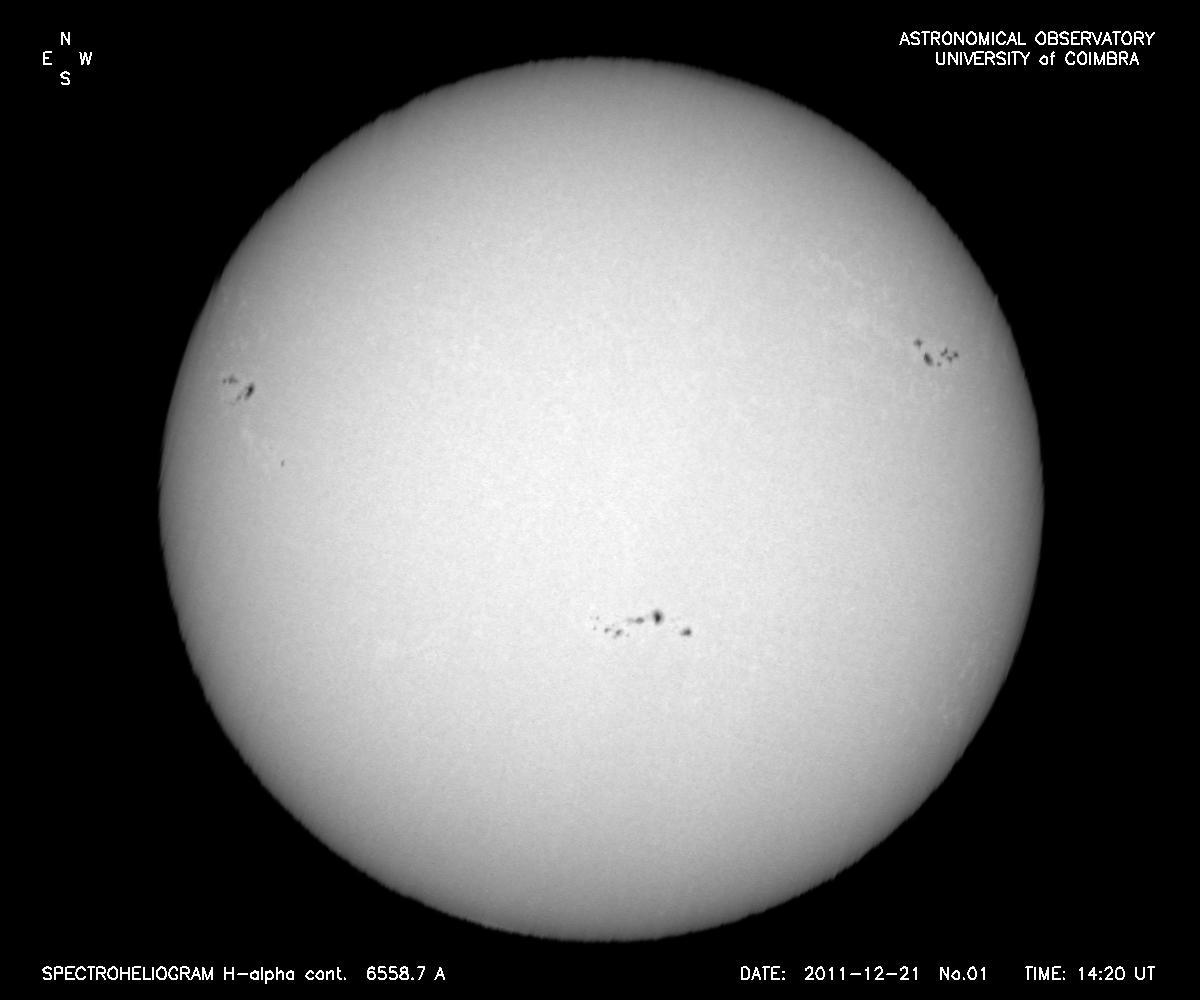}
	}
	\subfloat[\label{goodpi1}]{%
		\includegraphics[width=0.5\textwidth]{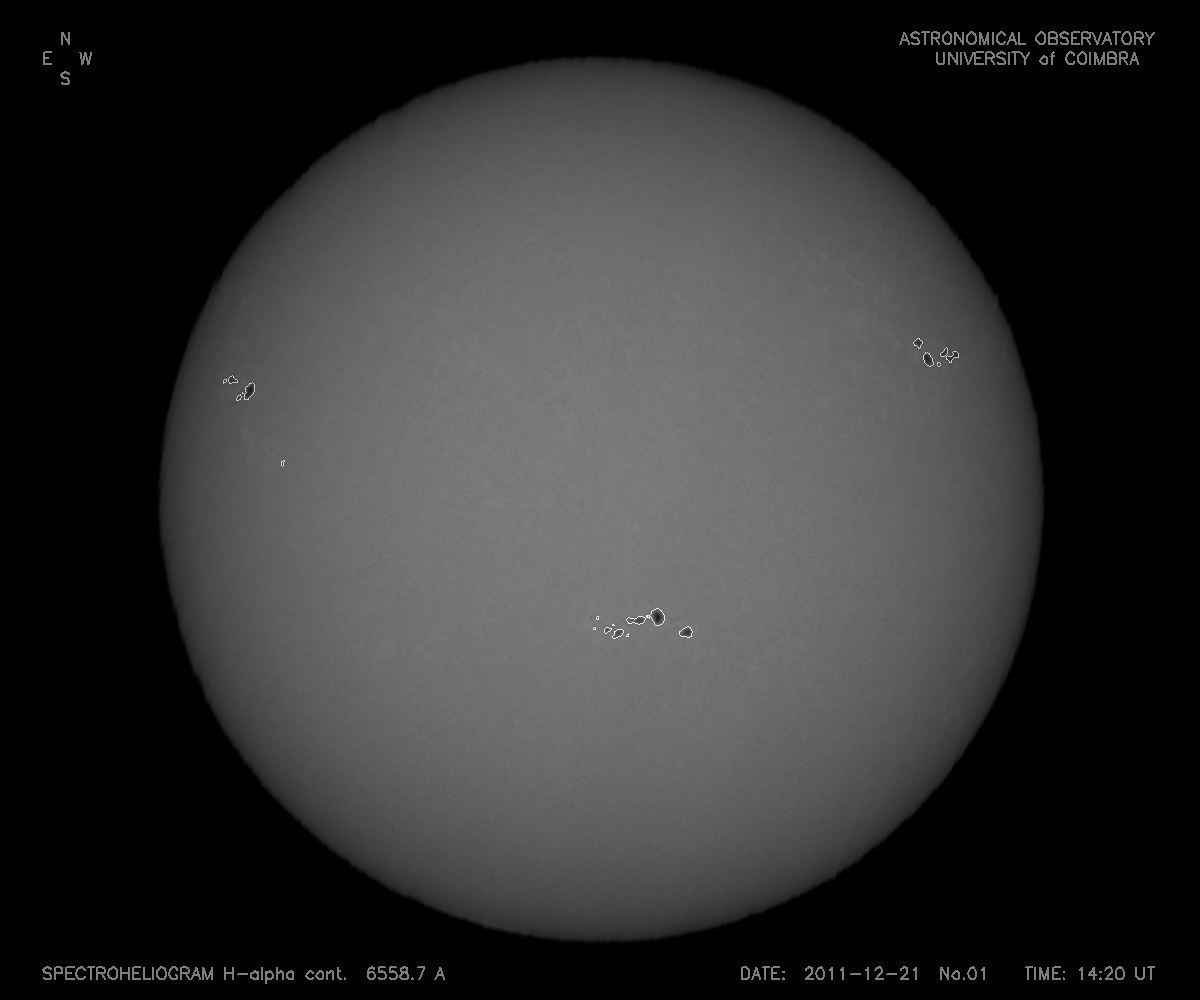}
	} \\
	\subfloat[\label{goodpioriginal2}]{%
		\includegraphics[width=0.5\textwidth]{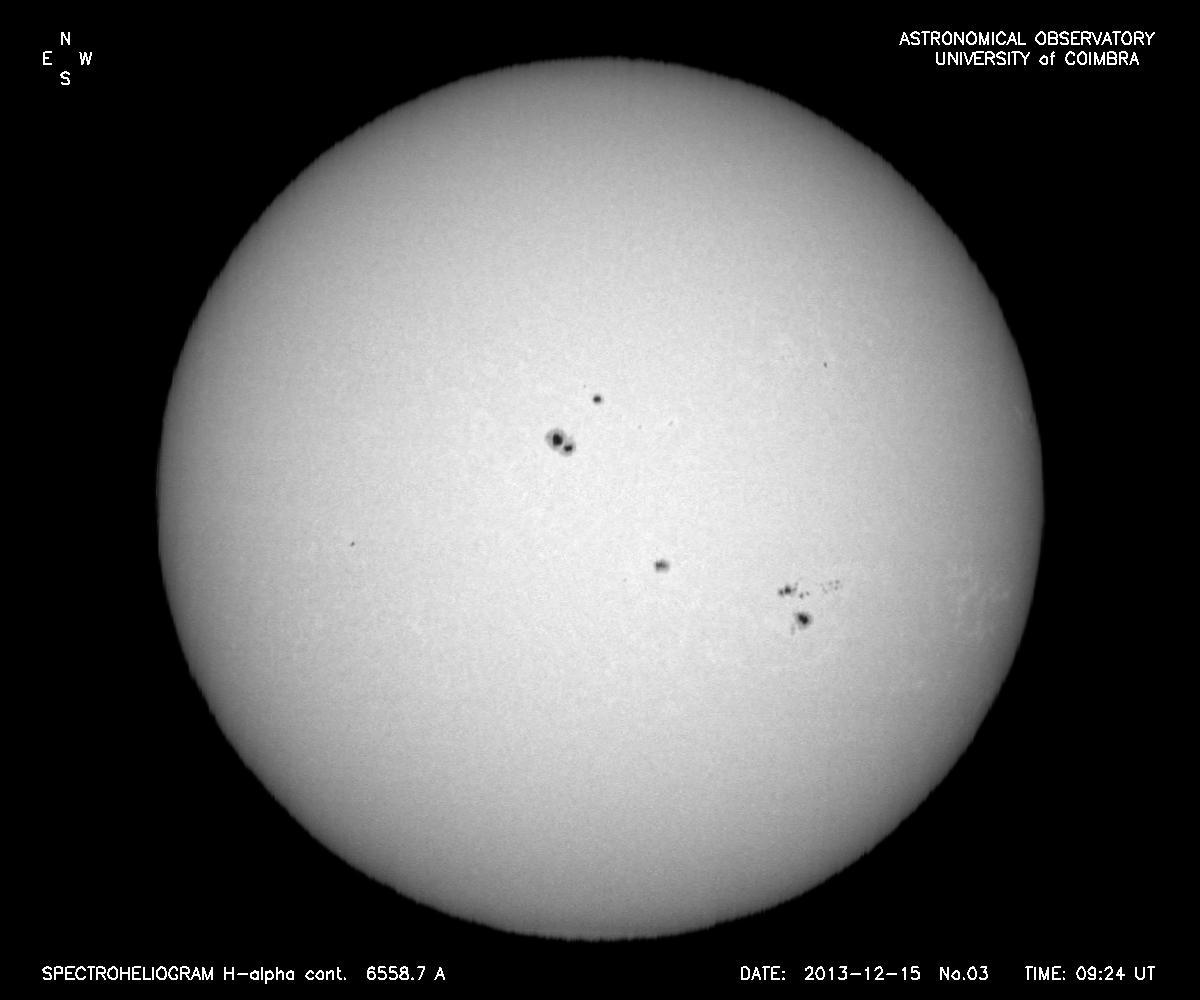}
	}
	\subfloat[\label{goodpi2}]{%
		\includegraphics[width=0.5\textwidth]{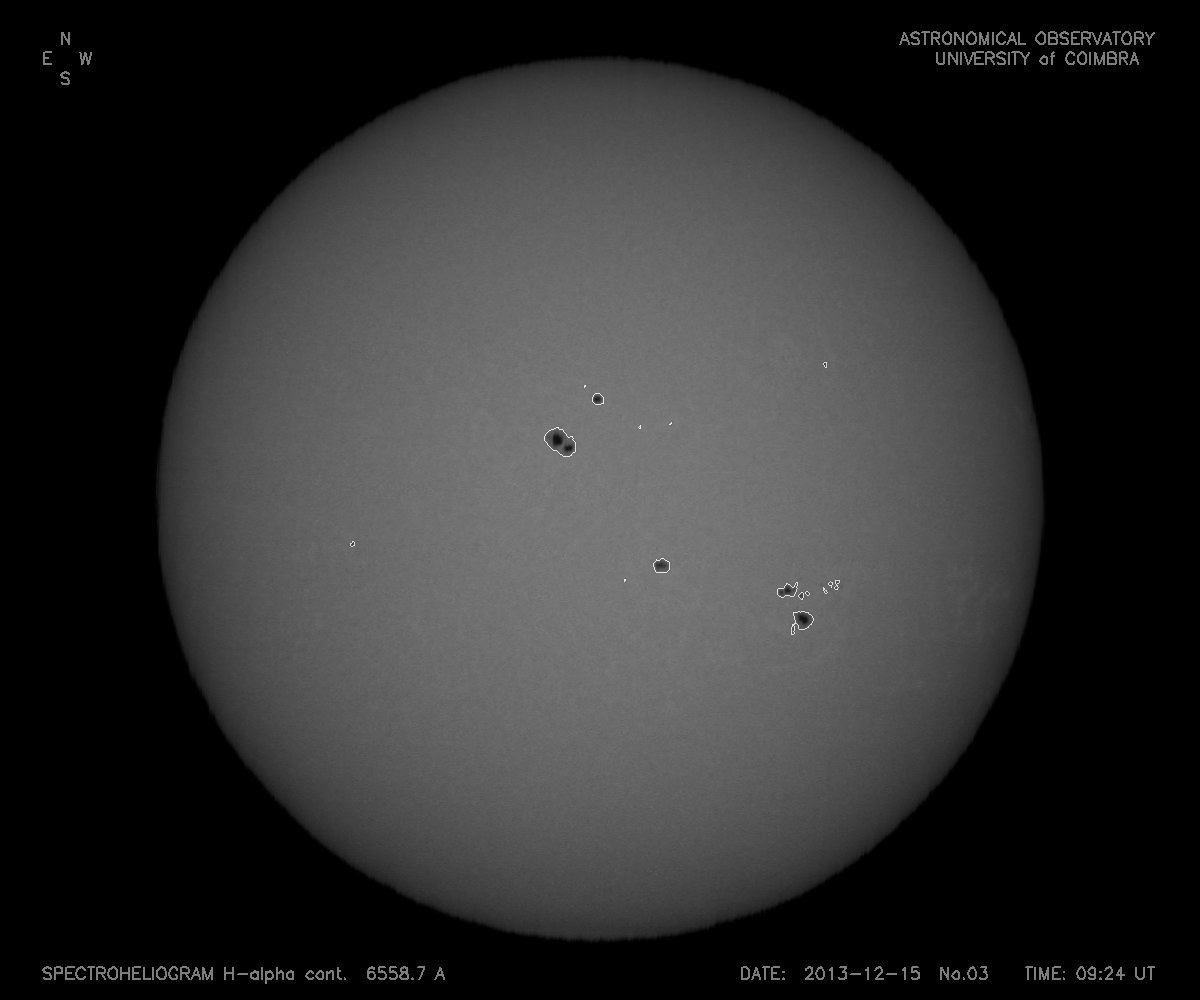}
	}
	\caption{Examples of algorithm of detection based on pixels intensities: (a) original image of 21/12/2011; (b) result for the  image of 21/12/2011; (c) original image of 15/12/2013; (d) result for the image of 15/12/2013.}
	\label{gooddetectionpi}
\end{figure}

Both methods developed present satisfactory results for most of the images as we can verify analyzing table \ref{evaluation}. Examples of good detections can be seen in Figs. \ref{gooddetectionmm} and \ref{gooddetectionpi}.


\begin{figure}
	\subfloat[ \label{cloudsoriginal1}]{%
		\includegraphics[width=0.5\textwidth]{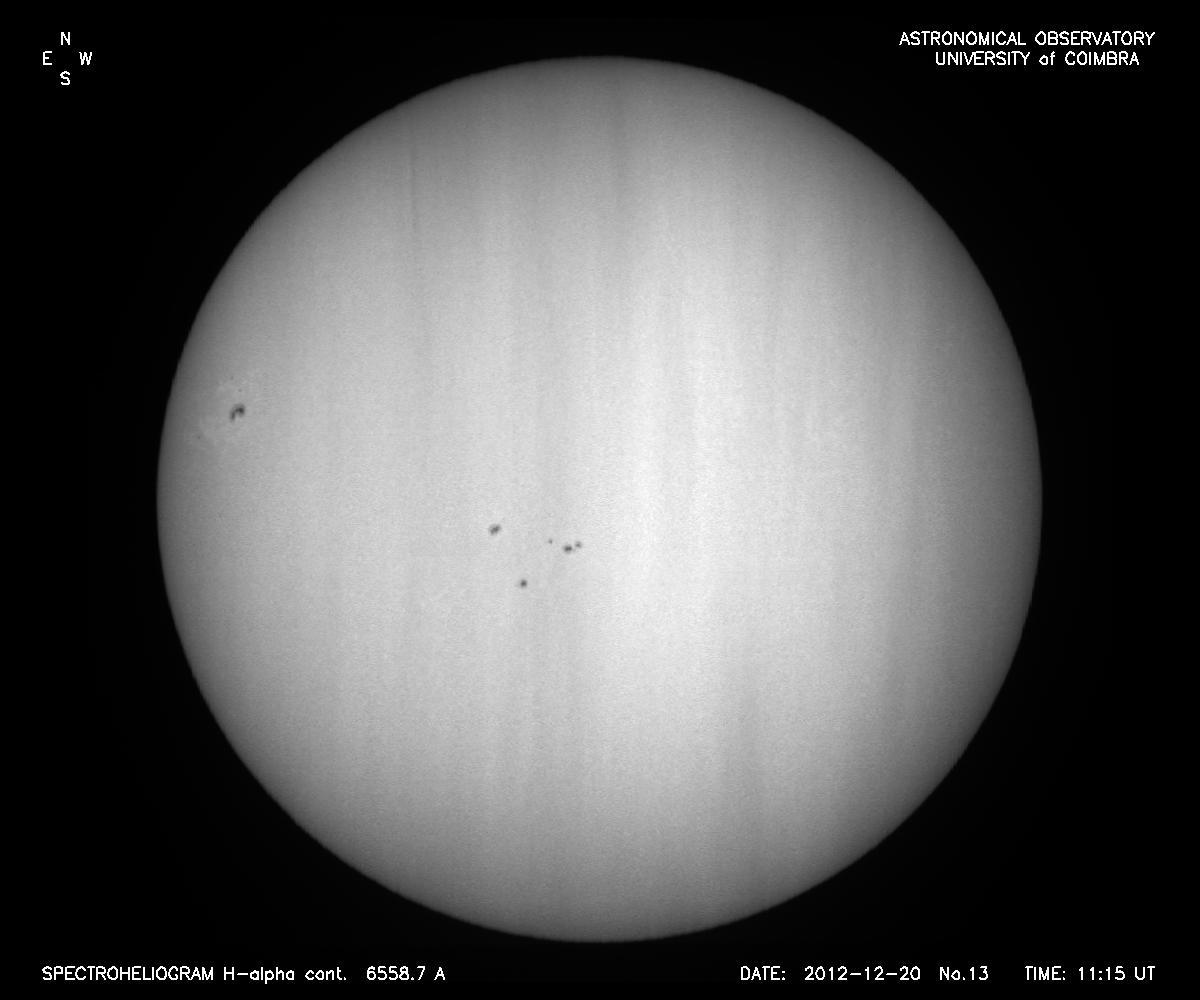}
	} \hfill
	\subfloat[\label{cloudsmm1}]{%
		\includegraphics[width=0.5\textwidth]{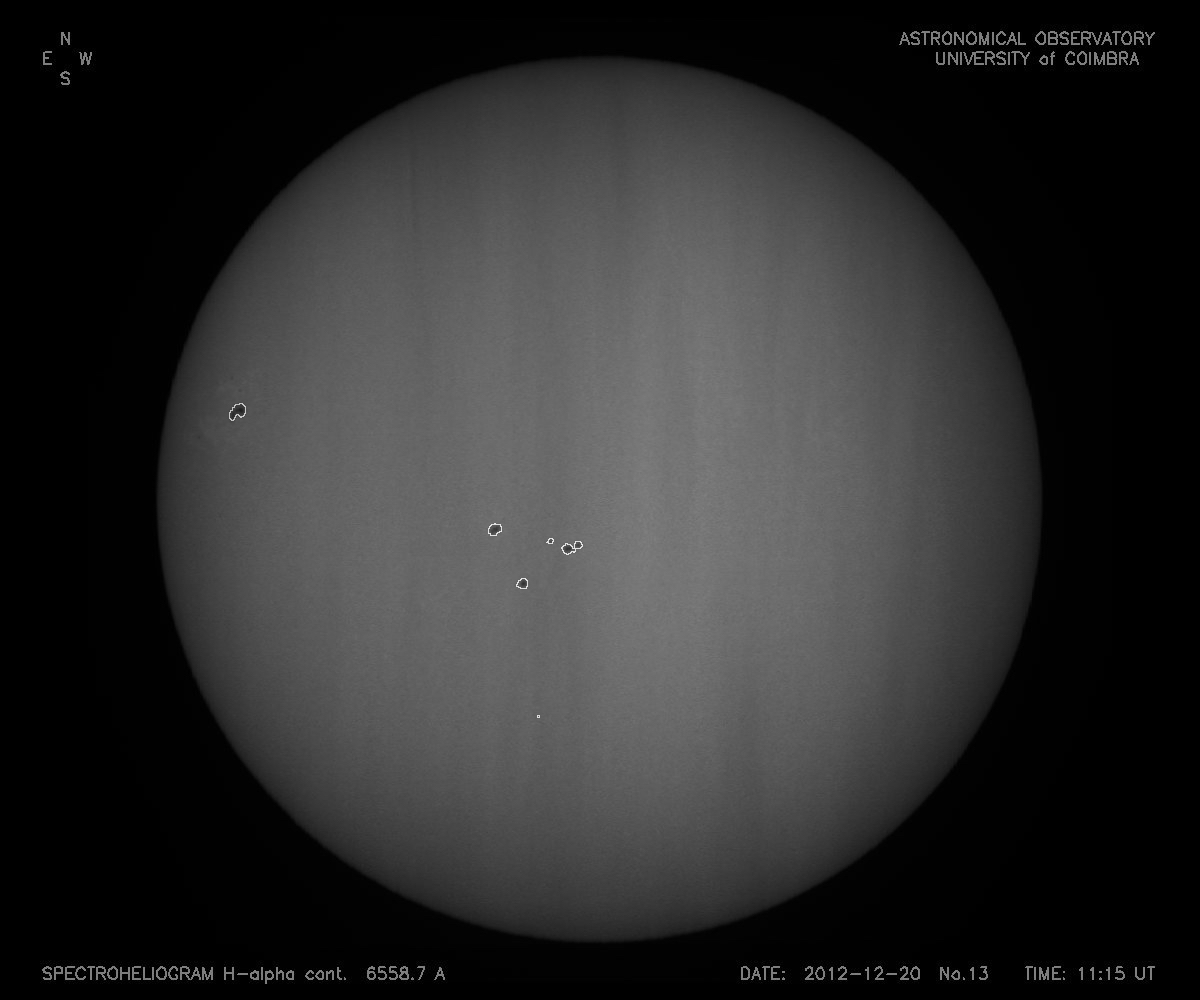}
	} \\ \hfill
	\subfloat[\label{ncloudspi1}]{%
		\includegraphics[width=0.5\textwidth]{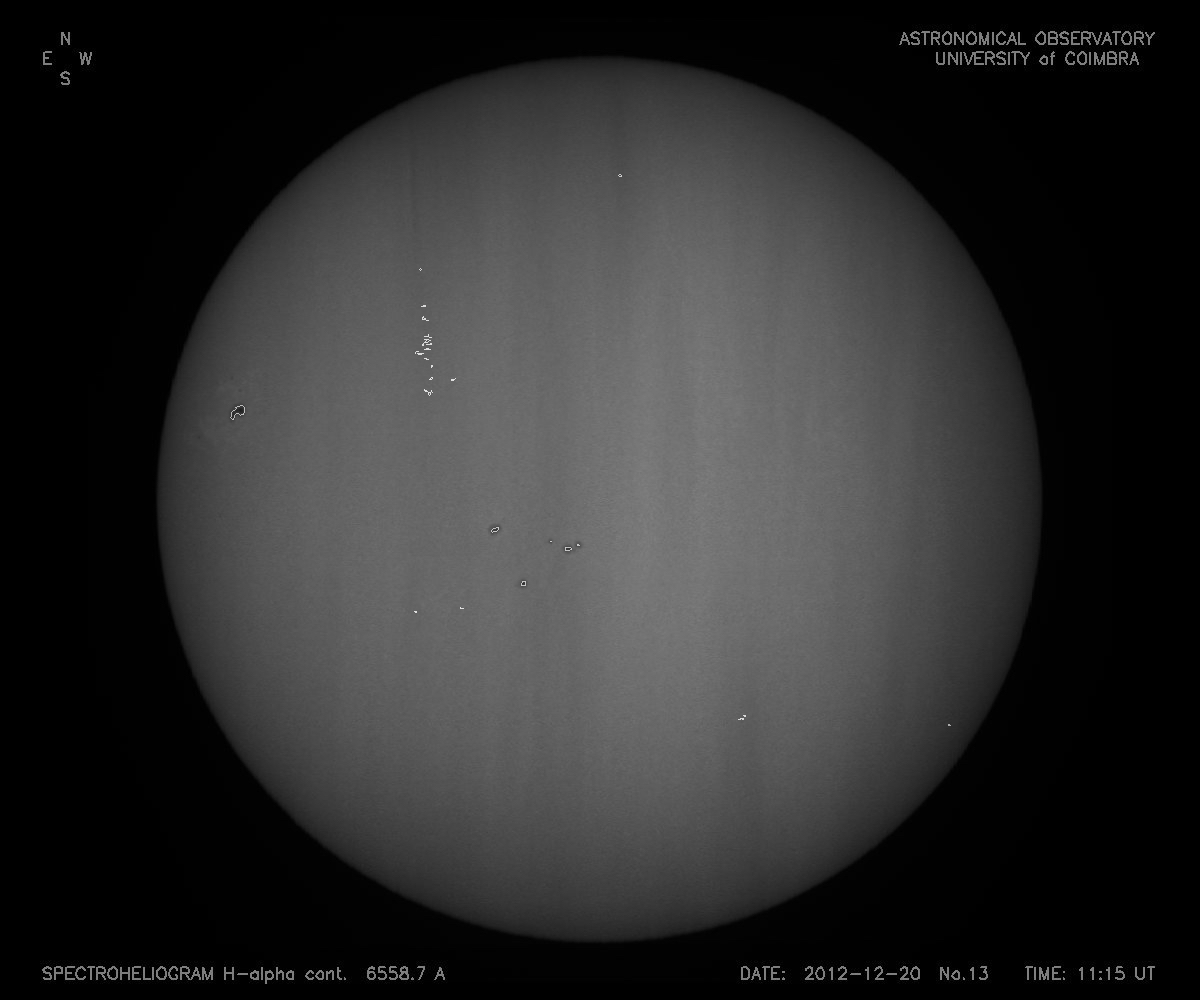}
	} 
	\subfloat[\label{cloudsoriginal2}]{%
		\includegraphics[width=0.5\textwidth]{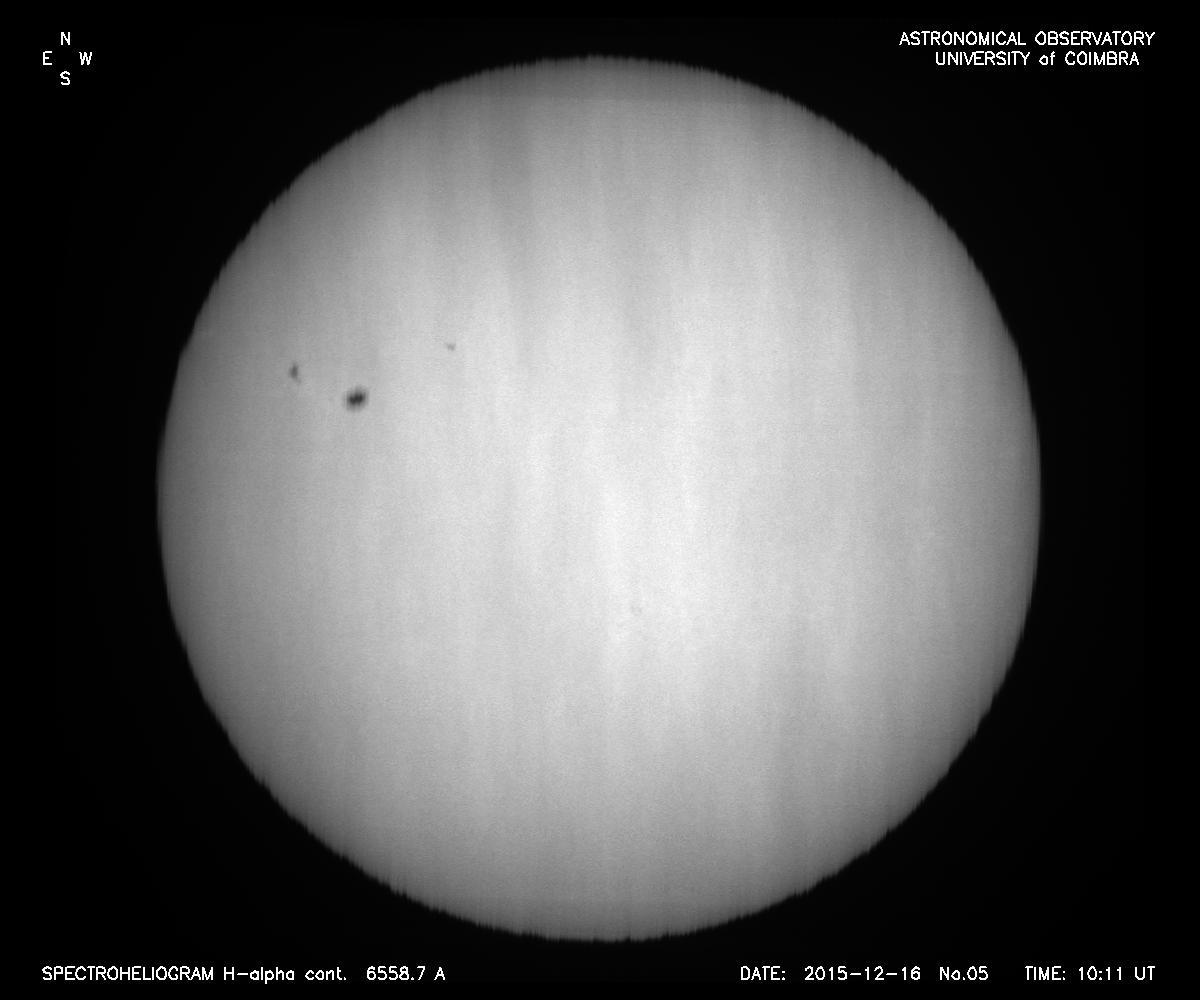}
	} \\ \hfill
	\subfloat[\label{cloudsmm2}]{%
		\includegraphics[width=0.5\textwidth]{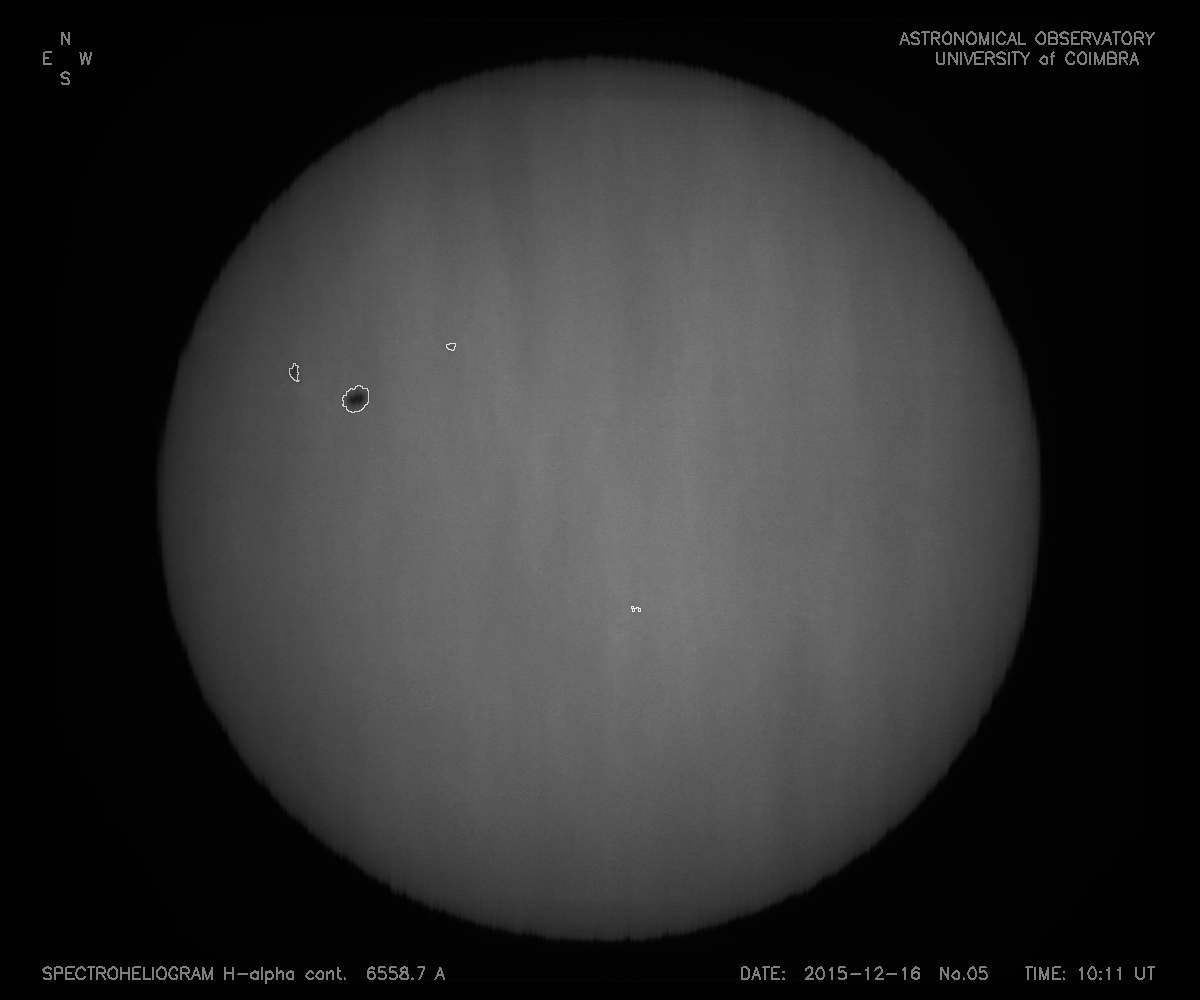}
	} 
	\subfloat[\label{cloudspi2}]{%
		\includegraphics[width=0.5\textwidth]{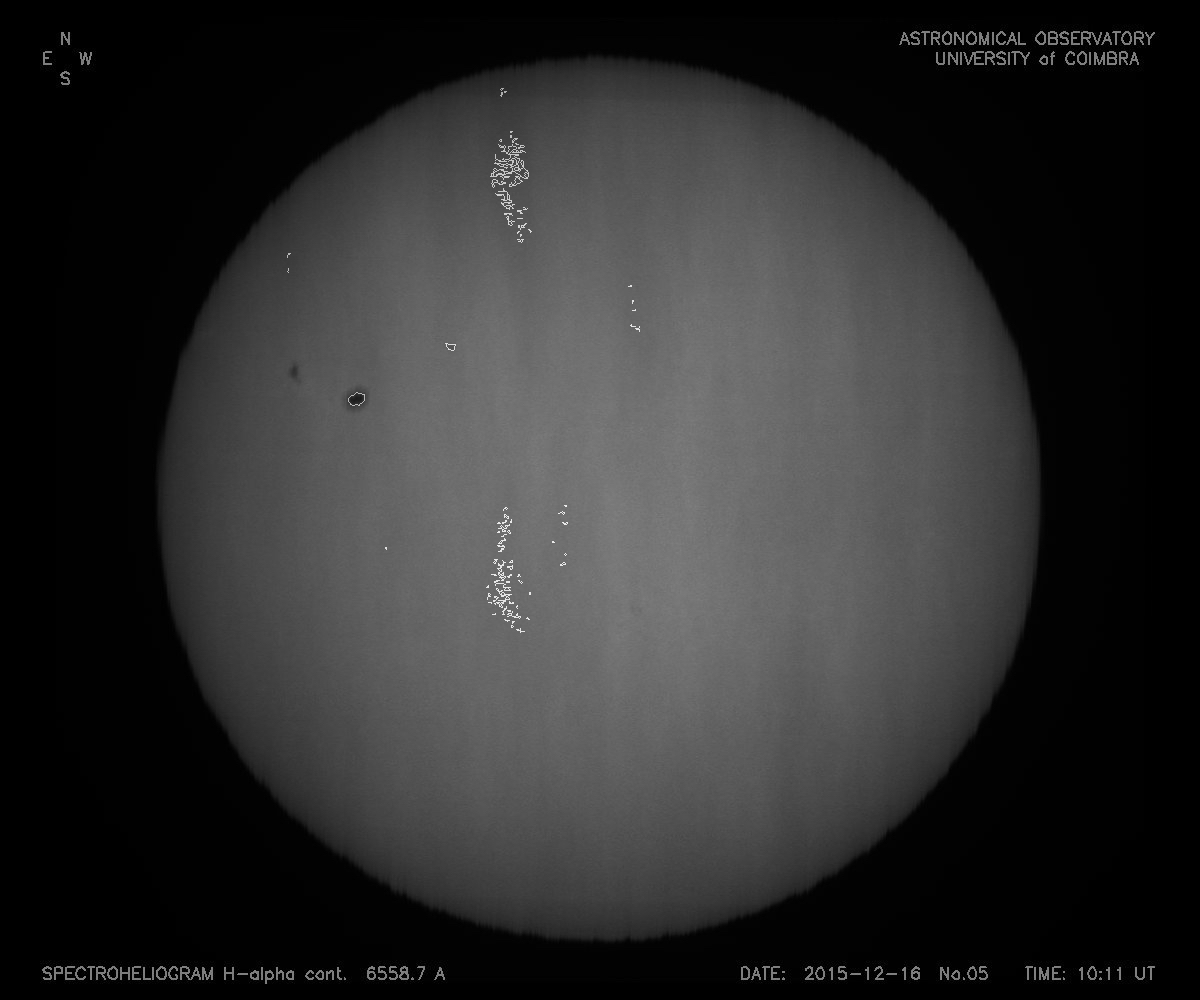}
	}
	\caption{Examples of detection results on spectroheliograms with clouds: (a) original image of 20/12/2012; (b) result of the morphological detection algorithm of the image of 20/12/2012; (c) result of the algorithm of detection based on pixels intensities of the image of 20/12/2012; (d) original image of 16/12/2015; (e) result of the morphological detection algorithm of the image of 16/12/2015; (f) result of the algorithm of detection based on pixels intensities of the image of 16/12/2015.}
	\label{clouds}
\end{figure}

However, there are images in which none of the methods give good results. This section lists some difficulties that can arise in the automatic detection algorithms on some spectroheliograms and analyzes the performance of the two automatic methods in the detection of sunspots in those images. 
Due the Earth’s atmosphere and meteorological factors, applying automatic detection methods to ground-based images present some specific hindrances. Despite this, the good performance of the algorithm based on mathematical morphology is essentially kept when applying to most of the images with strong atmospheric effects, whereas the method based on pixels intensities shows more problems with those images. Examples can be seen in Fig. \ref{clouds}.

\begin{figure}
	\subfloat[ \label{acquisitionoriginal1}]{%
		\includegraphics[width=0.5\textwidth]{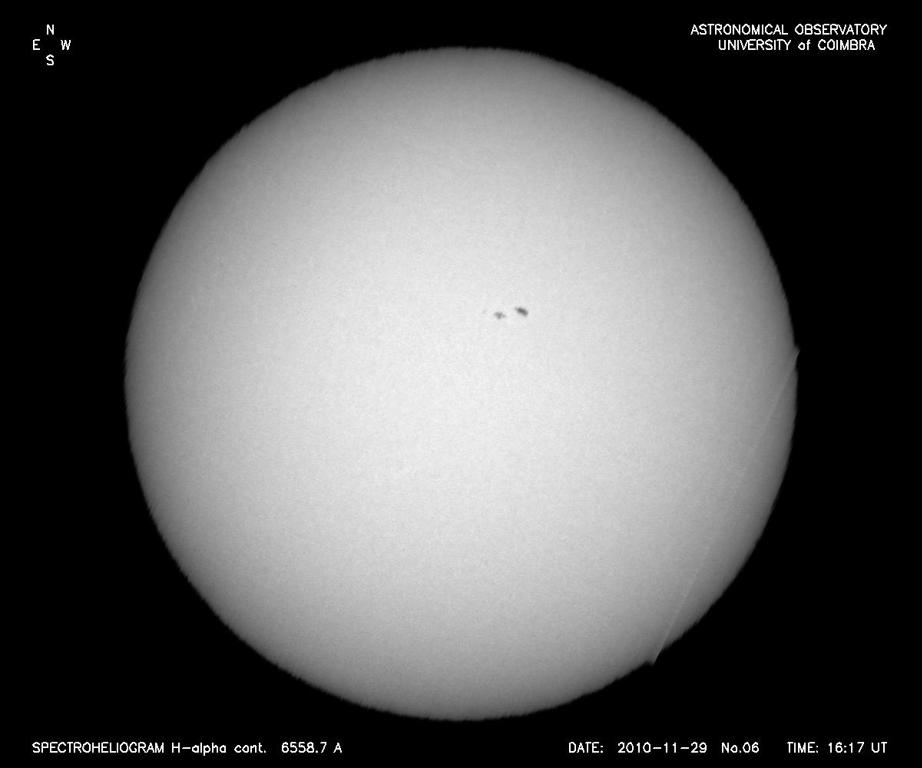}
	} 
	\subfloat[\label{acquisitionmm1}]{%
		\includegraphics[width=0.5\textwidth]{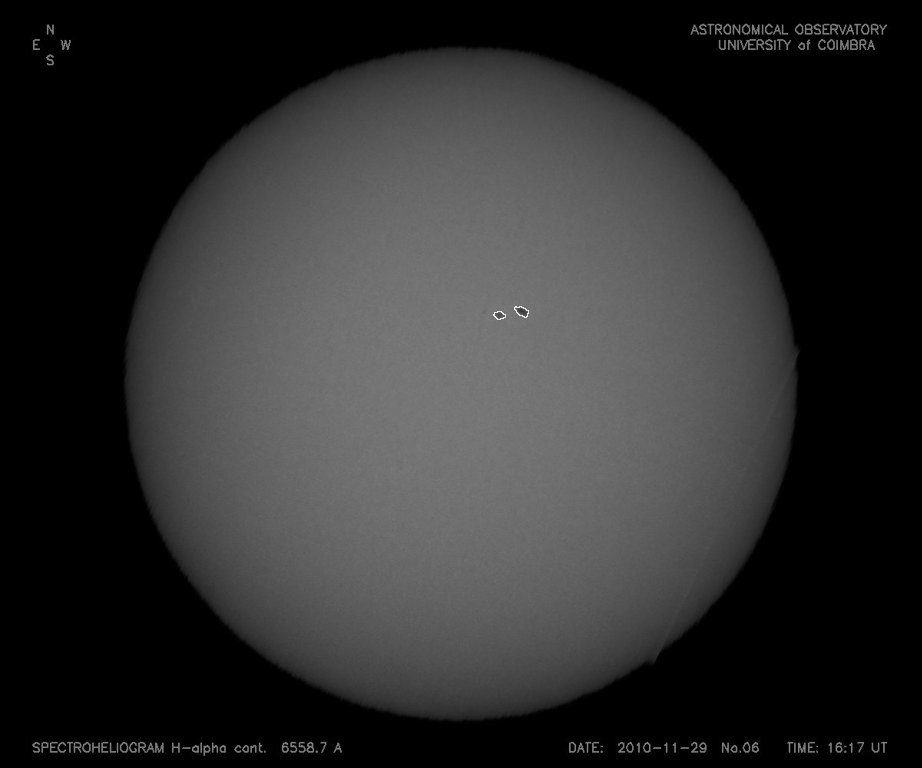}
	} \\	\hfill
	\subfloat[\label{acquisitionpi1}]{%
		\includegraphics[width=0.5\textwidth]{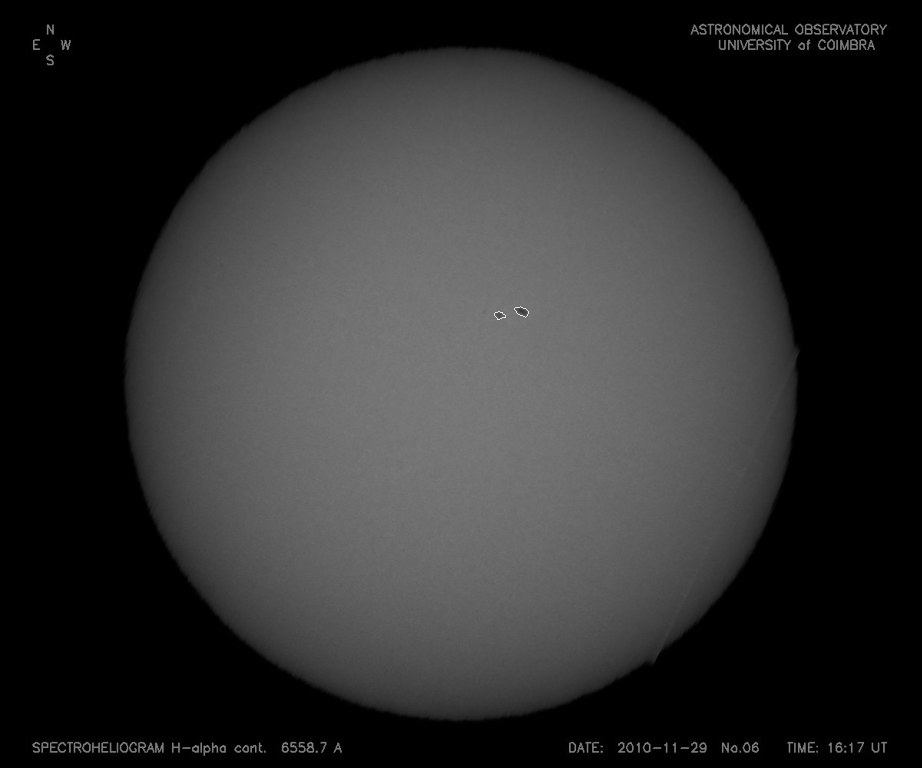}
	}
	\subfloat[\label{acquisitionoriginal2}]{%
		\includegraphics[width=0.5\textwidth]{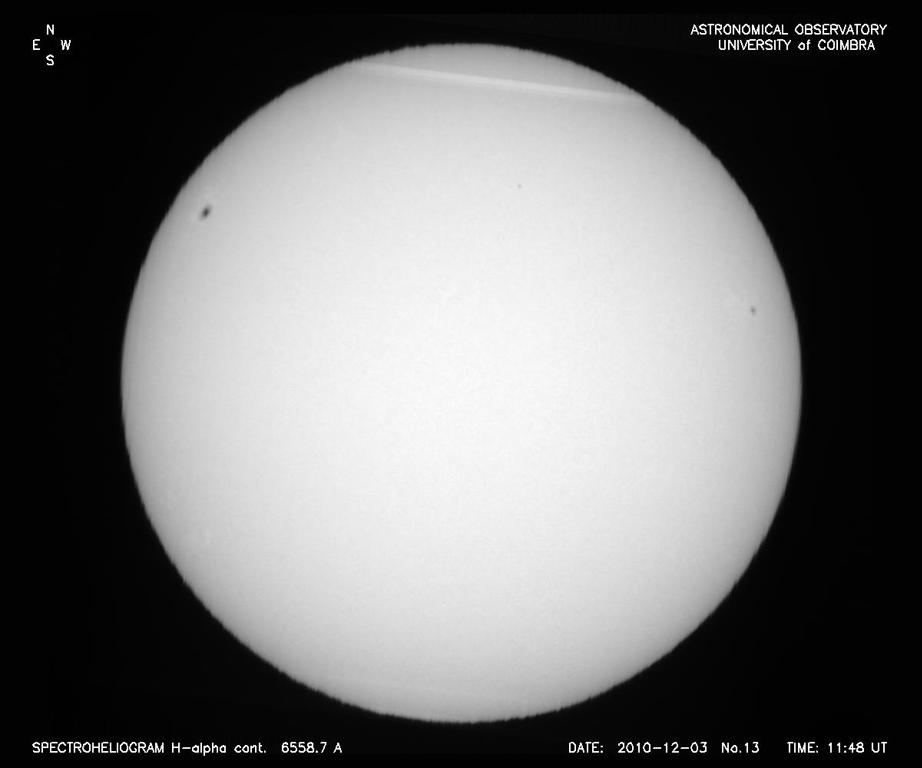}
	} \\ \hfill
	\subfloat[\label{acquisitionmm2}]{%
		\includegraphics[width=0.5\textwidth]{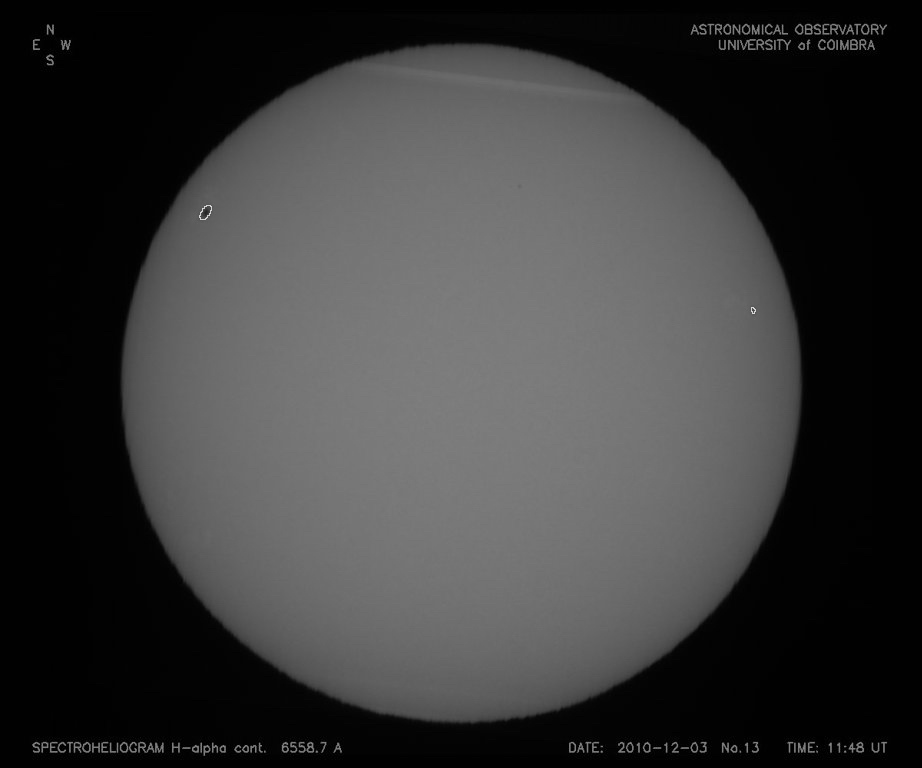}
	}
	\subfloat[\label{acquisitionpi2}]{%
		\includegraphics[width=0.5\textwidth]{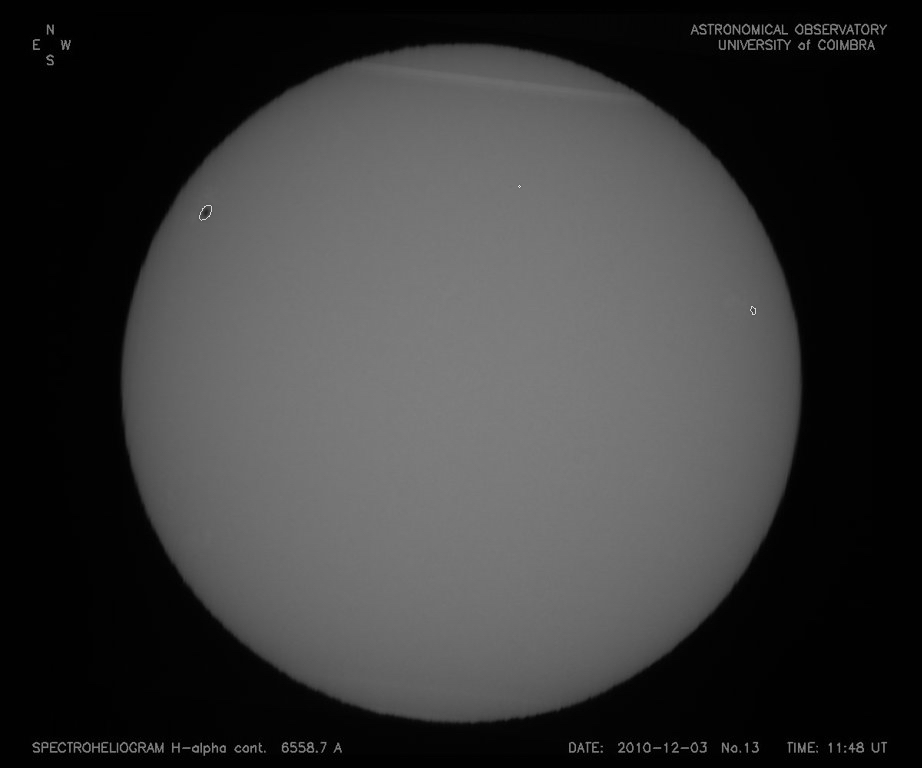}
	}
	\caption{Examples of detection results on spectroheliograms with acquisition erros: (a) original image of 20/12/2012; (b) result of the morphological detection algorithm of the image of 20/12/2012; (c) result of the algorithm of detection based on pixels intensities of the image of 20/12/2012; (d) original image of 16/12/2015; (e) result of the morphological detection algorithm of the image of 16/12/2015; (f) result of the algorithm of detection based on pixels intensities of the image of 16/12/2015.}
	\label{acquisition}
\end{figure}

Problems during the image acquisition process is another difficulty that can appear in some spectroheliograms. Although in these cases the solar disk may present some deformations, both methods perform well, as shown in Fig. \ref{acquisition}.

\begin{figure}[!ht]
	\subfloat[ \label{dustoriginal}]{%
		\includegraphics[width=0.5\textwidth]{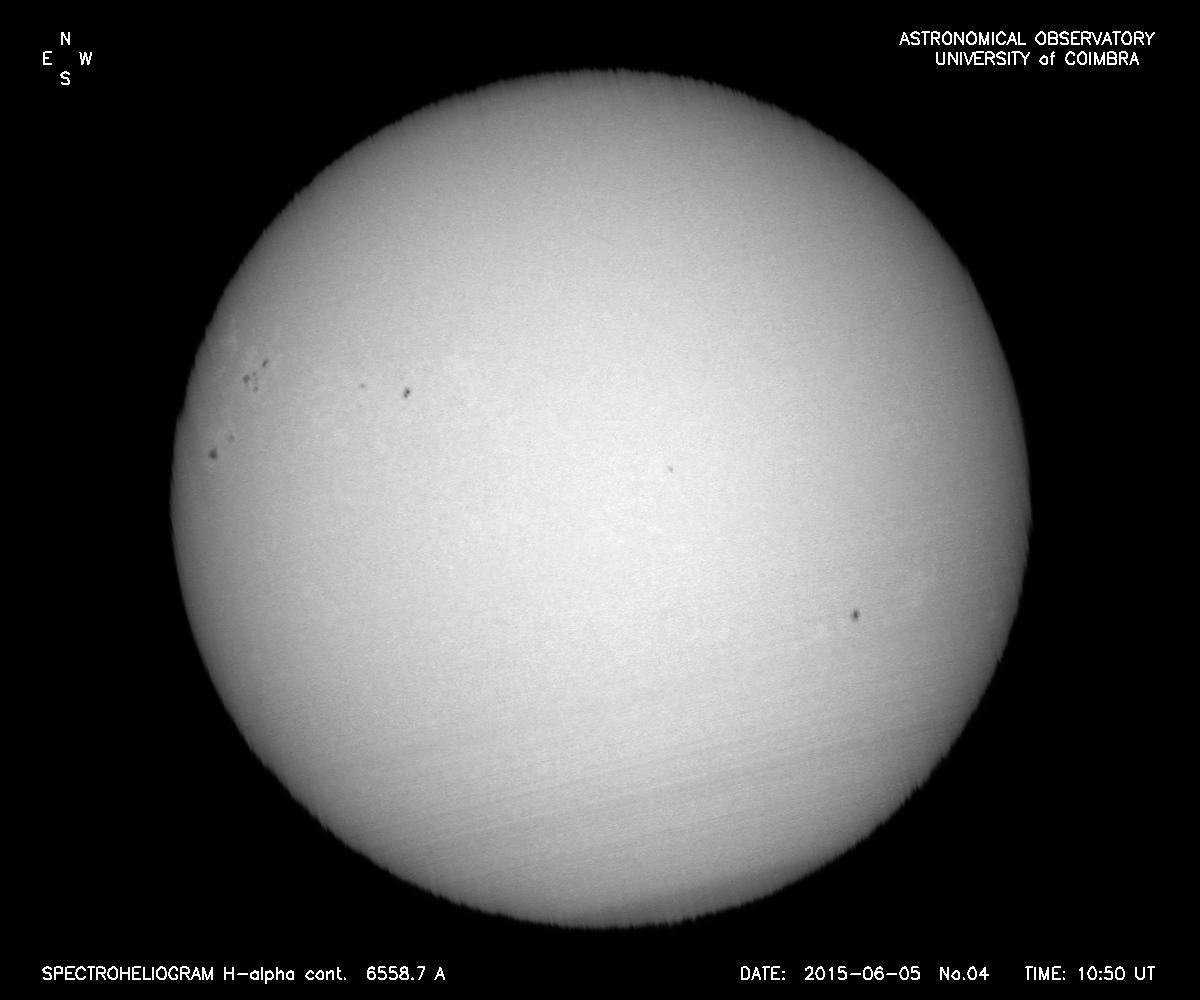}
	}  \hfill
	\subfloat[\label{dustmm}]{%
		\includegraphics[width=0.5\textwidth]{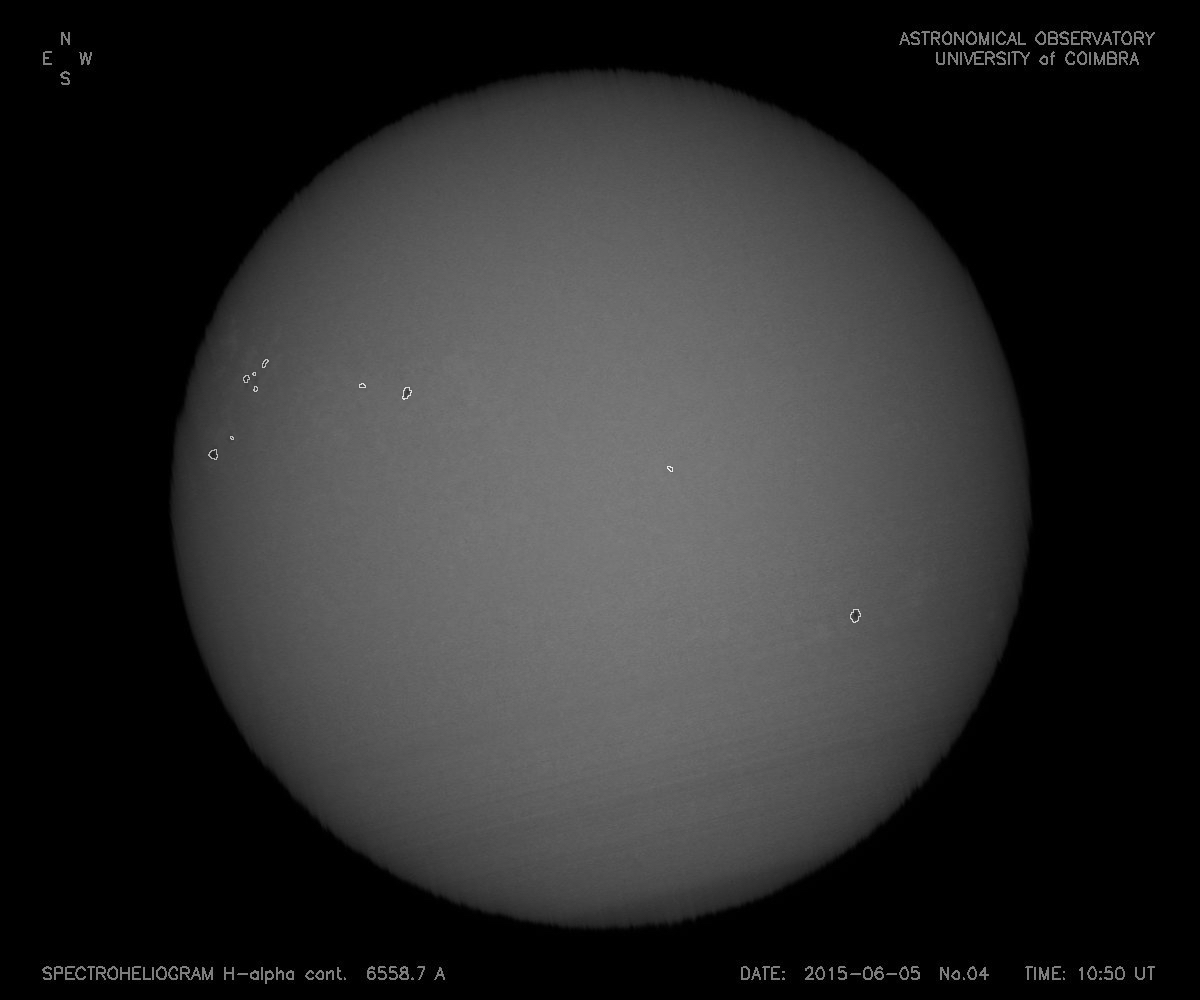}
	} \\  \\ \hfill

	\qquad\qquad\qquad\qquad\quad	
	\subfloat[\label{dustpi}]{%
		\includegraphics[width=0.5\textwidth]{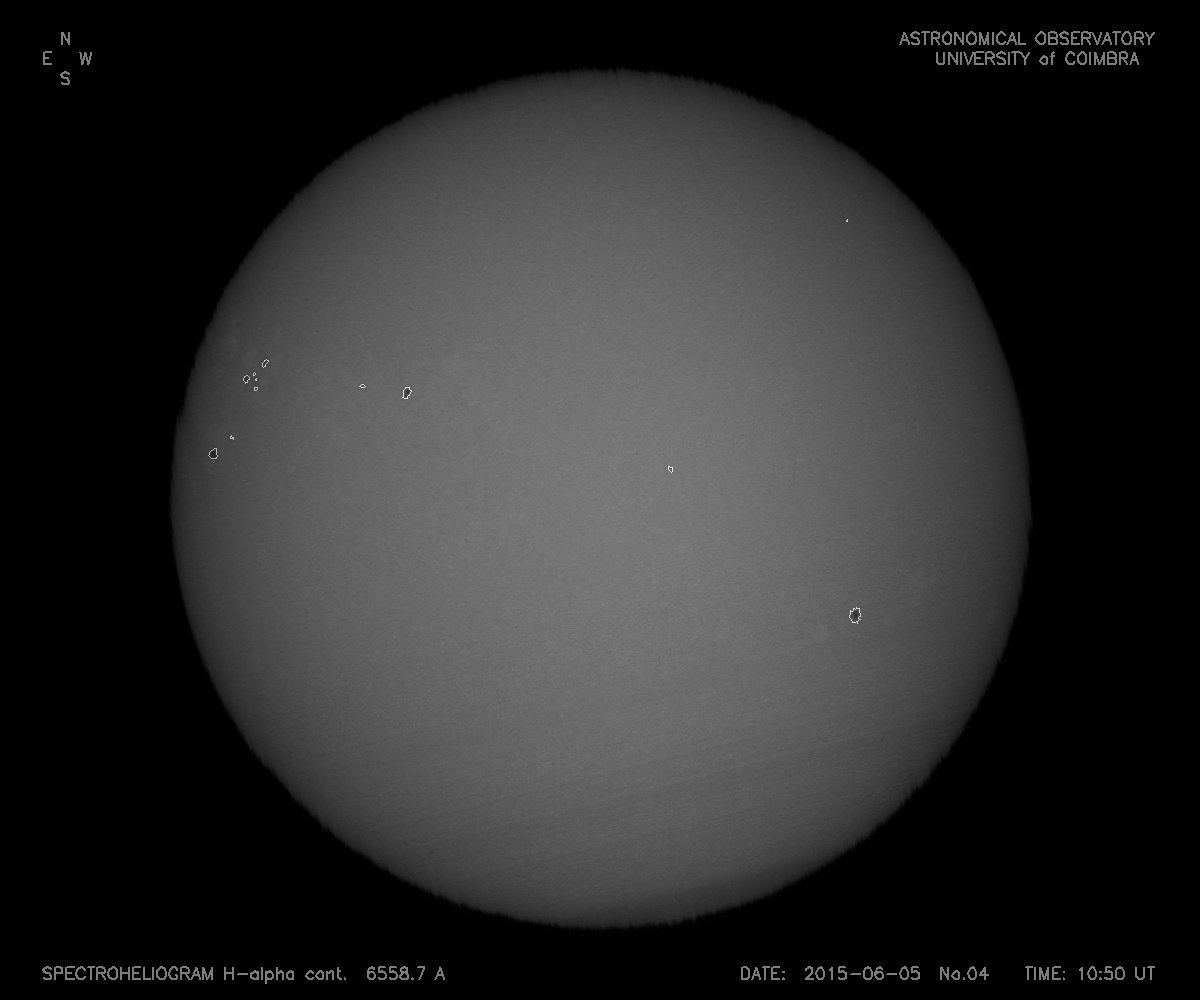}
	}
	\caption{Example of detection results on spectroheliograms with dust: (a) original image of 5/6/2015; (b) result of the morphological detection algorithm; (c) result of the algorithm of detection based on pixels intensities.}
	\label{dust}
\end{figure}

In the case of spectroheliograms with dust trapped in the slit (represented by almost horizontal lines in the image), which could not be removed or corrected by the image acquisition software, both methods proved to be efficient, as we can see in Fig. \ref{dust}.

\begin{figure}
	\subfloat[ \label{limboriginal1}]{%
		\includegraphics[width=0.5\textwidth]{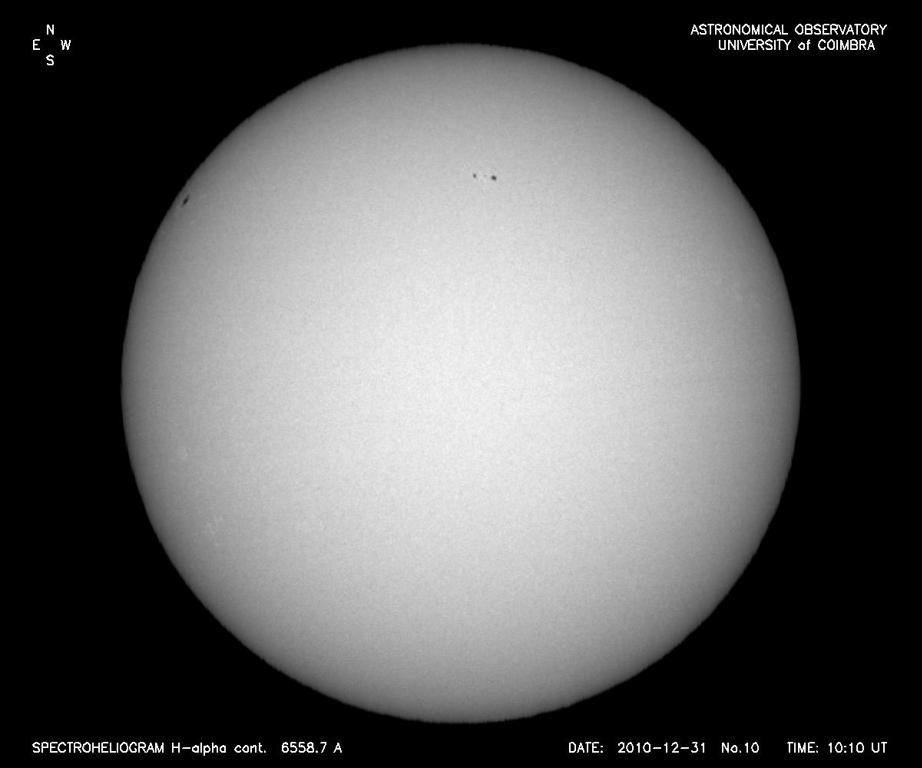}
	} 
	\subfloat[\label{limbmm1}]{%
		\includegraphics[width=0.5\textwidth]{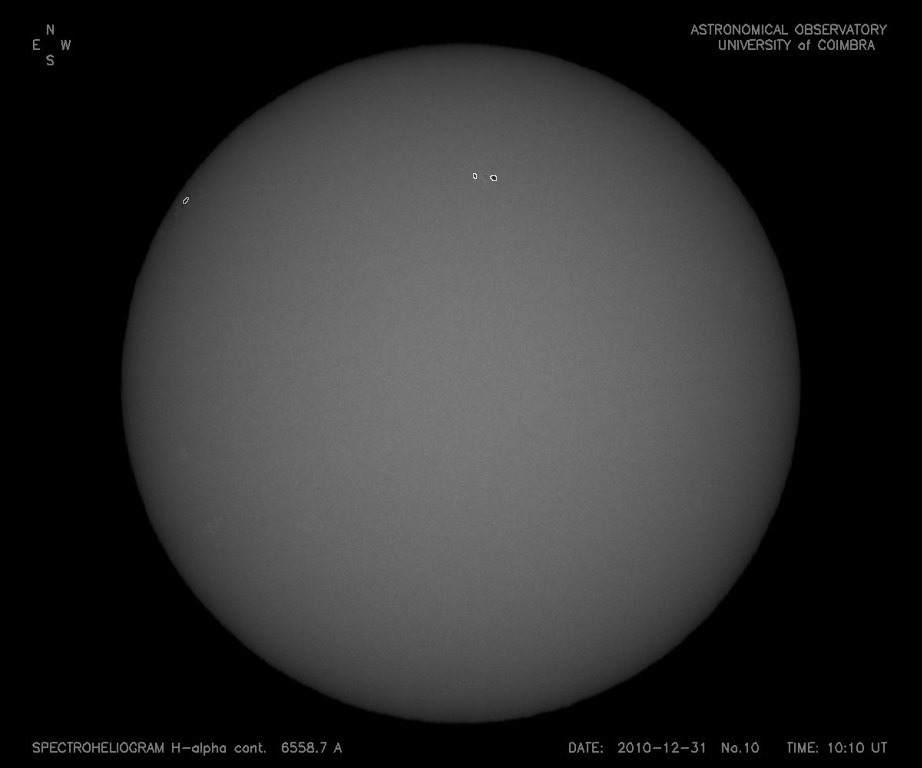}
	} \\ \hfill
	\subfloat[\label{limbpi1}]{%
		\includegraphics[width=0.5\textwidth]{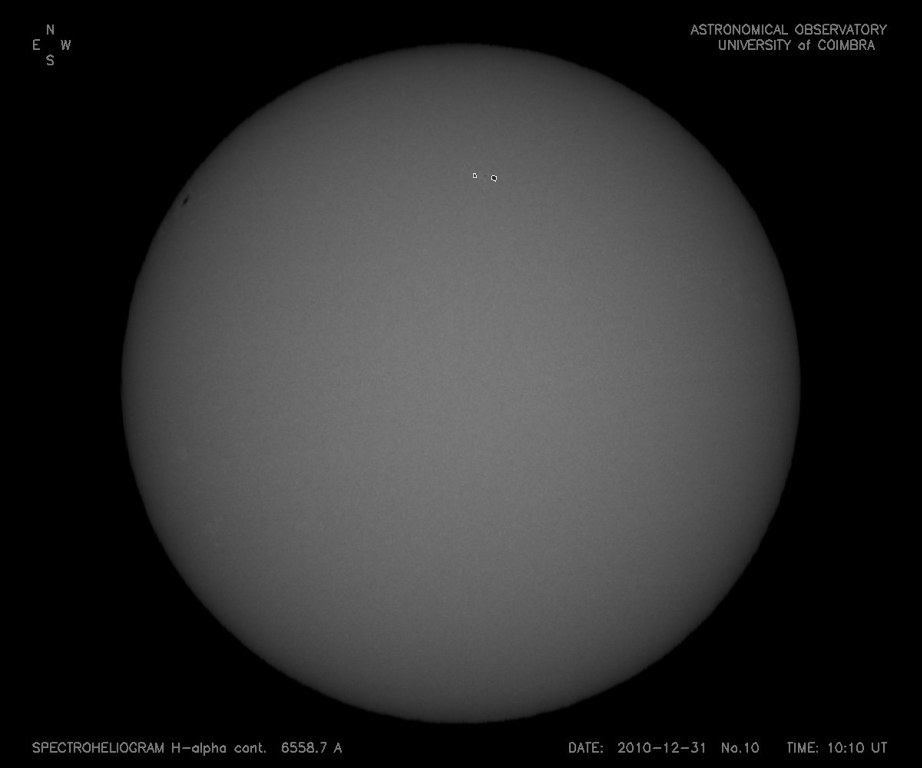}
	} 
	\subfloat[\label{limboriginal2}]{%
		\includegraphics[width=0.5\textwidth]{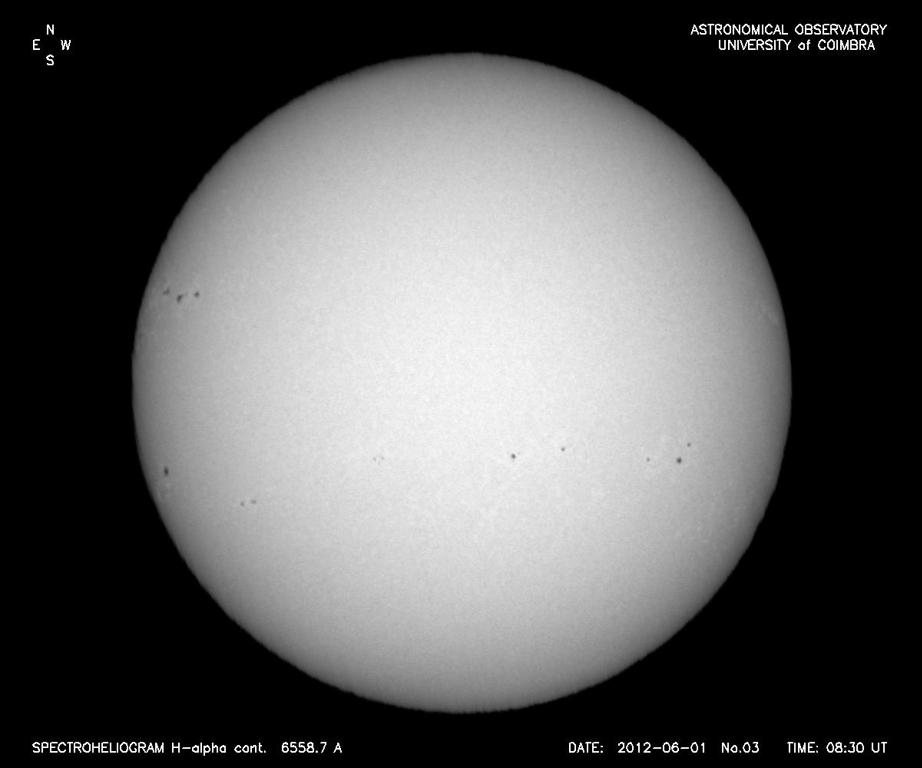}
	} \\ \hfill
	\subfloat[\label{limbmm2}]{%
		\includegraphics[width=0.5\textwidth]{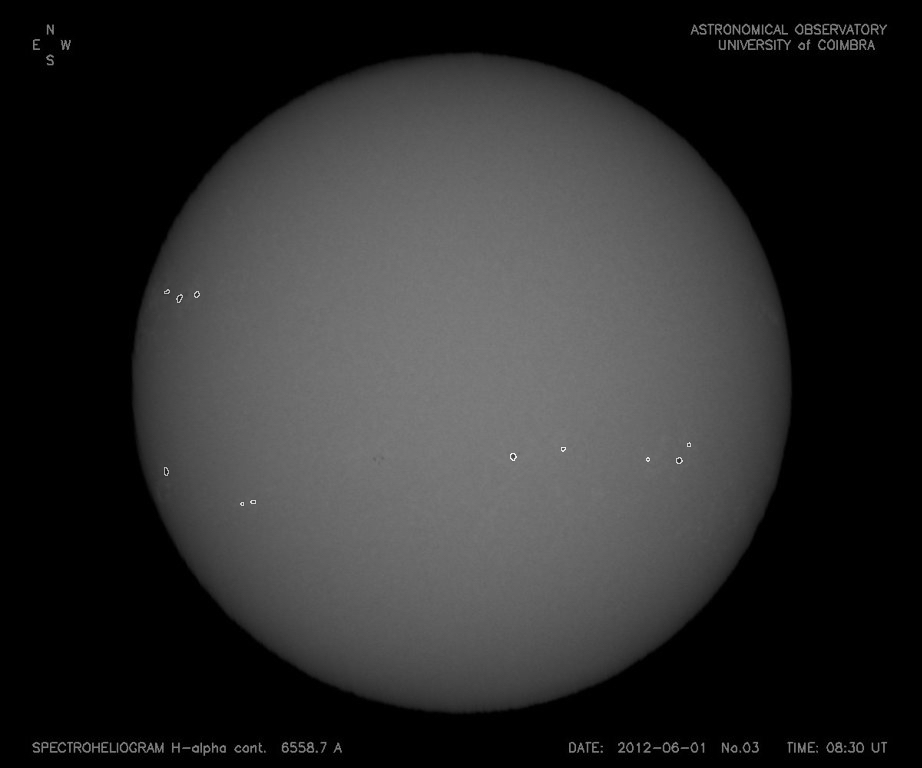}
	}
	\subfloat[\label{limbpi2}]{%
		\includegraphics[width=0.5\textwidth]{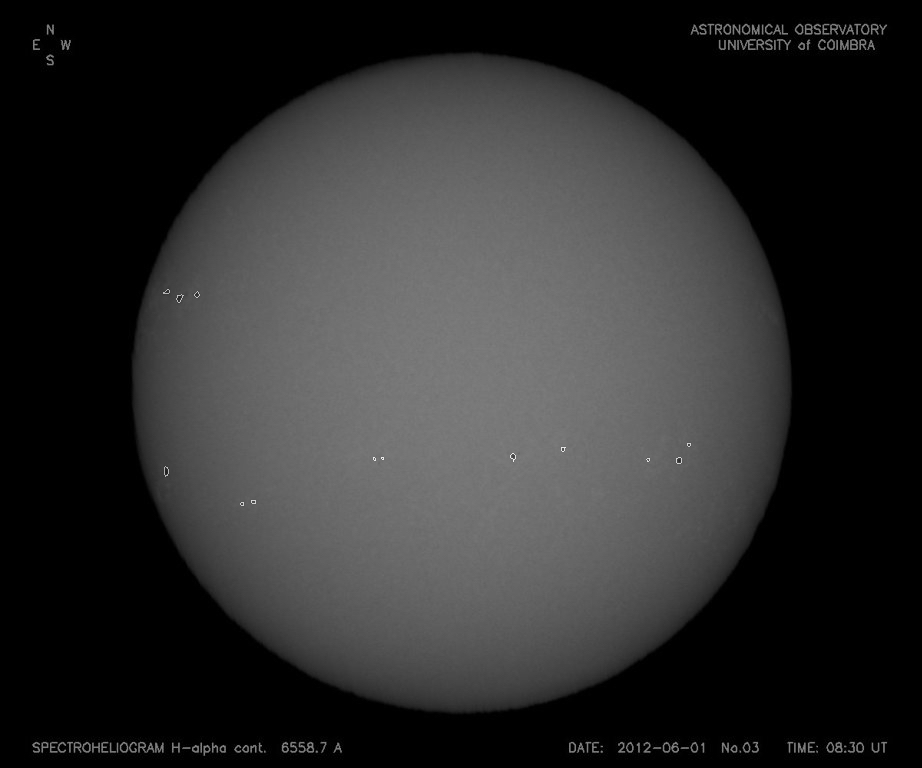}
	} \\	
	\caption{Examples of detection results on spectroheliograms with sunspots in the limb: (a) original image of 31/12/2010; (b) result of the morphological detection algorithm of the image of 31/12/2010; (c) result of the algorithm of detection based on pixels intensities of the image of 31/12/2010; (d) original image of 1/6/2012; (e) result of the morphological detection algorithm of the image of 1/6/2012; (f) result of the algorithm of detection based on pixels intensities of the image of 1/6/2012.}
	\label{limb}
\end{figure}

The method based on pixel intensity takes into account, in the pre-processing phase, the removing of limb-darkening so that sunspots on the solar disk’s limb can be detected. However, although not doing this removal, the method of mathematical morphology presents better results regarding the detection of spots in the limb, as we can see in the examples shown in Fig. \ref{limb}. This is due to the fact that mathematical morphology is based on the forms of the elements to be detected, and not their intensities. 

\begin{figure}[!ht]
	\subfloat[ \label{manyspotsmm1}]{%
		\includegraphics[width=0.5\textwidth]{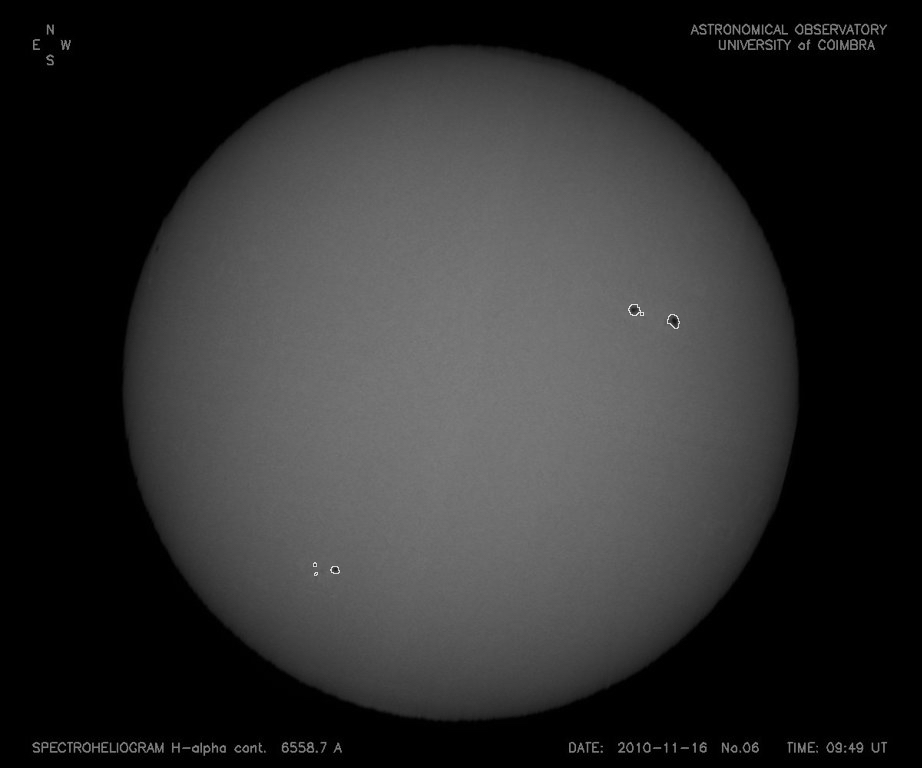}
	}  
	\subfloat[\label{manyspotsmm2}]{%
		\includegraphics[width=0.5\textwidth]{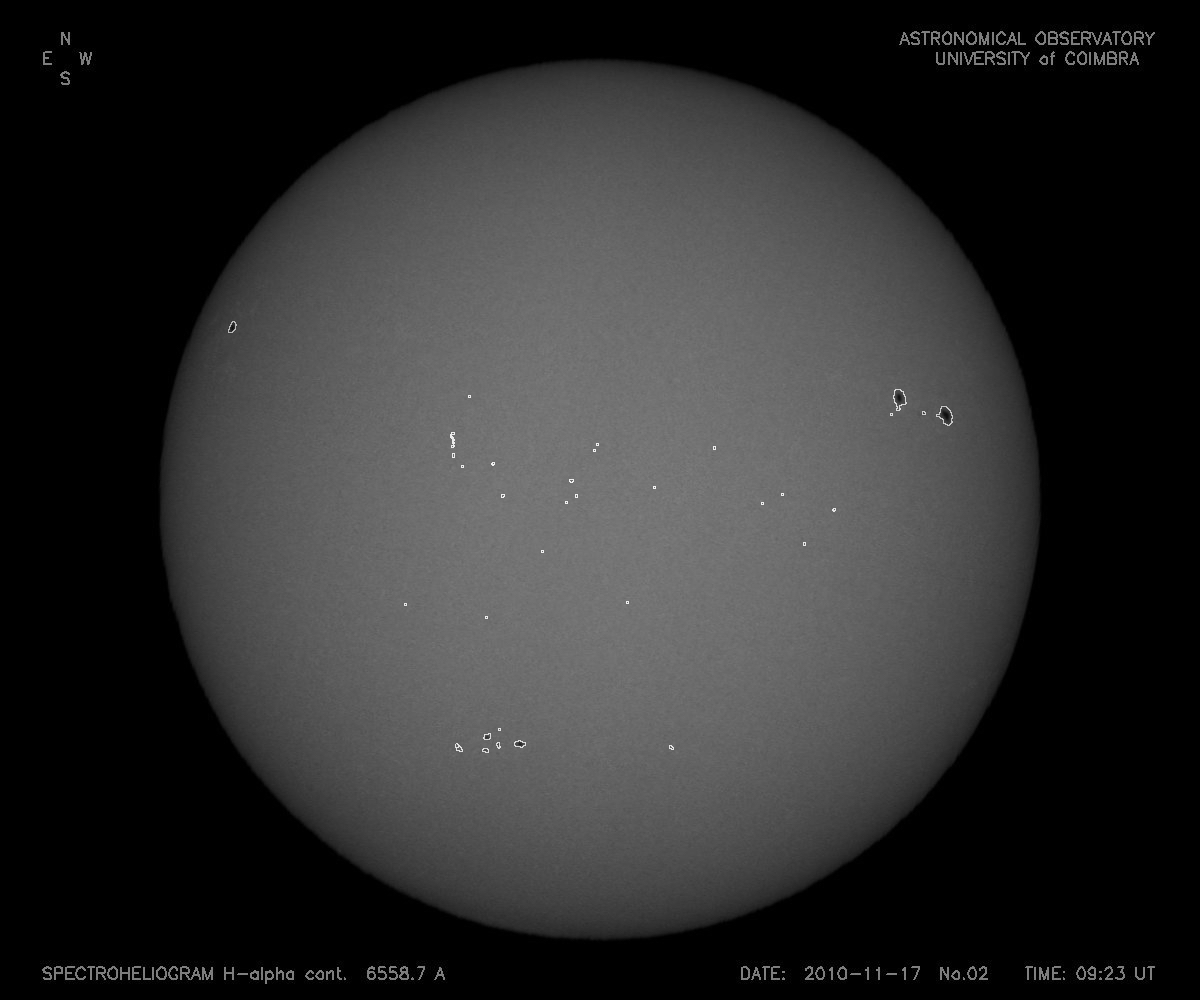}
	}  \\ \\ \hfill

	\qquad\qquad\qquad\qquad\quad	
	\subfloat[\label{manyspotsmm3}]{%
		\includegraphics[width=0.5\textwidth]{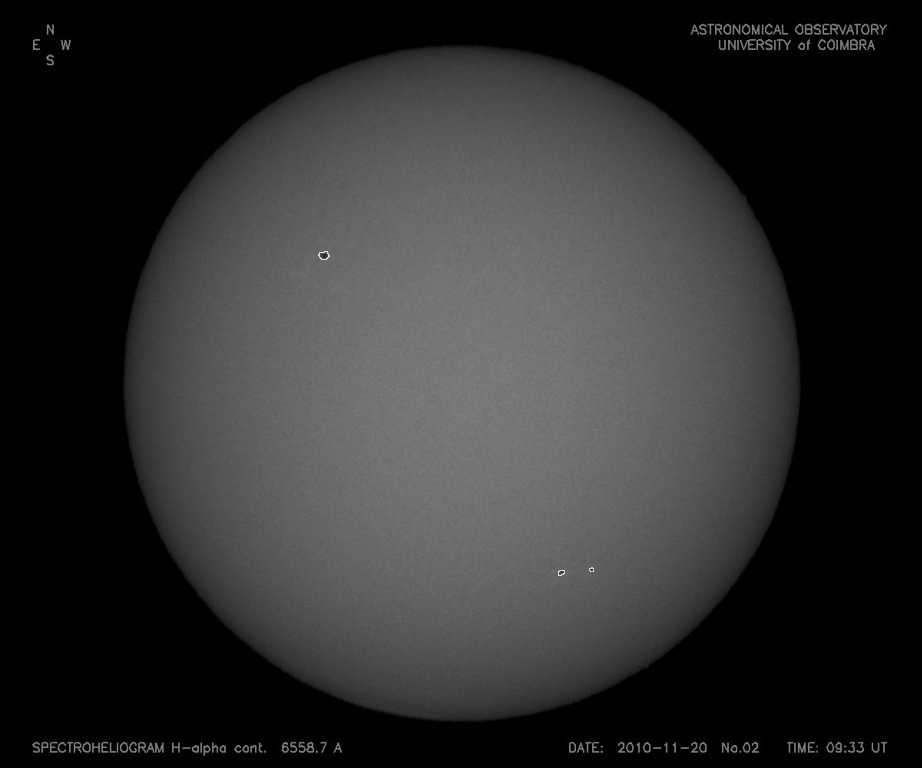}
	}
	\caption{Example of over detection and the results of the consecutive days using de method based on mathematical morphology: (a) result of the detection in the spectroheliogram of 16/11/2010; (b) result of the detection in the spectroheliogram of 17/11/2010 (over detection); (c) result of the detection in the spectroheliogram of 20/11/2010.}
	\label{mmmanyspots}
\end{figure}

\begin{figure}[!ht]
	\subfloat[ \label{manyspotspi1}]{%
		\includegraphics[width=0.5\textwidth]{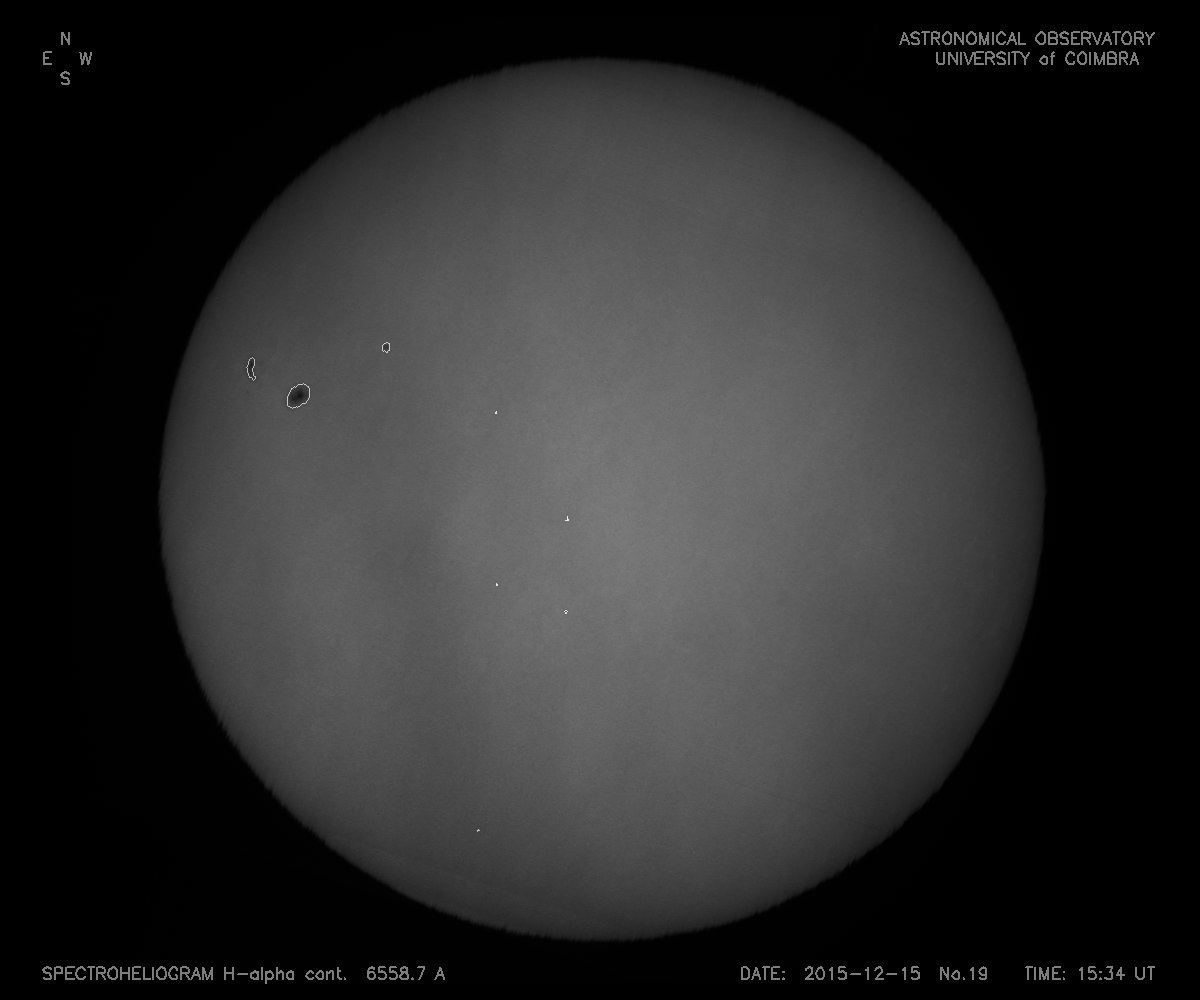}
	}  
	\subfloat[\label{manyspotspi2}]{%
		\includegraphics[width=0.5\textwidth]{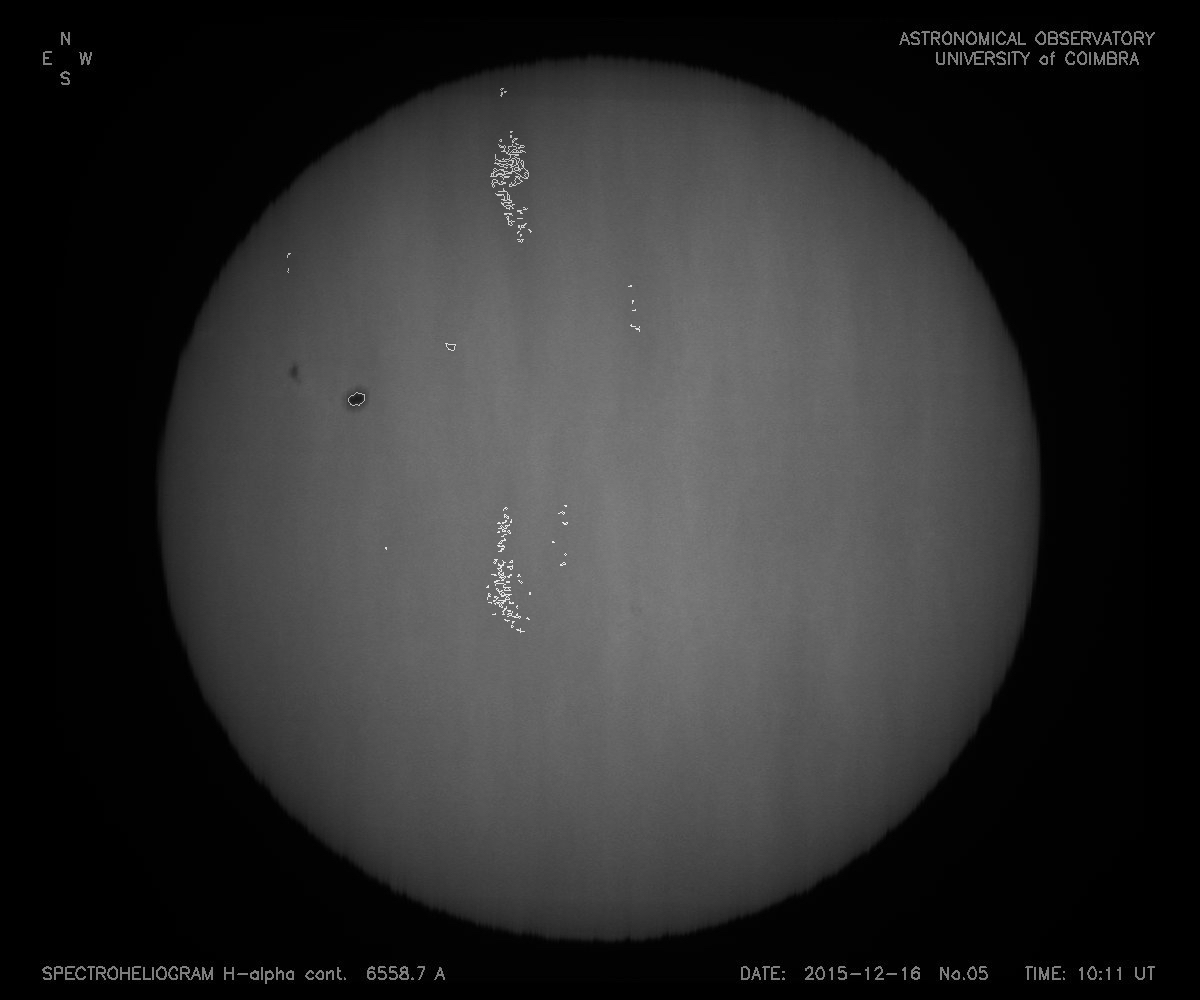}
	} \\ \\ \hfill

	\qquad\qquad\qquad\qquad\quad
	\subfloat[\label{manyspotspi3}]{%
		\includegraphics[width=0.5\textwidth]{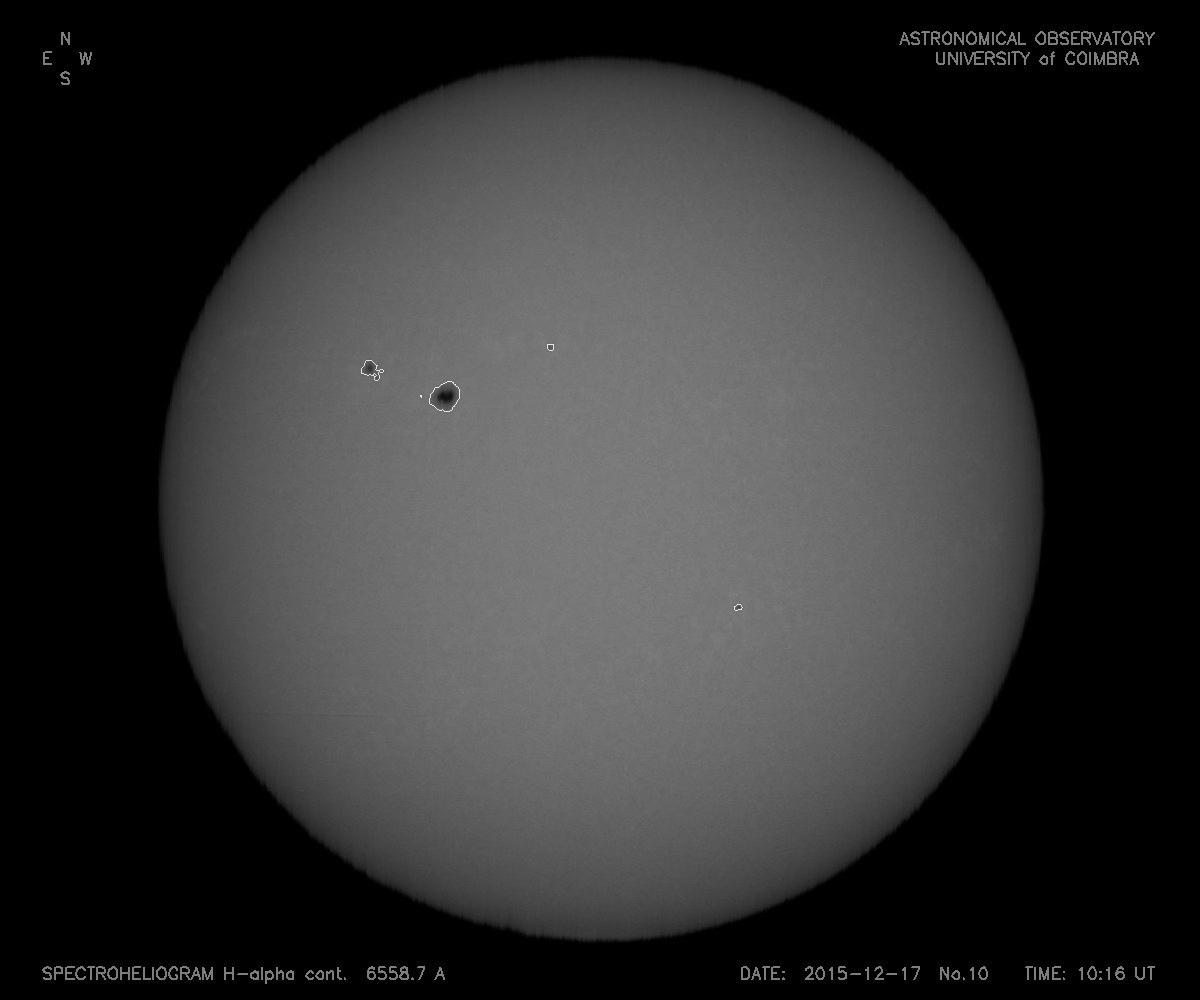}
	}
	\caption{Example of over detection and the results of the consecutive days using de method based on pixels intensities: (a) result of the detection in the spectroheliogram of 15/12/2015; (b) result of the detection in the spectroheliogram of 16/12/2015 (over detection); (c) result of the detection in the spectroheliogram of 17/12/2015.}
	\label{pimanyspots}
\end{figure}

There are a few cases on which an over detection occurs in both methods. Concerning the method based on mathematical morphology, this problem of over detection occurs due to the size of the structuring element in the top-hat transform. The value used was chosen in order to have a trade-off among all the images of the set considered, in order to reduce false positives. As future work we intend to solve these cases through a post processing. Concerning the method based on pixel intensity, the over detection problem is related with the presence of clouds on the images. However, given the slow rotation of the sun and the slow variation of the sunspots, the analysis of images acquired on consecutive days, or almost consecutive, is a good way to evaluate the results, as can be seen in Figs. \ref{mmmanyspots} and \ref{pimanyspots}.

Concerning the umbra-penumbra segmentation, the difference between the two methods is not very significant. The pixel intensity method evidences to be a little more efficient, with an OA of 87.89\%, whereas the method based on mathematical morphology presents an OA of 86.25\%. Fig. \ref{segmentationresults} shows an example of the segmentation made by the two methods. 

\begin{figure}
	\centering
	\subfloat[ \label{segoriginal}]{%
		\includegraphics[width=0.5\textwidth]{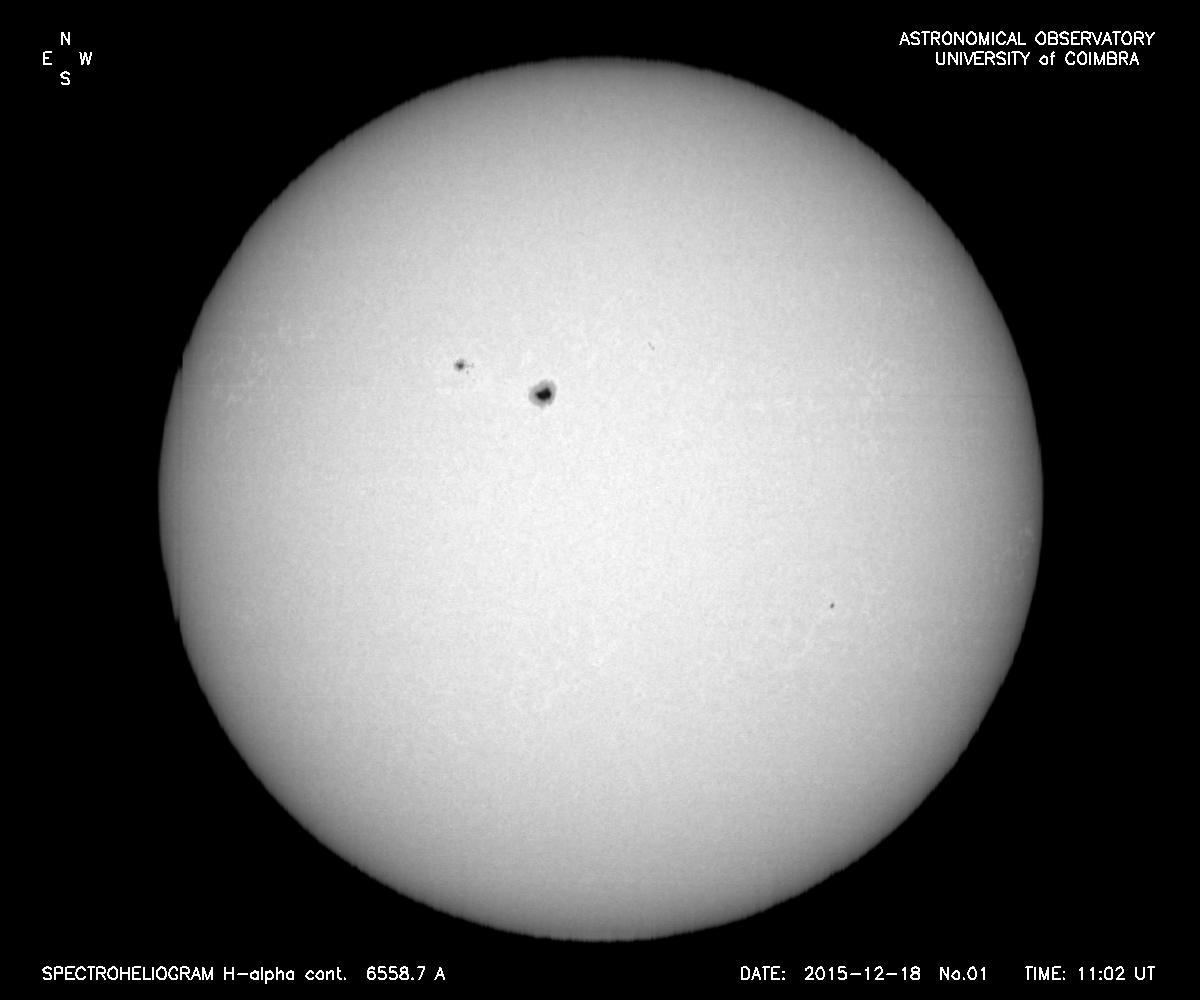}
	} \\ 
	\subfloat[\label{segmm}]{%
		\includegraphics[width=0.5\textwidth]{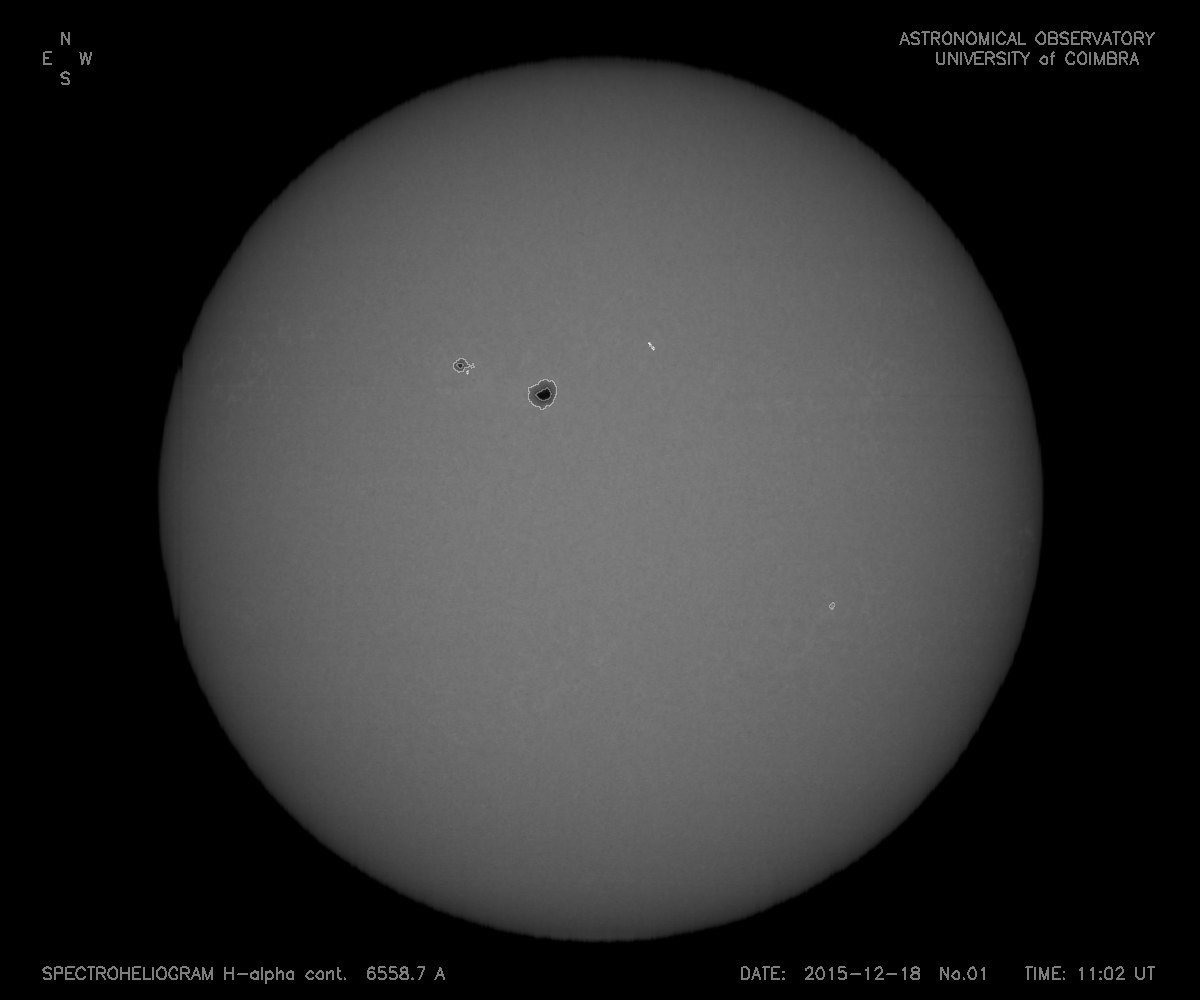}
	} \\ 
	\subfloat[\label{segpi}]{%
		\includegraphics[width=0.5\textwidth]{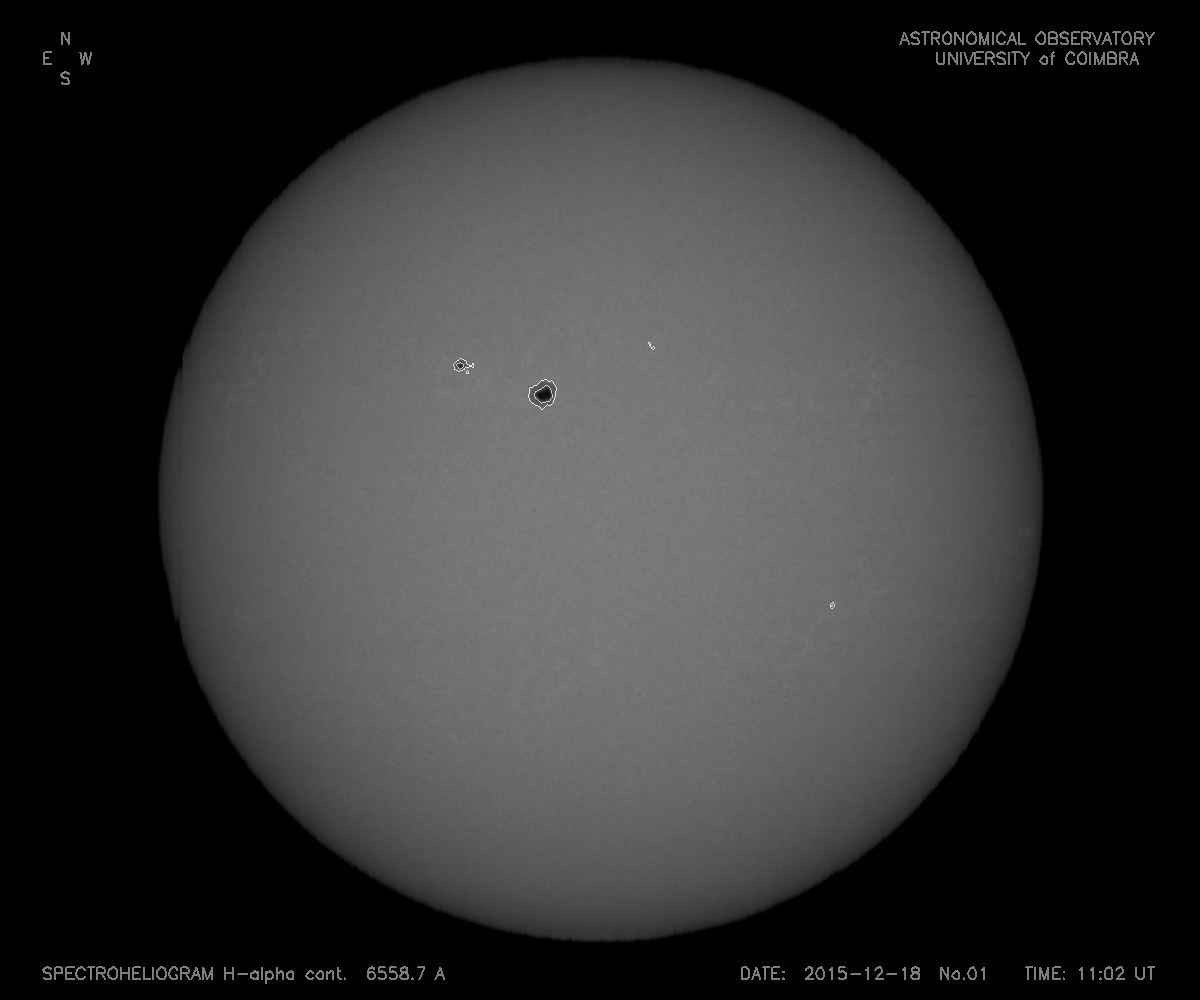}
	} \\
	\phantomcaption
\end{figure}

\begin{figure}
	\ContinuedFloat		
	\centering
	\subfloat[ \label{segoriginalcropped}]{%
		\includegraphics[width=0.5\textwidth]{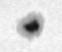}
	} \\
	\subfloat[\label{segmmcropped}]{%
		\includegraphics[width=0.5\textwidth]{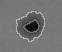}
	}\\
	\subfloat[\label{segpicropped}]{%
		\includegraphics[width=0.5\textwidth]{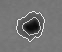}
	}
	\caption{Example of segmentation results: (a) original image of 18/12/2015; (b) result of the morphological detection algorithm of the original image; (c) result of the algorithm of detection based on pixels intensities of the original image; (d) original image of 18/12/2015 cropped; (e) result of the morphological detection algorithm of the cropped image ; (f) result of the algorithm of detection based on pixels intensities of the cropped image.}
	\label{segmentationresults}
\end{figure}

\subsection{Comparison with Other Sources}

The performance of both approaches was also compared with other sources, namely with the sunspots identified by other observatories which are published in solar catalogues. We compared our results with the same set of images from the Heliophysics Feature Catalogue and the Word Data Center SILSO.\\

\textit{Comparison with the Heliophysics Feature Catalogue}

\begin{figure}[!ht]
	\centering
	\subfloat[ \label{g1}]{%
		\includegraphics[width=0.8\textwidth]{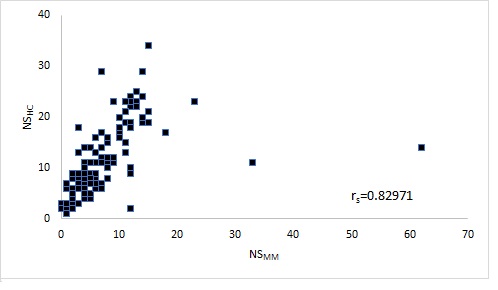}
	} \\
	\subfloat[\label{g2}]{%
		\includegraphics[width=0.8\textwidth]{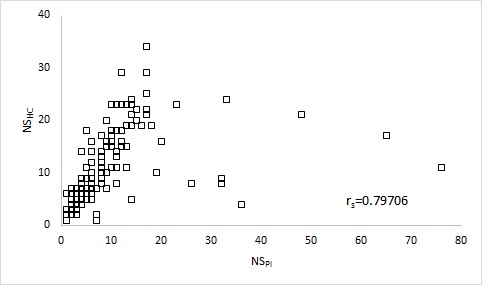}
	}
	\caption{Comparison with the Heliophysics Feature Catalogue: (a) MM method; (b) PI method. Spearman's correlation coefficients are shown.}
	\label{graph1}
\end{figure}

The Heliophysics Feature Catalogue is available at the Solar System Survey BASS2000 website http://bass2000.obspm.fr/. BASS2000 archives and distributes ground-based solar observations provided by various instruments, like the THEMIS telescope, the Nancay Radioheliograph, the Nancay Decametric Array, the spectroheliograms of Meudon and Coimbra, the Pic du Midi H-alpha Coronograph, and Christian Latouche solar imager CLIMSO of Pic du Midi, the Uccle Solar Equatorial Table USET of the Royal Observatory of Belgium, among others. The data used in this work were obtained at the BASS2000. The relations of the number of sunspots detected applying the mathematical morphology method to Coimbra’s images ($\textrm{NS}_{\textrm{MM}}$), and the pixel intensity method ($\textrm{NS}_{\textrm{PI}}$), with the number of sunspots available at the Heliophysics Feature Catalogue ($\textrm{NS}_{\textrm{HC}}$) are shown on Fig. \ref{graph1}. The Spearman-rank correlation between the number of detected sunspots by each of these methods and $\textrm{NS}_{\textrm{HC}}$ is strong and significant (respectively, 0.83 and 0.80; both significant at a 0.000001 level).

\textit{Comparison with the World Data Center SILSO}

\begin{figure}[!ht]
	\centering
	\subfloat[ \label{g3}]{%
		\includegraphics[width=0.8\textwidth]{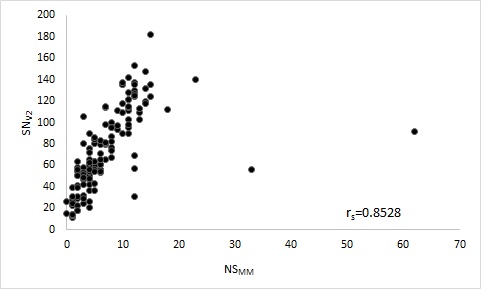}
	} \\
	\subfloat[\label{g4}]{%
		\includegraphics[width=0.8\textwidth]{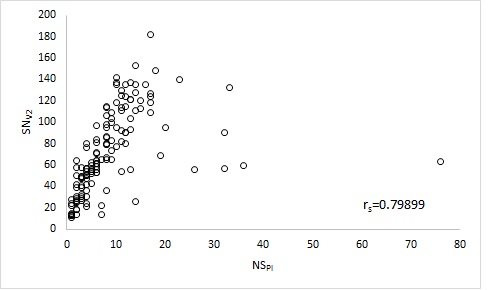}
	}
	\caption{Comparison with the World Data Center SILSO: (a) MM method; (b) PI method. The Spearman's correlation coefficients are shown.}
	\label{graph2}
\end{figure}

The World Data Center SILSO - Sunspot Index and Long-term Solar Observations - is a product developed by the Operational Directorate “Solar Physics and Space Weather”, also known internationally as the Solar Influences Data Analysis Center (SIDC). SIDC is a department of the Royal Observatory of Belgium. The data provided by SILSO includes, among others, the Sunspot Number (version 2.0)- hereafter $\textrm{SN}_{\textrm{V2}}$- calculated as $\textrm{SN}_{\textrm{V2}}$ = 10G + S, where G is the number of sunspot groups and S is the number of single sunspots registered. The data used in this work were obtained at the website http://www.sidc.be/silso/.  
The number of sunspots detected by the two algorithms ($\textrm{NS}_{\textrm{MM}}$ and $\textrm{NS}_{\textrm{PI}}$) developed in this work were compared with those included in the SILSO catalogue ($\textrm{SN}_{\textrm{V2}}$) (Fig. \ref{graph2}). Although the $\textrm{SN}_{\textrm{V2}}$ has another counting scheme relatively to the single number of sunspots detected using Coimbra’s images, a comparison between the two is possible because the variations should follow the same trend. A strong and significant correlation exists between our mathematical morphology (MM) and pixel intensity (PI) methods and the $\textrm{SN}_{\textrm{V2}}$ (respectively, 0.85 and 0.80; both significant at the 0.000001 level).

\textit{Comparison of temporal evolution between catalogues}

\begin{figure}[!ht]
	\centering
	\includegraphics[width=0.9\textwidth]{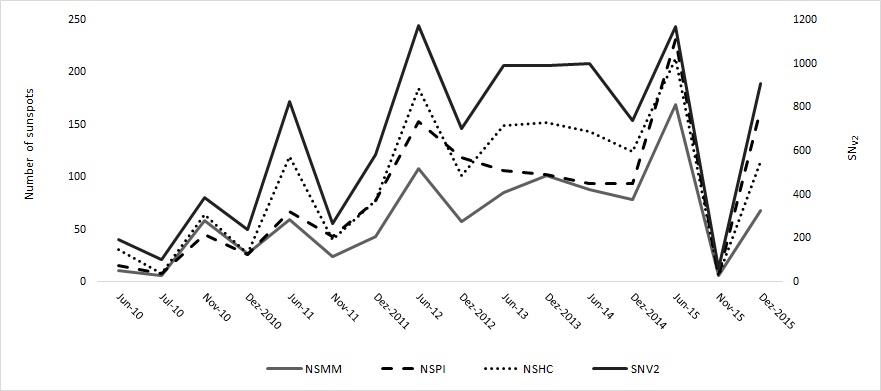}
	\caption{Temporal evolution of the number of sunspots detected by the MM method ($\textrm{NS}_{\textrm{MM}}$), and the PI method ($\textrm{NS}_{\textrm{PI}}$), and the Heliophysics Feature Catalogue ($\textrm{NS}_{\textrm{HC}}$), and the SILSO catalogue ($\textrm{SN}_{\textrm{V2}}$).}
	\label{graph3}
\end{figure}

Fig. \ref{graph3} depicts the temporal evolution of the sunspot number calculated using Coimbra’s images by both methods, and the data available in the Heliophysics Feature Catalogue, and in the SILSO catalogue. The temporal behavior of the two series computed using Coimbra’s images is similar to the one of the other catalogues. 

\section{Conclusions}
\label{conclusions}
The existence of large amounts of data needs to be processed and analyzed in a fast and efficient way. This is the case for the Observatory of Coimbra, which possesses a large collection of solar observations. Thus, in this work, two different approaches were developed in order to decide the best way to handle with the Coimbra collection and to detect, automatically, the sunspots: a method based on mathematical morphology (MM) and a method based on pixel intensity (PI). Both methods are significantly different, since one is based on the geometric shape of the sunspots and other is based on the digital level of the pixels. 

The ground-based observations, when compared with spatial missions, presents additional difficulties due to the atmospheric effects. Another important aspect, but common to all solar synoptic observations, is the limb darkening effect. Most of the automatic approaches found in the literature, consider the removal of limb darkening one of the firsts steps of pre-processing analysis. In this work, the method based on pixel intensity (PI) needs to take into account the removal of atmospheric effects and corrects for the limb darkening. Regarding the morphological approach (MM), since in its essence the most important aspect is the geometric shape of the objects, there is no need to pre-processing the images to remove noise and correct for limb darkening, making it more appealing. Moreover, the metrics used to evaluate the performance and robustness of the two approaches, demonstrate a better efficiency of the morphological method (MM). 

In order to investigate the reliability of the methods developed on this work, a comparison was made between the results obtained by these methods and those shown by other catalogues. The Heliophysics Features Catalogues and the SILSO catalogue were chosen for this analysis since they provide data about the number of sunspots. Comparing the number of sunspot detections by our methods with the numbers on the aforementioned catalogues shows very strong and significant correlations, despite some outliers, and the temporal evolution of the sunspot number detected shows the same behavior. These findings reinforce the robustness of both methods given their very high correlation with other catalogues. Also, the development of an automatic method allows to resume the production of the Coimbra Observatory solar catalogs, which was interrupted in 1986. 
 
\section*{Acknowledgements}
The authors acknowledge financial support by the Portuguese Government through the Foundation for Science and Technology - FCT, by CITEUC Funds (UID/Multi/00611/2019), by CMUC Funds (UID/MAT/00324/2019), and by FEDER - European Regional Development Fund through COMPETE 2020 - Operational Programme Competitiveness and Internationalization.  S. Carvalho's work has been funded by FCT grant SFRH/BD/107894/2015. T. Barata and A. Lourenço acknowledge financial support from the project ReNATURE - Valuation of Endogenous Natural Resources in the Central Region (CENTRO-01-0145-FEDER-000007). The authors are also very grateful to Pedro Pina, for all the suggestions concerning the morphological operators and the evaluation metrics used, and Adriana Garcia, for the precious help in the construction of the ground-truth images.

\section*{REFERENCES}
\bibliographystyle{model2-names}
\bibliography{bib_file}







\end{document}